\documentclass[a4paper, 12pt]{article}
\usepackage{style_new_designs}

\usepackage{enumitem}
\usepackage[colorlinks=true, allcolors=blue]{hyperref}
\usepackage{setspace} 

\date{\today}
\title{Covariate Adjustment in Stratified Experiments\footnote{I thank the anonymous referees for helpful suggestions during the revision process.}}
\author{Max Cytrynbaum\footnote{Yale Department of Economics. Correspondence: max.cytrynbaum@yale.edu}}

\begin{document}
\maketitle

\begin{abstract}
\singlespacing
This paper studies covariate adjusted estimation of the average treatment effect in stratified experiments. 
We work in a general framework that includes matched tuples designs, coarse stratification, and complete randomization as special cases.
Regression adjustment with treatment-covariate interactions is known to weakly improve efficiency for completely randomized designs. 
By contrast, we show that for stratified designs such regression estimators are generically inefficient, potentially even increasing estimator variance relative to the unadjusted benchmark.
Motivated by this result, we derive the asymptotically optimal linear covariate adjustment for a given stratification.
We construct several feasible estimators that implement this efficient adjustment in large samples.
In the special case of matched pairs, for example, the regression including treatment, covariates, and pair fixed effects is asymptotically optimal.
We also provide novel asymptotically exact inference methods that allow researchers to report smaller confidence intervals, fully reflecting the efficiency gains from both stratification and adjustment.
Simulations and an empirical application demonstrate the value of our proposed methods. \\
\end{abstract}



\onehalfspacing
\newpage

\section{Introduction} \label{section:introduction}

This paper studies covariate adjusted estimation of the average treatment effect (ATE) in stratified experiments. 
Researchers often make use of both stratified treatment assignment and ex-post covariate adjustment to improve the precision of experimental estimates. 
Indeed, out of a survey of over $50$ experimental papers published in the AER and AEJ between 2018-2023, we found that $57\%$ use stratified randomization, and $80\%$ used some form of ex-post covariate adjustment. 
An influential paper by \cite{lin2013} showed in a design-based setting that the regression estimator with full treatment-covariate interactions is always asymptotically weakly more efficient than difference of means estimation for completely randomized designs.  
\cite{negi2021} extended these results to estimation of the ATE using data sampled from a superpopulation. 
However, questions remain about the interaction between stratification and regression adjustment and the implications of combining these methods for both estimator efficiency and the power and validity of inference methods.  
To study these questions, we work in the stratified randomization framework of \cite{cytrynbaum2023}, which includes matched tuples designs (e.g. matched pairs), coarse stratification, and complete randomization as special cases. 

We show that the \cite{lin2013} interacted regression adjustment is generically inefficient in the family of linearly adjusted estimators, with asymptotic efficiency only in the limiting case of complete randomization. 
Motivated by this finding, we characterize the efficient linear covariate adjustment for a given stratified design, providing several new estimators that achieve the optimal variance. 

Our first result derives the optimal linear adjustment coefficient for a given stratification. 
We show that asymptotically the interacted regression estimator uses the wrong objective function, minimizing a marginal variance objective that is totally insensitive to the stratification. 
By contrast, the optimal adjustment coefficient minimizes a \emph{mean-conditional} variance objective, conditional on the covariates used to stratify. 
Intuitively, the efficient covariate adjustment is tailored to the stratification, ignoring fluctuations of the estimator that are predictable by the stratification covariates.
Section \ref{section:efficiency-semiparametric-regression} draws an interesting connection with partially linear regression (\cite{robinson88}), showing that efficient linear adjustment of a stratified design is asymptotically equivalent to doubly-robust semiparametric adjustment of an iid\ design.
Intuitively, stratification contributes the nonparametric component of the semiparametric adjustment function.  

Our second set of results develops feasible versions of the optimal linear adjustment derived in Section \ref{section:optimal-adjustment}. 
First, we show that if the conditional expectation of the adjustment covariates is linear in a known set of transformations of the stratification variables, then adding the latter to the interacted regression restores optimality.
Next we relax this assumption, providing four different regression estimators that are asymptotically efficient under weak conditions. 
For matched pairs experiments or in settings with limited treatment effect heterogeneity, the non-interacted regression with a full set of pair fixed effects is asymptotically efficient. 
More generally, we show asymptotic optimality of within-stratum (inconsistently) partialled versions of the Lin and tyranny-of-the-minority estimators (\cite{lin2013}).
We also define a ``group OLS'' estimator, extending a proposal of \cite{imbens2015} for matched pairs experiments to a larger class of designs. 
We show that this group OLS estimator is also asymptotically optimal. 

Our final contribution is to develop novel asymptotically exact inference methods for covariate adjusted estimation under stratified designs.
Confidence intervals based on the usual heteroskedasticity robust variance estimator are known to be conservative in stratified experiments (\cite{bai2021inference}).
By contrast, the coverage probabilities of our proposed confidence intervals converge to the specified nominal level, with no overcoverage. 
Our approach applies to a generic family of linear covariate adjustments and randomization schemes, including as special cases non-interacted regression adjustment, the \cite{lin2013} interacted regression, and all of the other estimators considered in this paper.
Simulations and an empirical application to the experiment in \cite{baysan2022} suggest that the usual robust confidence intervals can substantially overcover in stratified experiments, while our confidence intervals have close to nominal coverage. 

We present several extensions of our main results in the appendix. 
In the first, we consider estimation and inference in stratified experiments with noncompliance.
As a simple corollary of our results on ATE estimation, we characterize the optimal linearly adjusted Wald estimator for the LATE (\cite{imbens1994}), construct a feasible implementation of the efficient adjustment, and provide asymptotically exact inference methods. 
We also study efficient linear adjustment for finely stratified designs with non-constant treatment proportions, as in \cite{cytrynbaum2023}, and briefly consider the problem of efficient nonlinear adjustment.

There has been significant interest in treatment effect estimation under different experimental designs in the recent literature. 
Some papers studying covariate adjustment under stratified randomization include \cite{bugni2018inference}, \cite{fogarty2018b}, \cite{liu2020}, \cite{lu2022}, \cite{ma2020}, \cite{reluga2022}, \cite{wang2021}, \cite{ye2022}, \cite{zhu2022}, and \cite{chang2023}.
These works differ from our paper in at least one of the following ways: (1) studying inference on the sample average treatment effect (SATE) rather than the ATE in a superpopulation, (2) restricting to coarse stratification (stratum size going to infinity), or (3) proving weak efficiency gains but not optimality. 
In a finite population setting, \cite{zhu2022} shows asymptotic efficiency of a projection-based estimator numerically equivalent to the ``partialled Lin'' approach considered in Section \ref{section:partialled-lin}. 
In the same setting, \cite{lu2022} prove efficiency of a tyranny-of-the-minority style regression similar but not equivalent to one the considered in Section \ref{section:tyranny-of-the-minority}.
Both papers give conservative inference on the SATE, while we provide asymptotically exact inference on the ATE using a generalized pairs-of-pairs (\cite{abadie2008}) style approach. 
Remarks \ref{remark:partialled-lin-comparison} and \ref{remark:tom-comparison} in Section \ref{section:generic-efficiency} below provide a detailed comparison. 

Relative to the above papers, the superpopulation framework considered here creates some new technical challenges.
For example, as pointed out in \cite{bai2021inference}, matching units into data-dependent strata post-sampling produces a complicated dependence structure between the treatment assignments and random covariates.
We deal with this using a tight-matching condition (Equation \ref{equation:homogeneity}) and martingale CLT analysis similar to \cite{cytrynbaum2022local}. 
This setting also has analytical advantages, which allow us to establish new conceptual results.
For example, the population level characterization of the optimal adjustment coefficent in Section \ref{section:optimal-adjustment} allows us to give explicit necessary and sufficient conditions for the efficiency of several commonly used regression estimators.
The efficiency of interacted regression under a ``rich covariates'' condition, as well as the equivalence between optimal linear adjustment of stratified designs and doubly-robust semiparametric adjustment appear to be new observations in this literature. 
To the best of our knowledge, we give the first asymptotically exact inference on the ATE for general covariate adjusted estimators under finely stratified randomization. 

Independently, \cite{bai2023adjustment} study covariate adjustment under matched pairs randomization in a superpopulation framework. 
They also find that regression adjustment without pair fixed effects may be inefficient, while adding pair fixed effects restores efficiency. 
Relative to our work, they additionally study regularized regression adjustment under high-dimensional asymptotics, which we do not consider.
By contrast, we study more general forms of stratification, allowing coarse and fine stratification with arbitrary treatment proportions $\propfn \not = 1/2$.
For such designs, the strata fixed effects estimator may still be inefficient.
To fix this, we introduce novel forms of linear adjustment that are efficient under general stratified designs.

The rest of the paper is organized as follows. 
In Section \ref{section:framework} we define notation and introduce the family of stratified designs that we will consider throughout the paper.
Section \ref{section:results} gives our main results, characterizing optimal covariate adjustment and constructing efficient estimators.
Section \ref{section:inference} provides asymptotically exact inference on the ATE for generic linearly adjusted estimators.  
In Sections \ref{section:simulations} and \ref{section:empirical}, we study the finite sample properties of our method, including both simulations and an empirical application to the experiment in \cite{baysan2022}.
Section \ref{section:conclusion} concludes with some recommendations for practitioners.

\section{Framework and Stratified Designs} \label{section:framework}
For a binary treatment $d \in \{0, 1\}$, let $Y_i(1)$, $Y_i(0)$ denote the treated and control potential outcomes, respectively. 
For treatment assignment $\Di$, let $Y_i = Y_i(\Di) = \Di Y_i(1) + (1-\Di) Y_i(0)$ be the observed outcome.
Let $X_i$ denote covariates.
Consider data $(X_i, Y_i(1), Y_i(0))_{i=1}^n$ sampled i.i.d. from a superpopulation of interest. 
We are interested in estimating the average treatment effect in this population, $\ate = E[Y(1) - Y(0)]$. 
After sampling units $i = 1, \dots, n$, treatments $\Dn$ are assigned by stratified randomization. 
In particular, we use the ``local randomization'' framework introduced in \cite{cytrynbaum2022local}.

\begin{defn}[Local Randomization] \label{defn:local_randomization}
Let treatment proportions $\propfn = a/k$ with $\gcd(a,k) = 1$.\footnote{$\gcd(a,k)$ stands for greatest common divisor.}
Suppose that $n$ is divisible by $k$ for notational simplicity.
Partition the experimental units into $n/k$ groups $\group$ with $\{1, \dots, n\} = \bigcup_{\group} \group$ disjointly and $|\group| = k$. 
Let $\psi(X) \in \mr^{d_{\psi}}$ denote a vector of stratification variables. 
Suppose that the groups that satisfy a homogeneity condition with respect to $\psi(X)$ such that 
\begin{equation} \label{equation:homogeneity}
\frac{1}{n} \sum_{\group} \sum_{i,j \in \group} |\psi(X_i) - \psi(X_j)|_2^2 = \op(1).
\end{equation}
Require that the groups only depend on the stratification variables $\psin$ and data-independent randomness $\permn$, so that $\group = \group(\psin, \permn)$ for each $\group$.
Independently for each $|\group| = k$, draw treatment variables $(\Di)_{i \in \group}$ by setting $\Di = 1$ for exactly $a$ out of $k$ units, completely at random.
For a stratification satisfying these conditions, we denote $\Dn \sim \localdesigncond(\psi, \propfn)$.
\end{defn}

\begin{ex}[Matched Tuples]
Equation \ref{equation:homogeneity} requires units in a group to have similar $\psi(X_i)$ values and can be thought of as a tight-matching condition. 
\cite{cytrynbaum2023} provides an iterative pairing algorithm to match units into groups that provably satisfy this condition for any $k$.
Drawing treatments $\Dn \sim \localdesigncond(\psi, \propfn)$  produces a ``matched k-tuples'' design for $\propfn = a/k$. 
Matched pairs corresponds to the case $\propfn = 1/2$.
\end{ex}

\begin{ex}[Complete Randomization] \label{ex:complete-randomization}
We say variables $\Dn$ are completely randomized with treatment probability $\propfn$ if $\Dn$ is drawn uniformly from all vectors $d_{1:n}$ with $d_i=1$ for exactly proportion $\propfn$ of the units. 
Formally, $P(\Dn=d_{1:n}) = 1/\binom{n}{\propfn n}$ for all such vectors. 
We denote complete randomization by $\Dn \sim \crdist(\propfn)$. 
Complete randomization may be obtained in our framework by setting $\psi = 1$ and forming groups $|\group| = k$ at random, which automatically satisfies Equation \ref{equation:homogeneity}.
For example, assigning $2$ out of $3$ units in each group to treatment gives a ``random matched triples'' representation of complete randomization with $\propfn = 2/3$.
\end{ex}

\begin{remark}[Coarse Stratification]
Similarly, coarse stratification with large fixed strata $S(X) \in \{1, \dots, m\}$ can also be obtained in our framework by setting $\psi(X) = S(X)$ and matching units with identical $S(X)$ values into groups at random.
Because of this, our framework enables a unified asymptotic analysis for a wide range of stratifications.
\end{remark}

\textbf{Experiment Timing:} Suppose that the experimenter does the following 
\begin{enumerate}[label={(\arabic*)}, itemindent=.5pt, itemsep=.2pt] 
\item Samples units and observes their baseline covariates. 
\item Partitions the units into data-dependent groups $\group = \group(\psin, \permn)$ that satisfy Equation \ref{equation:homogeneity} for some stratification variables $\psi(X)$. 
\item Draws treatment assignments $\Dn \sim \localdesigncond(\psi, \propfn)$, observes outcomes $Y_i(\Di)$, and forms an estimate of the $\ate$, potentially adjusting for covariates $h(X)$.
\end{enumerate}

We are agnostic about the exact time at which the covariates are observed, subject to the constraints above.
For example, it could be that only $\psi(X)$ is observed at the design stage, while the full vector $X$ is collected later with the outcomes, and the experimenter chooses to adjust for $h(X) \subseteq X$.
Alternatively, the full vector $X$ could be observed at the design stage, but the experimenter chooses to only stratify on $\psi(X)$, and adjusts for $h(X) \subseteq X$ at step (3).
We may or may not have $\psi(X) \sub h(X)$.\footnote{Our asymptotic framework lets $h(X)$, $\psi(X)$ be fixed as $n \to \infty$.} \\

Consider the unadjusted estimator given by the coefficient $\est$ on $D$ in the regression $Y \sim 1 + D$.
Before discussing covariate adjustment, we first state a helpful variance decomposition for $\est$ that will be used extensively below.
Let $\catefn(X) = E[Y(1) - Y(0) | X]$ denote the conditional average treatment effect (CATE) and $\hk_d(X) = \var(Y(d) | X)$ the heteroskedasticity function.
Define the \emph{balance function} 
\begin{equation} \label{equation:balance_function}
\balancefn(X; \propfn) = E[Y(1)|X] \left (\frac{1-\propfn}{\propfn} \right)^{1/2} + \, E[Y(0)|X] \left(\frac{\propfn}{1-\propfn}\right)^{1/2}.
\end{equation}

We often denote $\balancefn = \balancefn(X; \propfn)$ in what follows.
\cite{cytrynbaum2022local} shows that if $\Dn \sim \localdesigncond(\psi, \propfn)$ then $\rootn(\est - \ate) \convwprocess \normal(0, V)$ with 

\begin{equation} \label{equation:localvariance}
V = \var(\catefn(X)) + E[\var(\balancefn | \psi)] + E\left[\frac{\hk_1(X)}{\propfn} + \frac{\hk_0(X)}{1-\propfn}\right]. 
\end{equation}

The variance $V$ is in fact the \cite{hahn1998} semiparametric variance bound\footnote{\cite{armstrong2022} shows that this variance bound also holds for stratified designs.} for the $\ate$ (with covariates $\psi(X)$), providing a formal sense in which stratification does nonparametric regression adjustment ``by design.''   
The middle term is the most important for our analysis below.  
For example, in this notation the difference in asymptotic efficiency between stratifications $\psi_1$ and $\psi_2$ (for fixed $\propfn$) is simply $E[\var(\balancefn | \psi_1)] - E[\var(\balancefn | \psi_2)]$. 
Note also that $E[\var(\balancefn | \psi)] \leq \var(\balancefn)$ for any $\psi$, showing how stratification removes the variance due to fluctuations that are predictable by $\psi(X)$. \\ 

Moving beyond the difference of means estimator $\est$, suppose that at the analysis stage, the experimenter has access to covariates $h(X)$, which may strictly contain $\psi(X)$.
One may try to further improve the efficiency of $\ate$ estimation by regression adjustment using these covariates, either using standard the regression $Y \sim 1 + D + h$ or the regression $Y \sim 1 + D + h + Dh$ (with de-meaned covariates) studied in \cite{lin2013}.
We study the interaction between covariate adjustment and stratification in Section \ref{section:optimal-adjustment} below, characterizing the optimal linear adjustment. 

\section{Main Results} \label{section:results}
\subsection{Efficient Linear Adjustment in Stratified Experiments} \label{section:optimal-adjustment}

In this section, we begin by studying the efficiency of commonly used covariate-adjusted estimators of the $\ate$ under stratified randomization.
\cite{lin2013} showed that in a completely randomized experiment, equivalent to $\Dn \sim \localdesigncond(\psi, \propfn)$ with $\psi=1$, regression adjustment with full treatment-covariate interactions is asymptotically weakly more efficient than difference of means estimation.
\cite{negi2021} extended this result to ATE estimation in the superpopulation framework that we use in this paper. 
Interestingly, we show that this result is atypical. 
For a general stratified experiment with $\psi \not = 1$, \cite{lin2013} style regression adjustment may be strictly inefficient relative to difference of means.
The problem is that the interacted regression solves the wrong optimization problem, minimizing a marginal variance objective when, due to the stratification, it should instead minimize a mean-conditional variance objective, conditional on the stratification variables $\psi$.
In fact, the Lin estimator is totally insensitive to the stratification, estimating the same adjustment coefficient for any stratified design $\Dn \sim \localdesigncond(\psi, \propfn)$.
Because of this, interacted regression is generically sub-optimal and in some cases can even be strictly less efficient than difference of means. 
Before proceeding, we state our main assumption. 

\begin{assumption}[Smoothness and Moment Conditions] \label{assumption:moment-conditions} Assume the following: 
\begin{enumerate}[label={(\roman*)}, itemindent=.5pt, itemsep=.4pt] 
\item The conditional expectations $E[h(X) | \psi]$ and $E[Y(d) | \psi]$ for $d \in \{0,1\}$ are Lipschitz continuous in the stratification variables $\psi$.
\item The moments $E[Y(d)^4] < \infty$ for $d \in \{0,1\}$ and $E[|h_t(X)|^4] < \infty$ for all $1 \leq t \leq \dim(h)$, $|\psi(X)|_2 < K < \infty$ a.s. and $\var(h) \succ 0$. 
\end{enumerate}
\end{assumption}

Now we are ready to define the Lin estimator and state our first result. 
Denote $\hi = h(X_i)$ and de-meaned covariates $\hitilde = \hi - \en[\hi]$, with $\en[\hi] \equiv n \inv \sum_{i=1}^n \hi$. 
The Lin estimator $\estlin$ is the coefficient on $\Di$ in the interacted regression 
\begin{equation} \label{equation:lin-estimator}
Y_i \sim 1 + \Di + \hitilde + \Di \hitilde.
\end{equation}
Define the within treatment arm covariate means $\bar h_1 = \en[\hi \Di] / \en[\Di]$ and $\bar h_0 = \en[\hi (1-\Di)] / \en[1-\Di]$. 
The Lin estimator $\estlin$ can be related to the difference of means estimator $\est$ as 
\begin{equation}
\estlin = \est - \coefflin'(\hbarone - \hbarzero).
\end{equation}
Here, the \emph{adjustment coefficient} $\coefflin$ is $\coefflin = (1-\propfn)(\wh a_1 + \wh a_0) + \propfn \wh a_0$, where $\wh a_0$ and $\wh a_1$ are the coefficients on $\hitilde$ and $\Di \hitilde$ in Equation \ref{equation:lin-estimator}. 
The following theorem characterizes the asymptotic properties of this estimator under stratified designs.

\begin{thm} \label{thm:lin}
Let Assumption \ref{assumption:moment-conditions} hold.
If $\Dn \sim \localdesigncond(\psi, \propfn)$ then the Lin estimator $\rootn(\estlin - \ate) \convwprocess \normal(0, \varlimit)$ with  
\begin{align*}
\varlimit = \var(\catefn(X)) + E\bigg[\var(\balancefn - \coefflinpop'h | \psi)\bigg] + E\left[\frac{\hk_1(X)}{\propfn} + \frac{\hk_0(X)}{1-\propfn}\right].
\end{align*}
The adjustment coefficient satisfies $\coefflin \convp \coefflinpop$ with $\coefflinpop = \argmin_{\gamma \in \mr^{\dimh}} \var(\balancefn - \gamma'h)$.  
\end{thm}

The variance $V$ differs from the variance of the unadjusted estimator only in the middle term, which changes from $E[\var(\balancefn | \psi)]$ in the unadjusted case to $E[\var(\balancefn - \coefflinpop'h | \psi)]$ for the interacted regression. 
Crucially, the second statement of Theorem \ref{thm:lin} shows that the adjustment coefficient $\coefflinpop$ attempts to minimize a marginal variance, instead of the mean-conditional variance that shows up in $V$ above. 
Because of this, the estimator may be inefficient for general stratifications $\psi \not = 1$, since in general 
\[
\coefflinpop = \argmin_{\gamma \in \mr^{\dimh}} \var(\balancefn - \gamma'h) \not = \argmin_{\gamma \in \mr^{\dimh}} E[\var(\balancefn - \gamma'h | \psi)] \equiv \gamma^*. 
\]

Observe that the Lin estimator is completely insensitive to the experimental design, estimating the same adjustment coefficient $\coefflinpop = \argmin_{\gamma} \var(\balancefn - \gamma'h)$ for any stratification variables $\psi(X)$. 
The following example shows that this can lead to strict inefficiency relative to difference of means estimation.

\begin{ex}[Random Assignment to Class Size] \label{ex:ineffadjustment}
Suppose $Y(d)$ are student test scores under random assignment to one of two class sizes $d \in \{0,1\}$.
Let $h(X)$ be parent's wealth and $\psi(X)$ previous year (baseline) test scores. 
Suppose parent's wealth is predictive of future test scores marginally so that $\cov(h, Y(d)) > 0$.
Then $\cov(h, \balancefn) > 0$ and the Lin coefficient is $\coefflinpop = \var(h) \inv \cov(h, \balancefn) > 0$.
However, if on average parent's wealth has no predictive power for test scores \emph{conditional} on a student's baseline scores (a proxy for ability) then $E[\cov(h, Y(d) | \psi)] = 0$.
In this case, regression adjustment for parent's wealth $h(X)$ in an experiment stratified on the earlier scores $\psi(X)$ will be strictly less efficient than unadjusted estimation since 
\begin{align*}
V_{lin} - V_{unadj} &= E[\var(\balancefn - \coefflinpop'h | \psi)] - E[\var(\balancefn | \psi)] \\
&= -2 \coefflinpop E[\cov(h, \balancefn | \psi)] + \coefflinpop ^2 E[\var(h | \psi)] = \coefflinpop ^2 E[\var(h | \psi)] > 0
\end{align*}
\end{ex}

An important special case occurs when the design is completely randomized ($\psi = 1$) or if the covariates and stratification variables are independent $h(X) \indep \psi(X)$.
In this case, the Lin estimator is weakly more efficient than difference of means since we have
\[
E[\var(\balancefn - \coefflinpop'h | \psi)] = \var(\balancefn - \coefflinpop'h) = \min_{\gamma} \var(\balancefn - \gamma'h)  \leq \var(\balancefn). 
\]

An analogue of Theorem \ref{thm:lin} also holds for the non-interacted regression estimator $Y_i \sim 1 + \Di + \hi$ under stratified designs $\Dn \sim \localdesigncond(\psi, \propfn)$.
The non-interacted estimator is known to be inefficient relative to difference of means even for completely randomized experiments unless $\propfn = 1/2$ or treatment effects are homogeneous.
For completeness, we give asymptotic theory and optimality conditions for this estimator under stratified randomization in Section \ref{section:naive-adjustment} in the appendix. 

We noted above that the Lin estimator $\estlin$ can be written in the canonical form $\estlin = \est - \coefflin'(\hbarone - \hbarzero)$. 
In fact, most commonly used adjusted estimators can be written in the standard form $\est_{adj} = \est - \wh \gamma'(\hbarone - \hbarzero)$ for some $\wh \gamma$, up to order $\Op(n \inv)$ factors. 
The following theorem describes the asymptotic properties of general covariate-adjusted estimators $\est_{adj}$ of this form. 
To avoid carrying around factors of $\propfn$ in our variance expressions, in what follows we scale adjusted estimators by the normalization constant $\propconstant = \sqrt{\propfn(1-\propfn)}$. 

\begin{thm} \label{thm:adjusted-efficiency}
Let Assumption \ref{assumption:moment-conditions} hold. 
Suppose $\wh \gamma \convp \gamma$ and consider the adjusted estimator 
\[
\est_{adj} = \est - \wh \gamma'(\hbarone - \hbarzero)\propconstant.
\]
If $\Dn \sim \localdesigncond(\psi, \propfn)$ then $\rootn(\est_{adj} - \ate) \convwprocess \normal(0, V(\gamma))$ with
\begin{equation} \label{equation:variance-augmented}
\varlimit(\gamma) = \var(\catefn(X)) + E\bigg[\var(\balancefn - \gamma'h | \psi)\bigg] + E\left[\frac{\hk_1(X)}{\propfn} + \frac{\hk_0(X)}{1-\propfn}\right]. 
\end{equation}
\end{thm}

We define a linearly-adjusted estimator to be asymptotically efficient if it globally minimizes the asymptotic variance $V(\gamma)$ in the previous theorem. 

\begin{defn}[Optimal Linear Adjustment] \label{defn:efficiency}
The estimator $\estadj = \est - \wh \gamma'(\hbarone - \hbarzero)\propconstant$ is \emph{efficient} for the design $\Dn \sim \localdesigncond(\psi, \propfn)$ and covariates $h(X)$ if $\wh \gamma \convp \gamma^*$ for an optimal adjustment coefficient 
\[
\gamma^* \in \argmin_{\gamma \in \mr^{\dimh}} E\bigg[\var(\balancefn - \gamma'h | \psi)\bigg].
\]
In particular, $V(\gamma^*) = \min_{\gamma \in \mr^{\dimh}} V(\gamma)$. 
\end{defn}

Note that efficiency is defined \emph{relative} to a design $\Dn \sim \localdesigncond(\psi, \propfn)$ and covariates $h(X)$.
Setting $\gamma = 0$ recovers unadjusted estimation, so any optimal estimator is in particular weakly more efficient than difference of means. 

\medskip

\textbf{Optimal Adjustment Coefficient}. If $E[\var(h | \psi)] \succ 0$, then the optimization in Definition \ref{defn:efficiency} is solved uniquely by a mean-conditional OLS coefficient
\begin{equation} \label{equation:optimal-coefficient}
\gamma^* = E[\var(h | \psi)] \inv E[\cov(h, \balancefn | \psi)]. 
\end{equation}

Intuitively, fine stratification makes treatment-control imbalances in the covariates $h(X)$ and the potential outcomes $Y(d)$ that are predictable by $\psi$ small enough that they do not contribute to first-order asymptotic variance. 
Because of this, the optimal covariate-adjusted estimator $\est - \gamma^*(\hbarone - \hbarzero)\propconstant$ ignores such fluctuations, minimizing the \emph{mean-conditional} variance objective $E[\var(\balancefn - \gamma'h | \psi)]$, instead of the marginal variance $\var(\balancefn - \gamma'h)$ targeted by the Lin estimator. 

\medskip

\textbf{Optimal Covariates}. 
Intuitively, the form of the variance in Equation \ref{equation:variance-augmented} suggests adjusting for variables $h$ that contain predictive information not already contained in $\psi$.
The (unknown) optimal covariates are $h^* = \balancefn$. 
In this case, $\gamma^* = 1$ makes the middle variance term identically zero, and $\estadj$ achieves the \cite{armstrong2022} semiparametric variance bound. 

\medskip

\textbf{Sample Average Treatment Effect}. Theorem \ref{thm:adjusted-efficiency} may be extended to covariate-adjusted estimation of the sample average treatment effect $\sate = \en[Y_i(1) - Y_i(0)]$. 
Defining the conditional treatment effect variance $\hkte(X) = \var(Y(1) - Y(0) | X)$, one can show that $\rootn(\est - \sate) \convwprocess \normal(0, V_S(\gamma))$ with 
\begin{equation} \label{equation:variance-sate}
V_S(\gamma) = E[\var(\balancefn - \gamma'h | \psi)] + E\left[\frac{\hk_1(X)}{\propfn} + \frac{\hk_0(X)}{1-\propfn} - \hkte(X) \right].
\end{equation}
In particular, the optimal adjustment for estimating the ATE and the SATE are the same, with $\gamma^*_{SATE} = \gamma^*$.

\begin{remark}[Non-Uniqueness] \label{remark:non-uniqueness}
In general, the optimal adjustment coefficient $\gamma^*$ may not be unique.
For example, if $h(x) = (z(\psi), w(x))$ with $z(\psi)$ a Lipschitz function of the stratification variables, then the variance objective is constant in the coefficient on $z(\psi)$ 
\[
E[\var(\balancefn - \gamma_z'z - \gamma_w'w | \psi)] = E[\var(\balancefn - \gamma_w'w | \psi)] \quad \quad \forall \gamma_z \in \mr^{d_z}.
\]
In fact, our analysis shows that the adjustment term $\gamma_z'(\bar z_1 - \bar z_0) = \op(\negrootn)$ for any coefficient $\gamma_z$ in this case. 
Intuitively, since the covariate $z(\psi)$ is already finely balanced by stratifying on $\psi(X)$, ex-post adjustment by $z(\psi)$ cannot improve first-order efficiency.
However, there may still be finite sample efficiency gains from such adjustments, if the covariates $z(\psi)$ are not completely balanced by the stratification.
Section \ref{section:further-adjustment} below provides methods to further adjust for covariates that are functions of the stratification variables.
\end{remark}

\subsubsection{Extensions}

Before continuing, we briefly mention some extensions to the framework above that are studied in detail in Appendices \ref{appendix:noncompliance}-\ref{appendix:nonlinear-adjustment}. 

\medskip

\textbf{Experiments with Noncompliance}. In settings with noncompliance, we may instead consider estimation and inference on the local average treatment effect (LATE) of \cite{imbens1994}. 
As a simple application of our main results, Section \ref{appendix:noncompliance} characterizes the optimal linear adjustment for estimating the LATE, constructs feasible efficient estimators, and provides asymptotically exact inference on the LATE under stratified randomization with ex-post covariate adjustment. 

\medskip

\textbf{Varying Treatment Proportions}. \cite{cytrynbaum2023} extends Definition \ref{defn:local_randomization} to allow fine stratification with non-constant assignment propensity $\propfn(\psi)$.  
Section \ref{appendix:varying-propensities} in the appendix characterizes the optimal adjustment coefficient for such designs and derives a feasible efficient estimator. 

\medskip

\textbf{Nonlinear Adjustment}. In some settings, it may be more natural to use nonlinear or nonparametric covariate adjustment to improve efficiency, for example in experiments with binary outcomes.   
Section \ref{appendix:nonlinear-adjustment} in the appendix characterizes the optimal adjustment over a general function space $\mc H$ for finely stratified designs with varying propensity $\propfn(\psi)$.   
Feasible estimation of the optimal nonlinear adjustment is an interesting problem that we leave for future work.

\subsection{Equivalence with Partially Linear Regression Adjustment} \label{section:efficiency-semiparametric-regression}
This section shows that optimal linear adjustment of a stratified design is as efficient as semiparametric \emph{partially linear} regression adjustment in an experiment with iid treatments, with adjustment function that is linear in $h(X)$ and nonparametric in $\psi(X)$. 
This suggests that experimenters stratify on a small set of covariates expected to be most predictive of outcomes at design-time, and (efficiently) adjust for the remaining covariates ex-post. 
See below for a more detailed discussion of stratification vs.\ adjustment. 

The main result of this section shows first-order asymptotic equivalence of the following (design, estimator) pairs 
\[
(\Dn \sim \localdesigncond(\psi, \propfn), \text{optimal linear}) \iff (\Di \simiid \bern(\propfn), \text{optimal semiparametric}).
\]

To define the latter, consider the within-arm partially linear regression models 
\begin{equation} \label{equation:semiparametric-regression}
(g_d^*, \gamma_d^*) = \argmin_{g \in L_2(\psi), \gamma \in \mr^{\dimh}} E[(Y(d) - g(\psi) - \gamma'h)^2]
\end{equation}
for $d \in \{0, 1\}$.
Define the partially linear adjustment function $F_d(x) = g_d^*(\psi(x)) + h(x)'\gamma_d^*$ and consider a \cite{robins95} style augmented inverse propensity weighting (AIPW) estimator 
\begin{align*}
\estsemiparam = \en[F_1(X_i) - F_0(X_i)] + \en\left[\frac{\Di (Y_i - F_1(X_i))}{\propfn}\right] - \en\left[\frac{(1-\Di) (Y_i - F_0(X_i))}{1-\propfn}\right].
\end{align*}
The next theorem shows that optimal linear adjustment of the design $\Dn \sim \localdesigncond(\psi, \propfn)$ is asymptotically equivalent to optimal semiparametric adjustment with nonparametric $\psi(X)$ and linear $h(X)$ components. 

\begin{thm} \label{thm:semiparam}
Require Assumption \ref{assumption:moment-conditions} and suppose $\Di \simiid \bern(\propfn)$. 
Then $\rootn(\estsemiparam - \ate) \convwprocess \normal(0, V^*)$ with 
\begin{align*}
V^* = \var(\catefn(X)) + \min_{\gamma \in \mr^{\dimh}} E\bigg[\var(\balancefn - \gamma'h | \psi)\bigg] + E\left[\frac{\hk_1(X)}{\propfn} + \frac{\hk_0(X)}{1-\propfn}\right].
\end{align*}
\end{thm}

The limiting variance $V^*$ is the same as the optimal linearly adjusted variance $V(\gamma^*)$ from Definition \ref{defn:efficiency}. 
Intuitively, stratification contributes the nonlinear component of the optimal model $F_d(x)$ above, while optimal adjustment contributes the linear component.  
The optimal adjustment coefficient $\gamma^* = \sqrt{\frac{1-\propfn}{\propfn}} \gamma_1 + \sqrt{\frac{\propfn}{1-\propfn}} \gamma_0$, for partially linear coefficients $\gamma_1, \gamma_0$ defined in Equation \ref{equation:semiparametric-regression} above.\medskip

\textbf{Stratification vs.\ Regression Adjustment.} 
Theorem \ref{thm:semiparam} shows that stratification provides nonparametric control over the fluctuations of the outcomes predictable by $\psi(X)$, while (linear) adjustment only provides linear control. 
In first-order asympotics, this suggests that we stratify on all available covariates, since the variance $V^*$ above is minimized by setting $\psi(X) = X$. 
However, this may perform poorly in finite samples due to a curse of dimensionality for stratification as $\dim(\psi)$ increases.
For example, \cite{cytrynbaum2023} shows the variance convergence rate $n \var(\est) = V + \Op(n^{-2/(\dim(\psi)+1)})$ for the variance $V$ in Equation \ref{equation:variance-augmented}, which may be slow even for moderate $\dim(\psi)$. 
Intuitively, this suggests stratifying on a small set\footnote{It is difficult to give concrete guidance for choosing $\dim(\psi)$, since the relevant quantities such as $E[\var(\balancefn | \psi)]$ are not estimable at design-time, before we have outcome data. 
The rate above suggests $\dim(\psi) = o(\log n)$ to achieve the variance $V$ in Equation \ref{equation:variance-augmented}.} of covariates $\psi(X)$ expected to be most predictive of outcomes at design time, and planning to optimally adjust for less predictive covariates $h(X)$ ex-post. \medskip

The next two sections show how to construct linearly-adjusted estimators for the design $\Dn \sim \localdesigncond(\psi, \propfn)$ that achieve the optimal variance $V^*$.

\subsection{Efficiency by Rich Strata Controls} \label{section:efficiency-tilted}
This section provides a ``rich covariates'' style condition on the relationship between adjustment covariates and stratification variables under which a simple parametric correction of the Lin estimator is fully efficient. 
The basic idea is to include rich functions $z(\psi)$ of the stratification variables in the adjustment set alongside the additional covariates we would like to adjust for ex-post.
The main result of this section shows that including $z(\psi)$ as covariates forces the Lin estimator to solve the mean-conditional variance minimization problem of Definition \ref{defn:efficiency}, restoring asymptotic optimality. 
An analogous result holds for the non-interacted regression estimator $Y \sim 1 + D + h$ if $\propfn = 1/2$.
As a simple application of this section's results, Example \ref{ex:coarse-stratification} shows that for coarsely stratified designs the Lin estimator with leave-one-out strata indicators is efficient. 

\medskip

Consider adjusting for covariates $h(X) = (w(X), z(\psi))$.
The main assumption of this section requires that the conditional mean $E[w | \psi]$ is well-approximated by known transformations $z(\psi)$ of the stratification variables. 

\begin{assumption} \label{assumption:rich-covariates}
There exist $c \in \mr^{d_w}$ and $\Lambda \in \mr^{d_w \times d_z}$ such that $E[w | \psi] = c + \Lambda z(\psi)$. 
\end{assumption}

Our next theorem shows that adding such transformations $z(\psi)$ to the adjustment set recovers full efficiency for the Lin estimator.

\begin{thm} \label{thm:efficiency-tilted}
Suppose Assumptions \ref{assumption:moment-conditions} and \ref{assumption:rich-covariates} hold.
Fix adjustment set $h(x) = (w(x), z(\psi))$. 
Then the Lin estimator $\estlin$ is fully efficient for the design $\Dn \sim \localdesigncond(\psi, \propfn)$.
In particular, $\rootn(\estlin - \ate) \convwprocess \normal(0, V^*)$ with 
\[
V^* = \var(\catefn(X)) + \min_{\gamma \in \mr^{\dimh}} E\bigg[\var(\balancefn - \gamma'h | \psi)\bigg] + E\left[\frac{\hk_1(X)}{\propfn} + \frac{\hk_0(X)}{1-\propfn}\right]. 
\]
Moreover, the asymptotic variance has  
\[
\min_{\gamma \in \mr^{\dimh}} E[\var(\balancefn - \gamma'h | \psi)] = \min_{\alpha \in \mr^{d_w}} E[\var(\balancefn - \alpha'w | \psi)].
\]
\end{thm}

In practice, Theorem \ref{thm:efficiency-tilted} suggests including flexible functions $z(\psi)$ of the stratification variables in the adjustment set. 
The proof is given in Section \ref{section:proofs:efficiency-tilted} of the supplement. 
The following corollary follows shows that if $\propfn = 1/2$ (matched pairs) or if treatment effect heterogeneity is limited then the non-interacted regression $Y \sim 1 + D + w + z$ with rich strata controls $z(\psi)$ is also asymptotically efficient.   

\begin{cor} \label{cor:naive-tilted}
Suppose additionally that $\propfn = 1/2$ or $E[\cov(Y(1)-Y(0), w | \psi)] = 0$.
Then the coefficient $\estnaive$ on $\Di$ in the regression $Y \sim 1 + D + w + z$ is asymptotically efficient. 
\end{cor}

The condition $E[\cov(Y(1)-Y(0), w | \psi)] = 0$ limits the explanatory power of covariates $w$ for treatment effect heterogeneity, conditional on the stratification variables.

\begin{remark}[Indirect Efficiency Gain]
The second statement of the theorem shows that optimal adjustment for $h(X)$ is as efficient as optimal adjustment for the subvector $w(X) \subseteq h(X) = (w(X), z(\psi))$. 
In this sense, the efficiency improvement due to including $z(\psi)$ is indirect.
Indeed, our analysis shows that $\est - \gamma_z'(\bar z_1 - \bar z_0) = \est + \op(\negrootn)$ for any $\gamma_z \in \mr^{d_z}$, so adjustment for $z(\psi)$ alone cannot affect the first-order asymptotic variance.
Intuitively, we are just using the inclusion of $z(\psi)$ as a device to ``tilt'' the coefficient on $w(X)$, forcing the Lin estimator to solve the correct mean-conditional variance optimization problem.
\end{remark}

The next example uses Theorem \ref{thm:efficiency-tilted} to show that including leave-one-out strata indicators as covariates in the Lin estimator restores asymptotic efficiency for coarsely stratified designs. 

\begin{ex}[Coarse Stratification] \label{ex:coarse-stratification}
Consider stratified randomization $\Dn \sim \localdesigncond(S, \propfn)$ with fixed strata $S(x) \in \{1, \dots, m\}$.
Let the adjustment covariates be $h(x) = (w(x), z(s))$ with leave-one-out strata indicators $z(S_i) = (\one(S_i = k))_{k=1}^{m-1}$.
In this case, Assumption \ref{assumption:rich-covariates} is automatically satisfied since we can write $E[w | S] = c + \Lambda z$ with $c = E[w | S=m]$ and $\Lambda_{jk} = (E[w_j | S=k] - E[w_j | S=m])_{jk}$.
Then by Theorem \ref{thm:efficiency-tilted}, the Lin estimator $\estlin$ with covariates $\hi = (\wi, \zi)$ is efficient.
In particular, we have $\rootn(\estlin - \ate) \convwprocess \normal(0, V^*)$ with optimal variance
\[
V^* = \var(\catefn(X)) + \min_{\gamma} E\bigg[\var(\balancefn - \gamma'w| S)\bigg] + E\left[\frac{\hk_1(X)}{\propfn} + \frac{\hk_0(X)}{1-\propfn}\right]. 
\]
Similarly, by Corollary \ref{cor:naive-tilted} if $\propfn = 1/2$ then including leave-one-out strata fixed effects in the non-interacted regression restores efficiency.
\end{ex}

\begin{remark}[Fine Stratification] \label{remark:fine-stratification}
Note that the argument in Example \ref{ex:coarse-stratification} only applies to coarse stratification, where the strata $S(x) \in \{1, \dots, m\}$ are data-independent and fixed as $n \to \infty$. 
For fine stratification $\Dn \sim \localdesigncond(\psi, \propfn)$ with continuous covariates $\psi(x)$, the strata are data-dependent and number of strata $m \asymp n$, so Theorem \ref{thm:partialled-lin} does not apply.  
Indeed, for matched pairs the Lin regression in Example \ref{ex:coarse-stratification} would have $n + 2 \dim(h) > n$ covariates, producing collinearity.
This collinearity problem occurs more generally, see Remark \ref{remark:fine-stratification-two} below for further discussion. 
\end{remark}

Leaving behind the rich covariates Assumption \ref{assumption:rich-covariates}, the next section provides new adjusted estimators that are fully efficient for any design in the class $\Dn \sim \localdesigncond(\psi, \propfn)$ under weak conditions.

\subsection{Generic Efficient Adjustment} \label{section:generic-efficiency}
In this section, we study several adjusted estimators that are asymptotically efficient under weak conditions for any stratified design $\Dn \sim \localdesigncond(\psi, \propfn)$.
For matched pairs designs, or in settings with limited treatment effect heterogeneity, the non-interacted regression including treatment, covariates, and pair fixed effects is efficient. 
More generally, we show that the following estimators are efficient under weak assumptions. 
\begin{enumerate}[label={(\arabic*)}, itemindent=.5pt, itemsep=.2pt] 
\item \textbf{PL} - A partialled Lin estimator with within-stratum (inconsistently) partialled covariates. 
\item \textbf{GO} - A ``Group OLS'' estimator, generalizing a proposal of \cite{imbens2015} for matched pairs designs.
\item \textbf{TM} - A tyranny-of-the-minority (ToM) estimator for stratified designs.
\end{enumerate}

The main new condition we impose in this section is that the adjustment covariates are not collinear, conditionally on the stratification variables.
This guarantees that the optimal adjustment coefficient $\gamma^*$ is unique with $\gamma^* = E[\var(h | \psi)] \inv E[\cov(h, \balancefn | \psi)]$, as discussed in Section \ref{section:optimal-adjustment}. 

\begin{assumption} \label{assumption:conditional-variance}
The conditional variance satisfies $E[\var(h | \psi)] \succ 0$. 
\end{assumption}

Note that this assumption rules out adjustment for functions $h(\psi)$ of the stratification variables.
To see why it is necessary, consider that, for example, in a regression with full strata fixed effects $Y \sim D + h + z^n$, covariates $\hi = h(\psii)$ would be asymptotically collinear with the strata fixed effects $z^n = (\one(i \in \group_j))_{j=1}^{n/k}$. 
More intuitively, the problem is that $h(\psi)$ has too little residual variation within local regions of $\psi(X)$ space defining the fine strata. 
We noted earlier that $\est - \alpha'(\bar h_1 - \bar h_0) = \est - \op(\negrootn)$ for any $\alpha \in \mr^{d_h}$, so such adjustment cannot improve first-order efficiency. 
Nevertheless, one may still wish to adjust for $h(\psi)$ to correct finite sample imbalances not controlled by the design.
Adjustment for such variables needs to be handled slightly differently, and we construct modified efficient estimators for this purpose in Section \ref{section:further-adjustment} below. 

\subsubsection{Strata Fixed Effects Estimator} \label{section:fixed-effects}
Recall that for $\propfn = a/k$, a finely stratified design $\Dn \sim \localdesigncond(\psi, \propfn)$ partitions the experimental units $\{1, \dots, n\}$ into $n/k$ disjoint groups $\group$.
Define the fixed effects estimator $\estnaivefe$ by the least squares equation
\begin{equation} \label{equation:naive-fe}
Y_i = \estnaivefe \Di + \coeffnaivefe' \hi + \sum_{j=1}^{n/k} \wh a_{j} \one(i \in \group_j) + e_i. 
\end{equation}

The next theorem shows that $\estnaivefe$ is fully efficient in the case of matched pairs or if treatment effect heterogeneity is limited, but may be inefficient in general.

\begin{thm} \label{thm:naive-fe}
Suppose Assumptions \ref{assumption:moment-conditions} and \ref{assumption:conditional-variance} hold.
The estimator has representation $\estnaivefe = \est - \coeffnaivefe'(\hbarone - \hbarzero) + \Op(n \inv)$.
If $\Dn \sim \localdesigncond(\psi, \propfn)$ then $\rootn(\estnaivefe - \ate) \convwprocess \normal(0, V)$ with variance 
\begin{align*}
V = \var(\catefn(X)) + E[\var(\balancefn - \coeffnaivefepop'h | \psi)] + E\left[\frac{\hk_1(X)}{\propfn} + \frac{\hk_0(X)}{1-\propfn}\right]. 
\end{align*}
and coefficient $\coeffnaivefepop = \argmin_{\gamma \in \mr^{\dimh}} E[\var(f - \gamma'h | \psi)]$ for target function 
\[
f(x) = \ceffn_1(x) \sqrt{\frac{\propfn}{1-\propfn}} + \ceffn_0(x)\sqrt{\frac{1-\propfn}{\propfn}}. 
\]
The function $f \not = \balancefn$ in general. 
If $\propfn = 1/2$, then $f = \balancefn$ and the fixed effects estimator is efficient. 
If $\propfn \not = 1/2$, it is efficient if and only if $E[\cov(h, Y(1) - Y(0) | \psi)] = 0$. 
\end{thm}
See Section \ref{section:proofs-generic-efficiency} for the proof.
Asymptotically exact inference for the $\ate$ using $\estnaivefe$ is available using the tools in Section \ref{section:inference}. 

\begin{remark}[Conditions for Efficiency] \label{remark:naive-fe-efficiency}
If $\propfn = 1/2$ then $f = \balancefn$ and $\estnaivefe$ is efficient.
More generally, $f(x) \not = \balancefn(x)$ and $\estnaivefe$ solves the wrong variance minimization problem, effectively targeting the wrong linear combination of outcomes. 
The necessary and sufficient condition $E[\cov(h, Y(1) - Y(0) | \psi)] = 0$ requires that treatment effect heterogeneity is not explained by the covariates $h(X)$, conditional on the stratification variables.  
\end{remark}

In the rest of this section, we develop estimators that are fully efficient for any finely stratified design, without imposing any assumptions on treatment effect heterogeneity or treatment proportions.

\subsubsection{Partialled Lin Estimator} \label{section:partialled-lin}
First, we define a partialled version of the Lin estimator. 
Let $\group(i)$ denote the group that unit $i$ belongs to and define the within-group partialled covariates 
\[
\hicheck = \hi - \frac{1}{k} \sum_{j \in \group(i)} \hj.
\]
For example, if $k = 2$ this is just the within-pair covariate difference $\hipartial = (1/2)(\hi - h_{m(i)})$, where $i$ is matched to $m(i)$. 
We can think of $\hicheck$ as an inconsistent but approximately unbiased signal for the non-parametrically residualized covariate $\hi - E[\hi | \psii]$.
Next, we use these partialled covariates in the Lin regression 
\begin{equation} \label{equation:lin-partialled}
Y \sim 1 + \Di + \hicheck + \Di \hicheck. 
\end{equation}
Define the \emph{partialled} Lin estimator $\estlinpartial$ to be the coefficient on $\Di$ in this regression.
For reference, similarly to the Lin regression we may write this in the standard form $\estlinpartial = \est - \coefflinpartial'(\hbarone - \hbarzero)\propconstant$ with adjustment coefficient $\coefflinpartial = (\wh a_1 + \wh a_0) \sqrt{\frac{1-\propfn}{\propfn}} + \wh a_0 \sqrt{\frac{\propfn}{1-\propfn}}$, where $\wh a_0$ and $\wh a_1$ are coefficients on $\hicheck$ and $\Di \hicheck$. \\

Our main result in Theorem \ref{thm:partialled-lin} below shows that the partialled Lin estimator $\estlinpartial$ is asymptotically efficient in the sense of Definition \ref{defn:efficiency}, with $\coefflinpartial \convp \gamma^*$ for the optimal adjustment coefficient $\gamma^*$.

\begin{remark}[Intuition for Optimality] \label{remark:intuition-partialled-lin}
Theorem \ref{thm:adjusted-efficiency} showed that an estimator $\est - \wh \gamma(\hbarone - \hbarzero)\propconstant$ is efficient if $\wh \gamma \convp \gamma^*$ and $\gamma^*$ solves the conditional-mean variance problem $\gamma^* \in \argmin_{\gamma} E[\var(\balancefn - \gamma'h | \psi)]$.
By using within-stratum partialled regressors $\hicheck$, we force the Lin estimator to only use covariate signal $\hi - E[\hi | \psii]$ that is mean-independent of the stratification variables.
\end{remark}

\begin{remark}[Treatment-Strata Interactions] \label{remark:fine-stratification-two}
As an alternative to $\estlinpartial$, one may attempt to use the Lin regression $Y_i \sim (1, \hi, g^n(i)) + \Di(1, \hi, g^n(i)) $ with leave-one-out strata fixed effects $g^n(i) = (\one(i \in \group_j))_{j=1}^{n/k-1}$. 
Unfortunately, this produces collinear regressors for $\propfn = a/k$ if either $a=1$ or $a=k-1$, which includes the case of matched pairs. 
To see the issue, one can show by Frisch-Waugh that in contrast to Equation \ref{equation:lin-partialled} above, this estimator partials covariates $\hi$ separately in each treatment arm, using $\check h_{i1} = \hi - a \inv \sum_{j \in \group(i)} \hj \Dj$ if $\Di = 1$ and $\check h_{i0} = \hi - (k-a) \inv \sum_{j \in \group(i)} \hj (1-\Dj)$ if $\Di = 0$. 
For instance, if $a = 1$ then this is $\hipartial = \hi - \hi = 0$ for all $i$, showing collinearity. 
In the case $1 < a < k-1$ where this estimator is feasible, it is asymptotically equivalent to the partialled Lin estimator.
However, finite sample properties will be worse due to noisier within-arm partialling.
\end{remark}

\begin{remark} \label{remark:partialled-lin-comparison}
A calculation shows that our estimator $\estlinpartial$ is numerically equivalent to a regression estimator proposed in \cite{zhu2022}, which the authors derive alternately through an optimal projection argument.  
They study estimation of the $\sate$ under stratified randomization in a finite population framework, providing conservative inference. 
They do not derive the exact form of the asymptotic variance, instead leaving it as an infinite sum, which they assume converges to some limit.
By contrast, we derive the exact form of the asymptotic variance under the data-adaptive stratifications in Definition \ref{defn:local_randomization}, enabling asymptotically exact inference on the $\ate$ using $\estlinpartial$.
\end{remark}

\subsubsection{Group OLS Estimator}
Next, we generalize an estimator proposed by \cite{imbens2015} for covariate adjustment in matched pairs experiments to more general stratified designs. 
For each group of units $\group = 1, \dots, n/k$ in the design $\Dn \sim \localdesigncond(\psi, \propfn)$, define the within-group difference of means of outcomes and covariates 
\[
\yg = \frac{1}{k} \sum_{i \in \group}  \frac{Y_i\Di}{\propfn} - \frac{1}{k} \sum_{i \in \group}  \frac{Y_i(1-\Di)}{1-\propfn} \quad \text{and} \quad  \hg = \frac{1}{k} \sum_{i \in \group}  \frac{\hi \Di}{\propfn} - \frac{1}{k} \sum_{i \in \group}  \frac{\hi (1-\Di)}{1-\propfn}. 
\]
For any group-indexed array $(x_g)_g$, denote $\eg[x_g] = \frac{k}{n} \sum_{\group} x_g$. 
Define the \emph{Group OLS} estimator $\estgroupols$ by the regression 
\begin{equation}
\yg = \estgroupols + \coeffgroupols' \hg + e_g 
\end{equation}
with $\eg[(1, \hg) e_g] = 0$.
For motivation, note that if $h = 0$ then this becomes $\yg = \estgroupols + e_g$  and $\estgroupols$ is just the unadjusted estimator $\estgroupols = \bar Y_1 - \bar Y_0$. 
More generally, the adjusted version can be written $\estgroupols = \eg[y_g] - \coeffgroupols'\eg[\hg] = \est - \coeffgroupols'(\bar h_1 - \bar h_0)$ with adjustment coefficient $\coeffgroupols = \var_g(h_g)\inv \cov_g(h_g, y_g)$. 
The estimators $\estgroupols$ and $\estlinpartial$ are numerically identical for the case of matched pairs, but not for $\propfn \not = 1/2$. 
The main result of this section shows that $\estgroupols$ is asymptotically equivalent to the partialled Lin estimator $\estlinpartial$, and both are asymptotically optimal.

\begin{remark}[Intuition for Efficiency]
The estimator $\estgroupols$ uses within-group differences of covariates $\hbaronegroup - \hbarzerogroup$ to predict within-group outcome differences $\bar Y_{1\group} - \bar Y_{0\group}$.  
Similar to the partialled Lin strategy, by doing this we only measure the variation in covariates and potential outcomes orthogonal to the stratification variables.
This forces least squares to compute a conditional variance-covariance tradeoff, solving the optimal adjustment problem in Definition \ref{defn:efficiency}.
In particular, the proof of Theorem \ref{thm:partialled-lin} shows that if $\Dn \sim \localdesigncond(\psi, \propfn)$ then the adjustment coefficient  
\[
\coeffgroupols = \var_{\group}(\hg)\inv \cov_{\group}(\hg, \yg) \convp \propconstant \argmin_{\gamma} E[\var(\balancefn - \gamma'h | \psi)]. 
\]
\end{remark}

\begin{remark}
\cite{imbens2015} propose $\estgroupols$ in the case of matched pairs $\propfn=1/2$. 
Their analysis uses a toy sampling model where the pairs themselves are drawn ``pre-matched'' from a superpopulation. 
By contrast, we model the experimental units as being sampled from a superpopulation, with units matched into data-dependent strata post-sampling.
This more realistic model complicates the analysis, producing different limiting variances and requiring different inference procedures.
In a design-based setting, \cite{fogarty2018b} shows that the \cite{imbens2015} estimator is weakly more efficient than difference of means for matched pairs designs. 
By contrast, we extend this estimator to a larger family of fine stratifications strictly containing matched pairs, and show that it is asymptotically optimal among linearly adjusted estimators. 
\end{remark}

\subsubsection{Tyranny-of-the-Minority (ToM) Estimator} \label{section:tyranny-of-the-minority}

Finally, we define tyranny-of-the-minority (ToM) adjustment, extending \cite{lin2013}.
To do so, define the adjustment coefficient
\begin{equation} \label{equation:tom_definition}
\coefftom = \var_n(\hicheck)\inv \left (\cov_n(\hicheck, Y_i | \Di=1) \sqrt{\frac{1-\propfn}{\propfn}} + \cov_n(\hicheck, Y_i | \Di=0) \sqrt{\frac{\propfn}{1-\propfn}} \right).
\end{equation}
Define the ToM estimator $\esttom = \est - \coefftom'(\hbarone - \hbarzero)\propconstant$. 
The main difference between the ToM and Partialled Lin adjustment coefficients is that $\coefftom$ estimates the conditional variance $E[\var(h |\psi)]$ only once, using the sample variance $\var_n(\hicheck)$ for the full experimental sample.
By contrast, partialled Lin estimates this term separately in each treatment arm, using $\var_n(\hicheck | \Di=1)$ and $\var_n(\hicheck | \Di=0)$. 
Because of this, we expect $\esttom$ to be more stable than $\estlinpartial$ in small experiments. 

\begin{remark} \label{remark:tom-comparison}
\cite{lu2022} propose an alternate ToM regression adjustment for stratified experiments.
To compare the approaches, for propensity $\propfn = a/k$ define the within-arm partialling $\check h_{i1} = \hi - a \inv \sum_{i \in \group} \Di \hi$ and $\check h_{i0} = \hi - (k-a)\inv \sum_{i \in \group} (1-\Di) \hi$.
Their estimator takes the form $\est_{LL} = \est - \wh \gamma_{LL}'(\hbarone - \hbarzero)$.
In our notation, their adjustment coefficient $\wh \gamma_{LL} = \wh S_{hh} \inv \wh S_{hY}$ has  
\begin{align*}
\wh S_{hh} &= \en\left[\frac{\Di \check h_{i1} \check h_{i1}'}{\propfn^2} \frac{a}{a-1} + \frac{(1-\Di) \check h_{i0} \check h_{i0}'}{(1-\propfn)^2} \frac{k-a}{k-a-1} \right] 
\end{align*}
and similarly for $\wh S_{hY}$.  
Their approach is infeasible if $a=1$ or $a=k-1$.
For example, this prohibits its use in matched pairs and matched triples experiments.

\end{remark}

\subsubsection{Main Result}
The main result of this section shows that all three estimators above are asymptotically equivalent and efficient in the sense of Definition \ref{defn:efficiency}. 

\begin{thm} \label{thm:partialled-lin}
Suppose Assumptions \ref{assumption:moment-conditions} and \ref{assumption:conditional-variance} hold. 
If $\Dn \sim \localdesigncond(\psi, \propfn)$, then $\estlinpartial - \estgroupols = \op(\negrootn)$ and $\estlinpartial - \esttom = \op(\negrootn)$.
We have $\rootn(\estlinpartial - \ate) \convwprocess \normal(0, V^*)$ with the optimal variance
\begin{align*}
V^* = \var(\catefn(X)) + \min_{\gamma} E[\var(\balancefn - \gamma'h | \psi)] + E\left[\frac{\hk_1(X)}{\propfn} + \frac{\hk_0(X)}{1-\propfn}\right]. 
\end{align*}
\end{thm}

Methods for asymptotically exact inference on the $\ate$ using these estimators are discussed in Section \ref{section:inference} below.
Our simulations and empirical results show that the partialled Lin, Group OLS, and ToM estimators behave very similarly in finite samples.

\subsection{Further Adjustment for Stratification Variables} \label{section:further-adjustment}
In this section, we provide modified versions of the previous estimators that allow further adjustment for covariates $z(\psi)$ that are functions of the stratification variables. 
As discussed above, this cannot improve first-order efficiency but may improve finite sample performance by correcting for any remaining imbalances in $\psi$ not controlled by the stratification. 

\medskip

Denote $\zi = z(\psii)$.
For each estimator $\est_k$ above with $k \in \{FE, PL, G, TM\}$, we define a modified estimator of the form $\wh \tau_k = \est_k - \wh \alpha_k'(\bar z_1 - \bar z_0) \propconstant$. 
For the fixed effects estimator, define $\estnaivefez$ to be the coefficient on $\Di$ in the regression $Y_i \sim (1, \Di, \hicheck, \zi)$.
For the partialled Lin estimator, define $\estlinpartialz$ to be the coefficient on $\Di$ in the regression $Y_i \sim (1, \hicheck, \zi) + \Di(1, \hicheck, \zi)$.
Define the modified ToM estimator to be as in Equation \ref{equation:tom_definition}, with $(\hicheck, \zi)$ in place of $\hicheck$.
Finally, define the modified group OLS estimator $\estgroupolsz = \estgroupols - \coeffgroupolsadjust'(\bar z_1 - \bar z_0) \propconstant$, with $\coeffgroupolsadjust = \coefflinpartialadjust$. 
Our next theorem shows that these estimators are asymptotically equivalent to the original versions of each estimator that do not adjust for $z(\psi)$.
However, the simulations in Sections \ref{section:simulations} and \ref{section:empirical} show that they may perform better in small experiments.

\begin{thm} \label{thm:z-estimators}
Suppose Assumptions \ref{assumption:moment-conditions} and \ref{assumption:conditional-variance} hold, as well as $\var(z) \succ 0$ and $E[|z|_2^2] < \infty$. 
Then if $\Dn \sim \localdesigncond(\psi, \propfn)$ we have $\wh \tau_k = \est_k + \op(\negrootn)$ for $k \in \{FE, PL, G, TM\}$. 
Each estimator has the form $\wh \tau_k = \est_k - \wh \alpha_k'(\bar z_1 - \bar z_0)\propconstant$ with $\coeffnaivefeadjust \convp \argmin_{\alpha} \var(f - \alpha'z)$ for $f$ as in Theorem \ref{thm:naive-fe} and $\coefflinpartialadjust, \coeffgroupolsadjust, \coefftomadjust \convp \argmin_{\alpha} \var(\balancefn - \alpha'z)$. 
\end{thm}

From the second statement of the theorem, we can interpret the modified estimators as taking a conservative approach that ignores stratification on $\psi$ and adjusts for imbalances in $z(\psi)$ as if the experiment were completely randomized. 

\section{Inference} \label{section:inference}
In this section, we provide asymptotically exact confidence intervals for the ATE in stratified experiments using generic linearly adjusted estimators. 
Overcoverage is known to be a problem for inference based on the usual Eicker-Huber-White (EHW) variance estimator in stratified experiments.
For example, \cite{bai2021inference} shows that the EHW variance estimators for $Y \sim 1 + D + h$ and the fixed effects regression $Y \sim D + h + z^n$ are asymptotically conservative for matched pairs designs if $h = 0$.
To the best of our knowledge, we give the first asymptotically exact inference methods for covariate-adjusted ($h \not = 0$) $\ate$ estimation under general stratified designs. 
Our main inference result applies to any estimator of the form $\est - \wh \gamma'(\hbarone - \hbarzero)\propconstant + \op(\negrootn)$.
In particular, this enables asymptotically exact inference on the $\ate$ using any of the estimators in this paper.
Our confidence intervals are shorter than those produced by EHW in the simulations and empirical application below, taking full advantage of the efficiency gains from both stratification and covariate adjustment. 

\medskip

To define our inference methods, consider such an estimator $\est(\wh \gamma) = \est - \wh \gamma'(\hbarone - \hbarzero)\propconstant$ with $\wh \gamma \convp \gamma$.
Define the augmented potential outcomes $Y_i^a(d) = Y_i(d) - \propconstant \wh \gamma' \hi$ for $d \in \{0, 1\}$ and the augmented outcome $Y^a_i = Y_i - \propconstant \wh \gamma' \hi$. 
Then apparently 
\begin{equation} \label{equation:inference-diffmeans-reduction}
\est(\wh \gamma) = \bar Y_1 - \bar Y_0 - \propconstant \wh \gamma'(\hbarone - \hbarzero)= \bar Y^a_1 - \bar Y^a_0. 
\end{equation}
Our strategy is to apply the inference results of \cite{cytrynbaum2023} for difference of means estimation $\est = \bar Y_1 - \bar Y_0$ to the difference of augmented potential outcomes $\bar Y^a_1 - \bar Y^a_0$. 
To do so, let $\groupset_n$ denote the set of groups in Definition \ref{defn:local_randomization}. 
For each $\group \in \groupset_n$ define the group centroid $\bar \psi_{\group} = |\group|\inv \sum_{i \in \group} \psii$.
Let $\groupmatching: \groupset_n \to \groupset_n$ be a bijective matching between groups satisfying $\groupmatching(\group) \not = \group$, $\groupmatching^2 = \identity$, and the homogeneity condition 
\begin{equation} \label{equation:group-matching-inference}
\frac{1}{n} \sum_{\group \in \groupset_n} |\bar \psi_{\group} - \bar \psi_{\groupmatching(\group)}|_2^2 = \op(1).  
\end{equation}
In practice, $\groupmatching$ is obtained by simply matching the group centroids $\bar \psi_{\group}$ into pairs using the \cite{derigs1988} matching algorithm. 
Let $\groupsetnu_n = \{\group \cup \groupmatching(\group): \group \in \groupset_n\}$ be the unions of paired groups formed by this matching. 
Define $a(\group) = \sum_{i \in \group} \Di$ and $k(\group) = |\group|$.
Finally, define the variance estimator components 
\begin{align*}
\varestone &= n \inv \sum_{\group \in \groupsetnu_n} \frac{1}{a(\group) - 1} \sum_{i \not = j \in \group} \frac{Y_i^a Y_j^a \Di \Dj (1-\propfn)}{\propfn^2} \\
\varestzero &= n \inv \sum_{\group \in \groupsetnu_n} \frac{1}{(k - a)(\group) - 1} \sum_{i \not = j \in \group} \frac{Y_i^a Y_j^a (1-\Di)(1-\Dj) \propfn}{(1-\propfn)^2}  \\
\varestcross &= n \inv \sum_{\group \in \groupset_n} \frac{k}{a(k-a)}(\group) \sum_{i,j \in \group} Y_i^a Y_j^a \Di (1-\Dj). 
\end{align*}

\noindent Next, define the variance estimator
\begin{equation} \label{equation:variance-estimator}
\wh V = \var_n \left( \frac{(\Di - \propfn) Y_i^a}{\propfn-\propfn^2} \right) - \varestone - \varestzero - 2 \varestcross.
\end{equation}

Our inference strategy begins with the sample variance of adjusted estimator, which is consistent for the asymptotic variance of $\estadj$ under an iid design, but too large under stratified designs. 
We correct this sample variance using the estimators above, which measure how well the stratification variables predict augmented outcomes in local regions of the covariate space. 
This section's main result shows that $\varest$ is consistent for the limiting variance of Theorem \ref{thm:adjusted-efficiency}, enabling asymptotically exact inference on the $\ate$ using adjusted estimators. 

\begin{thm}[Inference] \label{thm:inference}
Under the conditions of Theorem \ref{thm:adjusted-efficiency}, if $\Dn \sim \localdesigncond(\psi, \propfn)$, then $\varest = V + \op(1)$.
\end{thm}

By Theorem \ref{thm:inference} and our previous asymptotic results in Theorem \ref{thm:adjusted-efficiency}, the confidence interval $\wh C = [\est(\wh \gamma) \pm \varest^{1/2} c_{1-\alpha/2} / \rootn]$ with $c_{\alpha} = \Phi \inv(\alpha)$ is asymptotically exact in the sense that $P (\ate \in \wh C ) = 1-\alpha + o(1)$.

\section{Simulations} \label{section:simulations}
In this section, we use simulations to test the finite sample performance of the estimators studied above.
We consider quadratic outcome models of the form 
\[
Y_i(d) = \psii'Q_d\psii + \psii'L_d + c_d \cdot u(X_i) + \residual_i^d \quad \quad E[\residual_i^d | X_i] = 0
\]
for $d \in \{0,1\}$.
The component $u_i = u(X_i)$ represents covariate signal that is independent of the stratification variables $\psi(X_i)$. 
After implementing the design $\Dn \sim \localdesigncond(\psi, \propfn)$, we receive access to scalar covariates $\hi$ that are correlated with both $\psii$ and $Y_i(d)$. 
In particular, suppose that $\hi = \psii'Q_h \psii + \psii'L_h + u_i$ with $E[u_i | \psii] = 0$.
In the following simulations, we let $\psii \sim N(0, I_{m})$, $u_i \sim N(0, 1)$, and $\epsilon_i^d \sim N(0, 1/10)$ with $(\psii, \ui, \epsilon_i^d)$ jointly independent. 
We use treatment proportions $\propfn = 2/3$ unless otherwise specified. 
With $m \equiv \dim(\psi)$, let $A \in \mr^{m \times m}$ have $A_{ij} = 1$ for $i\not=j$ and $A_{ii} = 0$.
We simulate the following DGP's:
\begin{enumerate}[label={}, itemindent=.5pt, itemsep=.4pt] 
    \item \textbf{Model} 1: Quadratic coefficients $Q_h = (1/m^2) A$ and $Q_0=Q_1 = (1/m) A$.
    Linear coefficients $L_0 = \one_m$, $L_1 = 2\one_m$, $L_h = \one_m$.
    Regressor signal $c_1 = c_0 = -3$. 
    \item \textbf{Model} 2: As in Model 1 but $c_0 = -4$ and $c_1 = -1$.
    \item \textbf{Model} 3: As in Model 2 but $\propfn = 1/2$.
    \item \textbf{Model} 4: As in Model 1 but $c_0 = 2$ and $c_1 = 4$.
    \item \textbf{Model} 5: As in Model 1 but $c_0 = 2$ and $c_1 = 4$ and $\propfn = 1/2$.
    \item \textbf{Model} 6: As in Model 1 but $Q_h=(1/100)A$.
\end{enumerate}

We begin by comparing the efficiency properties of different linearly adjusted estimators. 
\textbf{Unadj} refers to simple difference of means (unadjusted). 
The \textbf{Lin} estimator is studied in Theorem \ref{thm:lin}.
\textbf{Naive} refers to the non-interacted regression $Y \sim (1, D, h)$, (Theorem \ref{thm:naive}).
\textbf{FE} refers to the fixed effects estimator (Theorem \ref{thm:naive-fe}) and \textbf{Plin} the partialled Lin estimator (Theorem \ref{thm:partialled-lin}).
\textbf{GO} refers to Group OLS and \textbf{ToM} refers to Tyranny-of-the-Minority estimation (Theorem \ref{thm:partialled-lin}).
\textbf{Strata Controls} refer to modified versions of each of the previous estimators that further adjust for parametric strata controls $z(\psi)$, as discussed in Section \ref{section:further-adjustment}.
In our simulations, we set $z(\psi) = \psi$. 
\textbf{Ad} refers to an adaptive\footnote{This estimator is pointwise asymptotically equivalent to $\estlinpartial$. 
Issues with post model-selection inference (e.g.\ \cite{Leeb2005}) appear to be less worrying here, since even under the fixed alternative $\gamma^* \not = \coefflinpop$, the Lin estimator is still $\rootn$-consistent and asymptotically unbiased.} estimator that sets $\estadj = \estlin$ if $\wh V(\coefflin) \leq \wh V(\coefflinpartial)$ and $\estadj = \estlinpartial$ otherwise,\footnote{
We could also use a cross-fit version of $\wh V(\gamma)$ to reduce bias. However, the in-sample criterion performed quite well in our simulations.} including parametric controls $z(\psi)=\psi$ in both cases.\\

\begin{table}[htbp]
\begin{adjustbox}{width=\columnwidth,center}
  \centering
    \begin{tabular}{cc|cccccccccccccc}
          & \multicolumn{1}{r}{} & \multicolumn{7}{c}{No Strata Controls}                & \multicolumn{6}{c}{Strata Controls $z(\psi)$} &  \\
\cmidrule{2-16}    $(n, \dim(\psi))$ & \multicolumn{1}{c}{Model} & Unadj & Naive & Lin   & FE    & Plin  & GO    & \multicolumn{1}{c|}{ToM} & Naive & Lin   & FE    & Plin  & GO    & ToM   & Ad \\
\cmidrule{2-16}          & 1     & 100   & 113   & 102   & 49    & 48    & 49    & 48    & 36    & 35    & 35    & 37    & 37    & 36    & 34 \\
          & 2     & 100   & 126   & 102   & 64    & 57    & 58    & 57    & 60    & 46    & 52    & 47    & 47    & 47    & 45 \\
    (600, 2) & 3     & 100   & 116   & 116   & 38    & 38    & 38    & 38    & 48    & 48    & 36    & 36    & 37    & 36    & 37 \\
          & 4     & 100   & 27    & 31    & 31    & 27    & 27    & 27    & 26    & 26    & 38    & 32    & 33    & 32    & 26 \\
          & 5     & 100   & 28    & 28    & 18    & 18    & 18    & 18    & 21    & 21    & 19    & 19    & 19    & 19    & 19 \\
          & 6     & 100   & 100   & 100   & 11    & 11    & 11    & 11    & 7     & 7     & 9     & 9     & 9     & 9     & 7 \\
\cmidrule{2-16}          & 1     & 100   & 114   & 103   & 44    & 44    & 44    & 44    & 35    & 34    & 31    & 33    & 33    & 33    & 32 \\
          & 2     & 100   & 126   & 102   & 60    & 56    & 56    & 56    & 61    & 47    & 50    & 47    & 46    & 47    & 45 \\
    (1200, 2) & 3     & 100   & 116   & 116   & 38    & 38    & 38    & 38    & 48    & 48    & 37    & 37    & 37    & 37    & 37 \\
          & 4     & 100   & 26    & 30    & 29    & 25    & 25    & 25    & 23    & 24    & 36    & 30    & 30    & 30    & 24 \\
          & 5     & 100   & 28    & 28    & 17    & 17    & 17    & 17    & 20    & 20    & 17    & 18    & 17    & 18    & 18 \\
          & 6     & 100   & 101   & 101   & 9     & 9     & 9     & 9     & 7     & 7     & 8     & 8     & 8     & 8     & 7 \\
\cmidrule{2-16}          & 1     & 100   & 142   & 127   & 85    & 84    & 84    & 84    & 25    & 24    & 41    & 46    & 55    & 46    & 24 \\
          & 2     & 100   & 145   & 123   & 94    & 86    & 87    & 86    & 45    & 34    & 57    & 54    & 62    & 54    & 34 \\
    (1200, 5) & 3     & 100   & 137   & 137   & 81    & 81    & 81    & 81    & 40    & 40    & 54    & 54    & 57    & 54    & 40 \\
          & 4     & 100   & 27    & 31    & 31    & 27    & 27    & 27    & 25    & 20    & 54    & 45    & 49    & 45    & 20 \\
          & 5     & 100   & 32    & 32    & 24    & 24    & 24    & 24    & 18    & 18    & 38    & 38    & 39    & 38    & 18 \\
          & 6     & 100   & 138   & 138   & 67    & 67    & 67    & 67    & 15    & 15    & 36    & 36    & 39    & 37    & 15 \\
\cmidrule{2-16}    &  $R_k$ & 73    & 65    & 60    & 17    & 15    & 15    & 15    & 5     & 2     & 10    & 8     & 10    & 8     & 0.2 \\
\cmidrule{2-16}    \end{tabular}%
\end{adjustbox}
\caption{Ratio of MSE's ($\%$), adjusted vs. unadjusted estimation.}
\label{table:simulation:efficiency}%
\end{table}%

Table \ref{table:simulation:efficiency} studies finite sample efficiency. 
We present the mean squared error (MSE) ratio, relative to unadjusted estimation, for each of the adjusted estimators above. 
The bottom line of the table reports the \emph{excess risk} $R_k$ of each estimator $k$ relative to the optimal estimator.
To define this, let $\mse_{k, s}$ be the relative MSE of estimator $k$ in simulation $s$. 
Then we set $R_k = (1/S)\sum_s (\mse_{k, s} - \min_j \mse_{j, s})$, averaging over all simulations in the table. 
All results are calculated using $2000$ Monte Carlo repetitions. \medskip 

In models 1, 2, and 3, both Naive and Lin style linear adjustment are strictly inefficient relative to unadjusted estimation. 
These models have marginal covariance $\cov(Y(d), h) > 0$ but conditional covariance $E[\cov(Y(d), h | \psi)] < 0$, conditional on the stratification variables. 
Because of this, the optimal adjustment coefficient $\gamma^* < 0$, while the Naive and Lin regressions estimate positive adjustment coefficients $\gamma_N, \gamma_L > 0$, leading to even worse performance than unadjusted estimation in some cases.
For Models 4 and 5, the \textbf{Naive} and \textbf{Lin} methods are competitive with the generic efficient methods from Section \ref{section:generic-efficiency}.
This is because in these cases we made it so that $\cov(Y(d), h) \approx E[\cov(Y(d), h | \psi)]$, so that ``by chance'' $\gamma^*$ is close to $\gamma_N$ and $\gamma_L$. 
However, the parametric coefficients $\gamma_N$ and $\gamma_L$ are estimated more precisely than the semiparametric object $\gamma^* = E[\var(h | \psi)] \inv E[\cov(h, \balancefn | \psi)]$.
For Model 6, \textbf{Lin} with $z(\psi) = \psi$ controls is (approximately) optimal by Theorem \ref{thm:efficiency-tilted}, since $E[w | \psi]$ is (approximately) linear in $\psi$. \medskip

Summarizing our findings, the \textbf{Lin}, \textbf{Plin}, and \textbf{Naive} estimators with parametric \textbf{Strata Controls} $z(\psi)=\psi$ had low excess risk across specifications, while the \textbf{Ad} estimator was the most efficient overall.
The \textbf{Naive} and \textbf{Lin} estimators without strata controls or within-stratum partialling had large MSE. 
The \textbf{Plin}, \textbf{GO}, and \textbf{ToM} estimators had similar MSE across model specifications. 
These generic methods performed the best in regimes with large $n$, small $\dim(\psi)$, and nonlinear $E[h | \psi]$. 
In these cases, the gap $\gamma_L - \gamma^*$ between the sub-optimal Lin coefficient and optimal coefficient $\gamma^*$ dominates the additional variability $\var(\wh \gamma^*) > \var( \wh \gamma_L)$ required to estimate $\gamma^*$ (this variability increases with $\dim(\psi)$). 
For example, \textbf{Plin} with $z(\psi)$ controls performs the best when $(n, \dim(\psi)) = (1200, 2)$, but \textbf{Lin} with $z(\psi)$ controls is much better when $\dim(\psi) = 5$.  
The \textbf{Ad} estimator used a variance pre-test to choose between \textbf{Plin} and \textbf{Lin} (including $z(\psi)$ controls), allowing it to perform well in both regimes. 

\begin{table}[htbp]
\begin{adjustbox}{width=\columnwidth,center}
  \centering
    \begin{tabular}{ccrrrrrrr|rrrrrrr}
          &       & \multicolumn{7}{c}{No Strata Controls}                & \multicolumn{6}{c}{Strata Controls $z(\psi)$}           &  \\
\cmidrule{2-16}          & Model & \multicolumn{1}{c}{Unadj} & \multicolumn{1}{c}{Naive} & \multicolumn{1}{c}{Lin} & \multicolumn{1}{c}{FE} & \multicolumn{1}{c}{Plin} & \multicolumn{1}{c}{GO} & \multicolumn{1}{c|}{ToM} & \multicolumn{1}{c}{Naive} & \multicolumn{1}{c}{Lin} & \multicolumn{1}{c}{FE} & \multicolumn{1}{c}{Plin} & \multicolumn{1}{c}{GO} & \multicolumn{1}{c}{ToM} & \multicolumn{1}{c}{Ad} \\
\cmidrule{2-16}          & 1     & 0     & 17    & 11    & -5    & -5    & -5    & -5    & -49   & -50   & -34   & -29   & -26   & -29   & -50 \\
          & 2     & 0     & 18    & 10    & -3    & -4    & -4    & -4    & -33   & -41   & -25   & -25   & -22   & -25   & -41 \\
    $\% \Delta$CI Length  & 3     & 0     & 16    & 16    & -6    & -6    & -6    & -6    & -36   & -36   & -24   & -24   & -24   & -24   & -36 \\
    vs. Unadj & 4     & 0     & -46   & -43   & -42   & -46   & -46   & -46   & -50   & -55   & -22   & -31   & -26   & -30   & -55 \\
          & 5     & 0     & -44   & -44   & -49   & -49   & -49   & -49   & -56   & -56   & -34   & -34   & -31   & -34   & -56 \\
          & 6     & 0     & 16    & 16    & -12   & -12   & -12   & -12   & -59   & -59   & -35   & -35   & -35   & -35   & -59 \\
\cmidrule{2-16}          & 1     & 0.95  & 0.95  & 0.95  & 0.96  & 0.96  & 0.96  & 0.96  & 0.95  & 0.95  & 0.96  & 0.96  & 0.95  & 0.96  & 0.95 \\
          & 2     & 0.95  & 0.95  & 0.95  & 0.95  & 0.96  & 0.96  & 0.96  & 0.95  & 0.96  & 0.95  & 0.96  & 0.95  & 0.96  & 0.96 \\
    Coverage  & 3     & 0.95  & 0.95  & 0.95  & 0.96  & 0.96  & 0.96  & 0.96  & 0.96  & 0.96  & 0.96  & 0.96  & 0.96  & 0.96  & 0.96 \\
 (Exact)         & 4     & 0.95  & 0.95  & 0.94  & 0.96  & 0.95  & 0.95  & 0.95  & 0.95  & 0.95  & 0.96  & 0.96  & 0.96  & 0.96  & 0.95 \\
          & 5     & 0.94  & 0.95  & 0.95  & 0.96  & 0.96  & 0.96  & 0.96  & 0.95  & 0.95  & 0.96  & 0.96  & 0.97  & 0.96  & 0.95 \\
          & 6     & 0.95  & 0.95  & 0.95  & 0.97  & 0.97  & 0.97  & 0.97  & 0.96  & 0.96  & 0.97  & 0.97  & 0.96  & 0.97  & 0.96 \\
\cmidrule{2-16}          & 1     & 0.99  & 0.95  & 0.96  & 0.97  & 0.99  &       &       & 0.99  & 0.98  & 0.99  & 0.97  &       &       &  \\
          & 2     & 0.99  & 0.95  & 0.95  & 0.94  & 0.99  &       &       & 0.98  & 0.93  & 0.98  & 0.96  &       &       &  \\
    Coverage  & 3     & 1.00  & 0.96  & 0.95  & 0.96  & 1.00  &       &       & 0.99  & 0.93  & 0.98  & 0.97  &       &       &  \\
 (EHW)         & 4     & 0.99  & 0.99  & 0.90  & 0.98  & 1.00  &       &       & 0.97  & 0.68  & 0.98  & 0.97  &       &       &  \\
          & 5     & 0.99  & 0.97  & 0.90  & 0.98  & 1.00  &       &       & 0.96  & 0.65  & 0.99  & 0.99  &       &       &  \\
          & 6     & 1.00  & 0.97  & 0.96  & 0.96  & 1.00  &       &       & 0.99  & 0.97  & 1.00  & 0.99  &       &       &  \\
\cmidrule{2-16}    \end{tabular}%
\end{adjustbox}
\caption{Properties of Inference}
\label{table:simulation:inference}%
\end{table}%

Table \ref{table:simulation:inference} reports finite sample efficiency and coverage properties of the asymptotically exact inference methods developed in Section \ref{section:inference}. 
We let $n=1200$ and $\dim(\psi) = 5$. 
The first panel shows $\%$ change in confidence interval length relative to unadjusted estimation.
All confidence intervals are computed using the method in Theorem \ref{thm:inference}. 
We see that the relative efficiency of different estimators are reflected by our inference methods. 
In particular, asymptotically exact inference allows researchers to report shorter confidence intervals when a more efficient adjustment method is used. 
In the second panel, we show coverage probabilities for our asymptotically exact confidence interval across a range of linearly adjusted estimators.  
The final panel shows coverage probabilities for confidence intervals based on the usual HC2 variance estimator, where applicable.
The HC2-based confidence intervals significantly overcover.

\section{Empirical Application} \label{section:empirical}

In this section we apply our methods to the experiment in \cite{baysan2022},\footnote{The data is available from \cite{baysan2022data}.} who estimates the effect of a political information campaign on support for a 2017 Turkish referendum removing checks and balances on executive power. 
The campaign was administered by the opposition Republican People's Party (CHP), who opposed the referendum.    
Randomization was performed at the neighborhood level, stratified on quartiles of CHP vote share in the previous 2015 elections.  
The main outcome is the ``No'' vote share in the 2017 referendum.\footnote{\cite{baysan2022} estimates effects of the campaign on vote share at both the ballot box and neighborhood level. 
We focus on the neighborhood level effects.}
Due to the cost of administering the campaign, $\propfn = 2/11$ out of $n=550$ total neighborhoods were treated. 
In the original analysis, \cite{baysan2022} performed non-interacted covariate adjustment (Theorem \ref{thm:naive}) for $h(X) = $ number of registered voters, number of valid votes, number of votes for the CHP in 2015, CHP vote share in 2015, voter turnout, and CHP vote share quartile fixed effects.

\begin{table}[htbp]
\begin{adjustbox}{width=\columnwidth,center}
  \centering
    \begin{tabular}{crrrrrrr|rrrrrr}
    \multicolumn{8}{c}{No Strata Controls}                        & \multicolumn{6}{c}{Strata Controls $z(\psi)$} \\
    \midrule
    Model & \multicolumn{1}{c}{Unadj} & \multicolumn{1}{c}{Naive} & \multicolumn{1}{c}{Lin} & \multicolumn{1}{c}{FE} & \multicolumn{1}{c}{Plin} & \multicolumn{1}{c}{GO} & \multicolumn{1}{c|}{ToM} & \multicolumn{1}{c}{Naive} & \multicolumn{1}{c}{Lin} & \multicolumn{1}{c}{FE} & \multicolumn{1}{c}{Plin} & \multicolumn{1}{c}{GO} & \multicolumn{1}{c}{ToM} \\
    \midrule
    $\estadj$ & -0.0054 & 0.0040 & 0.0047 & 0.0041 & 0.0021 & 0.0034 & 0.0021 & 0.0041 & 0.0040 & 0.0038 & 0.0037 & 0.0031 & 0.0019 \\
    SE    & 0.0088 & 0.0074 & 0.0074 & 0.0074 & 0.0078 & 0.0077 & 0.0081 & 0.0074 & 0.0073 & 0.0077 & 0.0076 & 0.0078 & 0.0083 \\
    HC2   & 0.0155 & 0.0075 & 0.0073 & 0.0075 & 0.0149 &       &       & 0.0075 & 0.0070 & 0.0736 & 0.0071 &       &  \\
    \bottomrule
    \end{tabular}%
\end{adjustbox}
\caption{Empirical Results}
\label{table:empirical}%
\end{table}%

In the first block of Table \ref{table:empirical}, we replicate the neighborhood-level analysis of \cite{baysan2022}.
$\estadj$ is the point estimate from each adjustment strategy, SE is the asymptotically exact standard error from Section \ref{section:inference}, and EHW is the usual robust standard error (HC2). 
Estimates in the ``strata controls $z(\psi)$'' section include quartile fixed effects, while the leftmost section does not.
The results in Section \ref{section:efficiency-tilted} show that Lin adjustment with quartile fixed effects is efficient in this case, and indeed this has the smallest estimated standard error.
The generic efficient estimators have slightly larger SE. 
The asymptotically exact standard errors from Section \ref{section:inference} are generally similar to or smaller than EHW, except for the Lin, FE, and Plin estimators with $z(\psi)$ controls. 
However, our simulation also showed that EHW standard errors may severely undercover in these cases.\footnote{We also note that \cite{bai2023tuples} have found the EHW standard error from a linear regression with block fixed effects to be potentially invalid in a related problem.}  

\begin{table}[htbp]
\begin{adjustbox}{width=\columnwidth,center}
  \centering
    \begin{tabular}{ccrrrrrrr|rrrrrr}
          & \multicolumn{8}{c}{No Strata Controls}                        & \multicolumn{6}{c}{Strata Controls $z(\psi)$} \\
\cmidrule{2-15}          & Model & \multicolumn{1}{c}{Unadj} & \multicolumn{1}{c}{Naive} & \multicolumn{1}{c}{Lin} & \multicolumn{1}{c}{FE} & \multicolumn{1}{c}{Plin} & \multicolumn{1}{c}{GO} & \multicolumn{1}{c|}{ToM} & \multicolumn{1}{c}{Naive} & \multicolumn{1}{c}{Lin} & \multicolumn{1}{c}{FE} & \multicolumn{1}{c}{Plin} & \multicolumn{1}{c}{GO} & \multicolumn{1}{c}{ToM} \\
\cmidrule{2-15}          & Est   & 0.0000 & 0.0001 & 0.0003 & 0.0001 & 0.0002 & 0.0003 & 0.0000 & 0.0001 & 0.0004 & 0.0001 & 0.0003 & 0.0002 & 0.0000 \\
    Coarse & SE    & 0.0085 & 0.0076 & 0.0075 & 0.0075 & 0.0077 & 0.0077 & 0.0078 & 0.0076 & 0.0074 & 0.0078 & 0.0076 & 0.0079 & 0.0079 \\
          & HC2   & 0.0144 & 0.0078 & 0.0078 & 0.0077 & 0.0141 &       &       & 0.0078 & 0.0077 & 0.0736 & 0.0080 &       &  \\
\cmidrule{2-15}          & Est   & -0.0001 & 0.0000 & 0.0000 & 0.0000 & 0.0000 & 0.0004 & -0.0001 & 0.0000 & 0.0002 & 0.0000 & 0.0002 & 0.0004 & -0.0001 \\
    Fine  & SE    & 0.0077 & 0.0081 & 0.0080 & 0.0076 & 0.0077 & 0.0077 & 0.0077 & 0.0075 & 0.0075 & 0.0075 & 0.0075 & 0.0077 & 0.0077 \\
          & HC2   & 0.0144 & 0.0141 & 0.0142 & 0.0078 & 0.0145 &       &       & 0.0078 & 0.0078 & 0.0733 & 0.0078 &       &  \\
\cmidrule{2-15}          & Est   & -0.0001 & -0.0001 & -0.0001 & 0.0001 & 0.0001 & 0.0001 & 0.0001 & 0.0000 & -0.0001 & 0.0001 & 0.0001 & 0.0001 & 0.0001 \\
    Fine  & SE    & 0.0072 & 0.0073 & 0.0073 & 0.0070 & 0.0070 & 0.0070 & 0.0070 & 0.0066 & 0.0066 & 0.0066 & 0.0066 & 0.0066 & 0.0066 \\
    $p=1/2$ & HC2   & 0.0113 & 0.0111 & 0.0111 & 0.0059 & 0.0114 &       &       & 0.0060 & 0.0060 & 0.0565 & 0.0061 &       &  \\
\cmidrule{2-15}          & Est   & 0.0000 & 0.0001 & 0.0002 & 0.0001 & 0.0002 & 0.0000 & 0.0003 & 0.0001 & 0.0002 & 0.0000 & 0.0002 & 0.0001 & 0.0001 \\
    Fine  & SE    & 0.0155 & 0.0155 & 0.0155 & 0.0155 & 0.0155 & 0.0155 & 0.0155 & 0.0146 & 0.0146 & 0.0146 & 0.0146 & 0.0147 & 0.0147 \\
    $\dim(\psi) = 3$ & HC2   & 0.0145 & 0.0145 & 0.0145 & 0.0089 & 0.0145 &       &       & 0.0079 & 0.0078 & 0.0740 & 0.0078 &       &  \\
\cmidrule{2-15}    \end{tabular}%
\end{adjustbox}
\caption{Simulated Designs}
\label{table:empirical_simulations}%
\end{table}%

Overall, changing the adjustment method did not have an economically meaningful effect on the conclusions of the study, and we recover the null effect of \cite{baysan2022} in all cases.  
The covariate $h_k = $ ``CHP vote share in 2015'' is highly predictive of $Y = $ ``CHP vote share in 2017,'' so adjusting for this variable ex-post provides a modest variance reduction even after stratifying on 2015 vote share quartiles.
However, the estimated optimal coefficient $\gamma^*_k \approx 0.27$ and Lin coefficient $\gamma_{L,k} \approx 0.31$ are quite similar, so (inefficient) Lin adjustment still performs quite well. 
The other covariates such as $h_j = $ ``voter turnout'' are very weak predictors of outcomes, so changing the adjustment coefficient on these variables doesn't matter much.

Next, we ask how each estimator would have performed in the experiment in \cite{baysan2022} under counterfactual randomization procedures, such as fine stratification.\footnote{Algorithms and inference methods for fine stratification with $\propfn \not =1/2$ have only been developed recently, e.g.\ \cite{bai2020pairs} and \cite{cytrynbaum2023}.}
To do so, we follow the nonparametric imputation strategy in \cite{bai2020pairs}, defining potential outcomes $\wh Y_i(d) = Y_i$ if $\Di=d$ and matching imputation $\wh Y_i(d) = Y_{j(i)}(d)$ with $j(i) = \argmin_{j: \Dj=d} |X_i - X_j|_2$ if $\Di \not = d$.
We let the matching variables $X_i$ include all controls used in the analysis of \cite{baysan2022}.
Given the imputed data $(X_i, \wh Y_i(0), \wh Y_i(1))_{i=1}^n$, we do the following simulation exercise: (1) draw treatment assignments $\Dn \sim \localdesigncond(\psi, \propfn)$, (2) reveal outcomes $\wh Y_i = \wh Y_i(\Di)$ and (3) form each estimator $\estadj$. 
We report average point estimates and standard errors over $N=2000$ Monte Carlo repetitions of this procedure. 

The first block of Table \ref{table:empirical_simulations} uses this imputation procedure to reproduce the empirical results in Table \ref{table:empirical}, stratifying by quartiles of CHP vote share and adjusting for exactly the same covariates. 
The standard errors are very similar to those in the empirical analysis, which provides some validation for this imputation exercise. 
In the second block of Table \ref{table:empirical_simulations}, we simulate a design with fine stratification on 2015 CHP vote share, rather than just stratifying by quartiles of the vote share as in \cite{baysan2022}.
We used a matched $11$-tuples design, letting $\Dn \sim \localdesigncond(\psi, \propfn)$ for $\propfn=2/11$ and $\psi = (\text{2015 CHP vote share})$. 
Covariates $h(X)$ are as above, with $z(\psi) = \psi$. 
In the third block, we simulate a matched pairs design $\Dn \sim \localdesigncond(\psi, 1/2)$. 
Note that $\propfn=1/2$ was infeasible in the original experiment due to the high cost of treatment.
The last block uses the design $\Dn \sim \localdesigncond(\psi_{alt}, \propfn)$ for $\psi_{alt} = (\text{CHP vote share}, \text{Num.\ of registered voters}, \text{Num.\ of valid votes})$, $\propfn=2/11$, and covariates $h = \text{Turnout}$. 

We make some brief observations about this simulation exercise.
First, note that the Naive and Lin adjustment are strictly less efficient than unadjusted estimation under simulated fine stratification, consistent with Theorems \ref{thm:lin} and Section \ref{thm:naive}. 
Lin and partialled Lin with $z(\psi)$ controls are the most efficient.  
Adjustment for extra covariates $h$ doesn't significantly improve efficiency relative to the baseline efficiency gain from finely stratifying on $\psi$ and adjusting for $z(\psi)$ ex-post. 
Using a matched pairs design $\propfn = 1/2$ improves efficiency, though the improvement is small considering that this design would require providing the information campaign to $175$ extra neighborhoods. 
Finally, fine stratification on $\psi_{alt}$ significantly reduces efficiency.
This is because the extra covariates are not very predictive of outcomes, but stratifying on these covariates force us to use worse matches on the important covariate $\psi = (\text{2015 CHP vote share})$. 

\section{Discussion and Recommendations for Practice} \label{section:conclusion}

Stratified randomization and covariate adjustment are both commonly used in the design and analysis of experiments. 
In general, experimenters should stratify on a few variables $\psi(X)$ expected to be most predictive of outcomes at design-time, and plan to adjust for imbalances in the remaining covariates $h(X)$ ex-post, as discussed in Section \ref{section:efficiency-semiparametric-regression}. 
Our analysis showed that under stratified randomization, the usual regression adjusted estimators can be inefficient. 
Motivated by this, we provide feasible alternatives that are asymptotically optimal in the class of linearly adjusted estimators. 
We conclude by giving some recommendations for empirical practice based on the theory, simulations, and empirical results above. \medskip 

We recommend that applied researchers use either (1) the Lin estimator with parametric strata controls $z(\psi)$ (e.g.\ $z(\psi) = \psi)$ or (2) the partialled Lin estimator with parametric controls $z(\psi)$, since these estimators performed the best across our simulations and empirical application.    
Lin with parametric controls $z(\psi)$ is efficient under a rich covariates condition (Section \ref{section:efficiency-tilted}), while partialled Lin is generically efficient (Section \ref{section:further-adjustment}).
Both estimators are robust to treatment effect heterogeneity, while the strata fixed effects estimator (Theorem \ref{thm:naive-fe}) is not unless $\propfn = 1/2$.

\medskip

In our simulations, partialled Lin had good finite sample performance in regimes where $n$ was large relative to $\dim(\psi)$, especially when $E[h|\psi]$ was very nonlinear.
Lin with $z(\psi)=\psi$ controls performed better when $\dim(\psi)$ was large relative to $n$, or if $E[h | \psi]$ was approximately linear. 
To decide which regime we are in, we suggest model selection using a variance pre-test, choosing Lin if $\wh V(\wh \gamma_L) \leq \wh V(\coefflinpartial)$ and partialled Lin otherwise. 
This adaptive estimator (\textbf{Ad} in Section 6) was efficient in both regimes and had good coverage properties.
We leave a more general study of such post model-selection estimators in this context to future work. \medskip

Regardless of the adjustment strategy, we recommend using the asymptotically exact confidence intervals provided in Section \ref{section:inference}.
Our simulations showed close to nominal coverage for these confidence intervals across all considered estimators. 
By contrast, confidence intervals based on the HC2 robust variance estimator often had significant overcoverage.

\bibliography{design_references_adjustment.bib}

\begin{thebibliography}{33}
\providecommand{\natexlab}[1]{#1}
\providecommand{\url}[1]{\texttt{#1}}
\expandafter\ifx\csname urlstyle\endcsname\relax
  \providecommand{\doi}[1]{doi: #1}\else
  \providecommand{\doi}{doi: \begingroup \urlstyle{rm}\Url}\fi

\bibitem[Abadie and Imbens(2008)]{abadie2008}
Alberto Abadie and Guido~W. Imbens.
\newblock Estimation of the conditional variance in paired experiments.
\newblock \emph{Annales d'Economie et de Statistique}, pages 175--187, 2008.

\bibitem[Ansel et~al.(2018)Ansel, Hong, and Li]{ansel2018}
Jason Ansel, Han Hong, and Jessie Li.
\newblock Ols and 2sls in randomized and conditionally randomized experiments.
\newblock \emph{Jahrb{\"u}cher f{\"u}r National {\"o}konomie und Statistik},
  2018.

\bibitem[Armstrong(2022)]{armstrong2022}
Tim Armstrong.
\newblock Asymptotic efficiency bounds for a class of experimental designs.
\newblock \emph{arXiv preprint arXiv:2205.02726}, 2022.

\bibitem[Bai(2022)]{bai2020pairs}
Yuehao Bai.
\newblock Optimality of matched-pair designs in randomized controlled trials.
\newblock \emph{American Economic Review}, 2022.

\bibitem[Bai et~al.(2021)Bai, Romano, and Shaikh]{bai2021inference}
Yuehao Bai, Joseph~P. Romano, and Azeem~M. Shaikh.
\newblock Inference in experiments with matched pairs.
\newblock \emph{Journal of the American Statistical Association}, 2021.

\bibitem[Bai et~al.(2024{\natexlab{a}})Bai, Guo, Shaikh, and
  Tabord-Meehan]{bai2023noncompliance}
Yuehao Bai, Hongchang Guo, Azeem~M. Shaikh, and Max Tabord-Meehan.
\newblock Inference in experiments with matched pairs and imperfect compliance.
\newblock \emph{arXiv preprint arXiv:2307.13094}, 2024{\natexlab{a}}.

\bibitem[Bai et~al.(2024{\natexlab{b}})Bai, Jiang, Romano, Shaikh, and
  Zhang]{bai2023adjustment}
Yuehao Bai, Liang Jiang, Joseph~P. Romano, Azeem~M. Shaikh, and Yichong Zhang.
\newblock Covariate adjustment in experiments with matched pairs.
\newblock \emph{Journal of Econometrics}, 2024{\natexlab{b}}.

\bibitem[Bai et~al.(2024{\natexlab{c}})Bai, Tabord-Meehan, and
  Liu]{bai2023tuples}
Yuehao Bai, Max Tabord-Meehan, and Jizhou Liu.
\newblock Inference for matched tuples and fully blocked factorial designs.
\newblock \emph{Quantitative Economics}, 2024{\natexlab{c}}.

\bibitem[Baysan(2022{\natexlab{a}})]{baysan2022}
Ceren Baysan.
\newblock Persistent polarizing effects of persuasion: Experimental evidence
  from turkey.
\newblock \emph{American Economic Review}, November 2022{\natexlab{a}}.

\bibitem[Baysan(2022{\natexlab{b}})]{baysan2022data}
Ceren Baysan.
\newblock Data and code for: Persistent polarizing effects of persuasion:
  Experimental evidence from turkey.
\newblock Nashville, TN: American Economic Association, 2022. Ann Arbor, MI:
  Inter-university Consortium for Political and Social Research, 2022-10-19.
  https://doi.org/10.3886/E172061V1, December 2022{\natexlab{b}}.

\bibitem[Bugni et~al.(2018)Bugni, Canay, and Shaikh]{bugni2018inference}
Federico~A. Bugni, Ivan~A. Canay, and Azeem~M. Shaikh.
\newblock Inference under covariate-adaptive randomization.
\newblock \emph{Journal of the American Statistical Association}, 2018.

\bibitem[Chang(2023)]{chang2023}
Haoge Chang.
\newblock Design-based estimation theory for complex experiments.
\newblock \emph{arXiv preprint arXiv:2311.06891}, 2023.

\bibitem[Cytrynbaum(2022)]{cytrynbaum2022local}
Max Cytrynbaum.
\newblock Designing representative and balanced experiments by local
  randomization.
\newblock \emph{arXiv preprint arXiv:2111.08157v1}, 2022.

\bibitem[Cytrynbaum(2023)]{cytrynbaum2023}
Max Cytrynbaum.
\newblock Optimal stratification of survey experiments.
\newblock \emph{arXiv preprint arXiv:2111.08157}, 2023.

\bibitem[Derigs(1988)]{derigs1988}
Ulrich Derigs.
\newblock Solving non-bipartite matching problems via shortest path techniques.
\newblock \emph{Annals of Operations Research}, 13:\penalty0 225--261, 1988.

\bibitem[Fogarty(2018)]{fogarty2018b}
Colin~B. Fogarty.
\newblock Regression-assisted inference for the average treatment effect in
  paired experiments.
\newblock \emph{Biometrika}, 105\penalty0 (4), 2018.

\bibitem[Hahn(1998)]{hahn1998}
Jinyong Hahn.
\newblock On the role of the propensity score in efficient semiparametric
  estimation of average treatment effects.
\newblock \emph{Econometrica}, 1998.

\bibitem[Imbens and Angrist(1994)]{imbens1994}
Guido Imbens and Joshua~D. Angrist.
\newblock Identification and estimation of local average treatment effects.
\newblock \emph{Econometrica}, 62, 1994.

\bibitem[Imbens and Rubin(2015)]{imbens2015}
Guido~W. Imbens and Donald~B. Rubin.
\newblock \emph{Causal Inference for Statistics, Social, and Biomedical
  Sciences: An Introduction}.
\newblock Cambridge University Press, 2015.

\bibitem[Jiang et~al.(2024)Jiang, Linton, Tang, and
  Zhang]{jiang2023noncompliance}
Liang Jiang, Oliver~B. Linton, Haihan Tang, and Yichong Zhang.
\newblock Improving estimation efficiency via regression-adjustment in
  covariate-adaptive randomizations with imperfect compliance.
\newblock \emph{Review of Economics and Statistics}, 2024.

\bibitem[Leeb and Potscher(2005)]{Leeb2005}
Hannes Leeb and Benedikt~M. Potscher.
\newblock Model selection and inference: Facts and fiction.
\newblock \emph{Econometric Theory}, 2005.

\bibitem[Lin(2013)]{lin2013}
Winston Lin.
\newblock Agnostic notes on regression adjustments to experimental data:
  Reexamining freedman's critique.
\newblock \emph{The Annals of Applied Statistics}, 7\penalty0 (1):\penalty0
  295--318, 2013.

\bibitem[Liu and Yang(2020)]{liu2020}
Hanzhong Liu and Yuehan Yang.
\newblock Regression-adjusted average treatment effect estimates in stratified
  randomized experiments.
\newblock \emph{Biometrika}, 2020.

\bibitem[Lu and Liu(2024)]{lu2022}
Xin Lu and Hanzhong Liu.
\newblock Tyranny-of-the-minority regression adjustment in randomized
  experiments.
\newblock \emph{Journal of the American Statistical Association}, June 2024.

\bibitem[Ma et~al.(2022)Ma, Tu, and Liu]{ma2020}
Wei Ma, Fuyi Tu, and Hanzhong Liu.
\newblock Regression analysis for covariate-adaptive randomization: A robust
  and efficient inference perspective.
\newblock \emph{Statistics in Medicine}, 2022.

\bibitem[Negi and Wooldridge(2021)]{negi2021}
Akanksha Negi and Jeffrey~M. Wooldridge.
\newblock Revisiting regression adjustment in experiments with heterogeneous
  treatment effects.
\newblock \emph{Econometric Reviews}, 2021.

\bibitem[Reluga et~al.(2024)Reluga, Ye, and Zhao]{reluga2022}
Katarzyna Reluga, Ting Ye, and Qingyuan Zhao.
\newblock A unified analysis of regression adjustment in randomized
  experiments.
\newblock \emph{Electronic Journal of Statistics}, 2024.

\bibitem[Ren(2023)]{ren2023noncompliance}
Jiyang Ren.
\newblock Model-assisted complier average treatment effect estimates in
  randomized experiments with non-compliance and a binary outcome.
\newblock \emph{Journal of Business and Economic Statistics}, 2023.

\bibitem[Robins and Rotnitzky(1995)]{robins95}
James~M. Robins and Andrea Rotnitzky.
\newblock Semiparametric efficiency in multivariate regression models with
  missing data.
\newblock \emph{Journal of the American Statistical Association}, 90:\penalty0
  122--129, 1995.

\bibitem[Robinson(1988)]{robinson88}
Peter~M. Robinson.
\newblock Root-n-consistent semiparametric regression.
\newblock \emph{Econometrica}, 56\penalty0 (4), 1988.

\bibitem[Wang et~al.(2021)Wang, Wang, and Liu]{wang2021}
Xinhe Wang, Tingyu Wang, and Hanzhong Liu.
\newblock Rerandomization in stratified randomized experiments.
\newblock \emph{Journal of the American Statistical Association}, 2021.
\newblock Working Paper.

\bibitem[Ye et~al.(2022)Ye, Shao, Yi, and Zhao]{ye2022}
Ting Ye, Jun Shao, Yanyao Yi, and Qingyuan Zhao.
\newblock Toward better practice of covariate adjustment in analyzing
  randomized clinical trials.
\newblock \emph{Journal of the American Statistical Association}, 00\penalty0
  (0), 2022.

\bibitem[Zhu et~al.(2024)Zhu, Liu, and Yang]{zhu2022}
Ke~Zhu, Hanzhong Liu, and Yuehan Yang.
\newblock Design-based theory for lasso adjustment in randomized block
  experiments with a general blocking scheme.
\newblock \emph{arXiv preprint arXiv:2109.11271}, 2024.

\end{thebibliography}
\clearpage

\appendix
\renewcommand{\thesection}{A}
\pagenumbering{arabic}\renewcommand{\thepage}{\arabic{page}}

\begin{center}
{\Large Supplement to ``Covariate Adjustment in Stratified Experiments''}
\vskip 24pt
{\large Max Cytrynbaum}
\end{center}

\section{Appendix} \label{section:proofs}

\subsection{Experiments with Noncompliance} \label{appendix:noncompliance}
In this section, we extend our main results to the case of experiments with imperfect compliance.
The theorems in this section are simple corollaries of our main results. 
For completeness, full proofs are provided in Section \ref{section:proofs-noncompliance}.

Previously, \cite{ansel2018} studied covariate adjustment in experiments with noncompliance and iid or coarsely stratified treatment assignment. 
\cite{bai2023noncompliance} study matched pairs experiments with noncompliance. 
See also \cite{jiang2023noncompliance} and \cite{ren2023noncompliance} for nonlinear adjustment in coarsely stratified experiments and completely randomized experiments with noncompliance, respectively. 

Let $z \in \{0, 1\}$ denote a binary instrument.
Let $D(z)$ be the potential treatments and $Y(d, z) = Y(d)$ the potential outcomes, satisfying exclusion.
Define the intention-to-treat (ITT) potential outcomes $W_i(z) = Y_i(D_i(z))$, so that $Y_i = Z_i W_i(1) + (1-Z_i) W_i(0)$ and $D_i = Z_i D_i(1) + (1-Z_i) D_i(0)$. 
Impose monotonicity $D(1) \geq D(0)$ and positive compliance $\ated = P(D(1) > D(0)) > 0$. 
Define the ITT effect $\atew = E[W(1) - W(0)]$. 
Under these assumptions, the parameter $\atel \equiv \atew / \ated = E[Y(1)-Y(0)|D(1) > D(0)]$ is the local average treatment effect (LATE) (\cite{imbens1994}).
To estimate $\atel$, we consider adjusted Wald estimators of the form
\begin{equation} \label{equation:wald-estimator}
\wh \tau_{adj} = \frac{\bar W_1 - \bar W_0 - \wh \gamma_W'(\hbarone - \hbarzero)\propconstant}{\bar D_1 - \bar D_0 - \wh \gamma_D'(\hbarone - \hbarzero)\propconstant}    
\end{equation}
To analyze $\wh \tau_{adj}$, we require that Assumption \ref{assumption:moment-conditions} holds for both potential outcomes $W(z)$ and $D(z)$ and covariates $h(X)$, and also impose Assumption \ref{assumption:conditional-variance}.
Suppose the adjustment coefficients $(\wh \gamma_W, \wh \gamma_D) = (\gamma_W, \gamma_D) + \op(1)$. 
Our first result is a consequence of Theorem \ref{thm:adjusted-efficiency}. 
To state the result, we define the modified potential outcomes $Q(z) = W(z) - \atel D(z)$ for $z \in \{0,1\}$ and modified adjustment coefficient $\gamma_Q = \gamma_W - \atel \gamma_D$.

\begin{thm} \label{thm:adjusted-efficiency-late}
If $\Zn \sim \localdesigncond(\psi, \propfn)$ then $\rootn(\wh \tau_{adj} - \atel) \convwprocess \normal(0, V(\gamma_Q) / \ated^2)$ with 
\[
V(\gamma_Q) = \var(\catefn_Q) + E\big[\var(\balancefn_Q - \gamma_Q'h| \psi)\big] + E\left[\frac{\hk_{1Q}(X)}{\propfn} + \frac{\hk_{0Q}(X)}{1-\propfn}\right]. 
\]
\end{thm}
The terms $\catefn_Q(X) = E[Q(1)-Q(0)|X]$, similarly for $\balancefn_Q$ and $\hk_{zQ}$, substituting the potential outcomes $Q(z)$ for $Y(d)$ in each formula.

\medskip

\textbf{Optimal Adjustment}. Let $\wh \gamma_Q = \wh \gamma_W - \atel \wh \gamma_D$ and define the adjustment scheme $\wh \tau_{adj}$ to be efficient if $\wh \gamma_Q \convp \gamma_Q^* \in \argmin_{\gamma} V(\gamma)$.
We construct efficient adjusted Wald estimators using the generic efficient estimators of Section \ref{section:generic-efficiency}.
Let $\est_k^W$ and $\est_k^D$ for $k \in \{PL, GO, TM\}$ be any of the generic efficient estimators of Section, plugging in outcomes $W$ or $D$ in place of $Y$.
For example, $\est_{PL}^W$ is the coefficient on $Z_i$ in the regression $W_i \sim (1, \hicheck) + Z_i(1, \hicheck)$ and $\est_{PL}^D$ the coefficient on $Z_i$ in $D_i \sim (1, \hicheck) + Z_i(1, \hicheck)$. 
Define the $\late$ estimators $\estadjlatek = \est^W_k / \est^D_k$ for $k \in \{PL, GO, TM\}$.
Our next theorem is a consequence of the efficiency results in Section \ref{section:generic-efficiency}.
\begin{thm} \label{thm:generic-efficiency-late}
Suppose $\Zn \sim \localdesigncond(\psi, \propfn)$.
For each $k \in \{PL, GO, TM\}$, the estimator $\estadjlatek$ is efficient with $\rootn(\estadjlatek - \atel) \convwprocess \normal(0, V^*)$ for $V^* = \min_{\gamma} V(\gamma)$.
\end{thm}
Finally, we provide asymptotically exact inference on $\atel$ using the adjusted estimators $\estadjlatek$ above. 
Define the augmented outcomes $Q_i^a = W_i - \estadjlatek D_i - \hi'(\wh \gamma_W - \estadjlatek \wh \gamma_D)$.
Let $\varestone^q$, $\varestzero^q$, and $\varestcross^q$ be the variance estimators in Equation \ref{equation:variance-estimator}, plugging in $Q_i^a$ in place of $Y_i^a$.
Define the variance estimator
\begin{equation} \label{equation:variance-estimator-late}
\wh V = \frac{1}{(\est_k^D)^2} \left [\var_n \left( \frac{(\Di - \propfn) Q_i^a}{\propfn-\propfn^2} \right) - \varestone^q - \varestzero^q - 2 \varestcross^q \right ]
\end{equation}

\begin{thm} \label{thm:inference-late}
Suppose $\Zn \sim \localdesigncond(\psi, \propfn)$.
Then $\wh V = V^* + \op(1)$. 
\end{thm}
Theorems \ref{thm:adjusted-efficiency-late} and \ref{thm:inference-late} show that the confidence interval $\wh C = [\estadjlatek \pm \varest^{1/2} c_{1-\alpha/2} / \rootn]$ with $c_{\alpha} = \Phi \inv(\alpha)$ is asymptotically exact in the sense that $P (\atel \in \wh C ) = 1-\alpha + o(1)$.

\subsection{Varying Propensities} \label{appendix:varying-propensities}
In this section, we extend our results to fine stratification with varying propensities $\propfn(\psi)$. 
To that end, let $\propfn(\psi) \in \{a_l / k_l: l \in L\}$ with $|L| < \infty$ a finite index set.
\cite{cytrynbaum2023} extends Definition \ref{defn:local_randomization} to non-constant $\propfn(\psi)$ by the following double stratification procedure:  
\begin{enumerate}[label={(\arabic*)}, itemindent=.5pt, itemsep=.4pt] 
\item Partition the units $\{1, \dots, n\}$ into propensity strata $S_l \equiv \{i: \propfn(X_i) = a_l / k_l\}$. 
\item In each propensity stratum $S_l$, draw samples $(D_i)_{i \in S_l} \sim \localdesigncond(\psi, a_l / k_l)$.
\end{enumerate}
To implement this, we run the algorithm of \cite{cytrynbaum2023} to match units into groups of $k_l$ separately in each propensity stratum $S_l$, drawing treatment assignments $(\Di)_{i \in \group} \sim \crdist(a_l/k_l)$ independently for each $\group \in \mc G_l$.
Define $\estadj(\gamma)$ to be the AIPW estimator of Section \ref{section:efficiency-semiparametric-regression}, with linear models $f_d(X_i) = \gamma_d'h(X_i)$ for $d \in \{0, 1\}$, so that 
\[
\estadj(\gamma) = (\gamma_1 - \gamma_0)'\en[\hi] + \en\left[\frac{\Di (Y_i - \gamma_1'\hi)}{\propfn(\psii)}\right] - \en\left[\frac{(1-\Di) (Y_i - \gamma_0'\hi)}{1-\propfn(\psii)}\right].
\]
Define $\gamma = (\gamma_0, \gamma_1)$ and weighted covariates $\hpropi = \left (\hi \sqrt{\frac{\propfni}{1-\propfni}}, \hi \sqrt{\frac{1-\propfni}{\propfni}} \right)$. 
Under assumption \ref{assumption:moment-conditions}, Theorem \ref{thm:adjusted-efficiency} may be extended to show that if $\wh \gamma \convp \gamma$ and $\Dn \sim \localdesigncond(\psi, \propfn(\psi))$ then $\rootn(\estadj(\wh \gamma) - \ate) \convwprocess \normal(0, V(\gamma))$ with variance 
\[
V(\gamma) = \var(\catefn(X)) + E\left [\var\left (\balancefn - \gamma' h^{\propfn} \big | \psi \right ) \right] + E\left[\frac{\hk_1(X)}{\propfn(\psi)} + \frac{\hk_0(X)}{1-\propfn(\psi)}\right]. 
\]
The optimal adjustment coefficient is $\gamma^* = E[\var(\hpropi | \psii)]\inv E[\cov(\hpropi, \balancefni | \psii)]$ if the condition $E[\var(\hpropi | \psii)] \succ 0$ is satisfied.
Let $k_i$ denote the size of the group that unit $i$ belongs to.
Extending the work in Section \ref{section:generic-efficiency}, the estimator 
\[
\wh \gamma = \en\left [\hpropchecki (\hpropchecki)' \frac{k_i}{k_i-1} \right]\inv \en\left[\hpropchecki Y_i^{TM} \frac{k_i}{k_i-1} \right]
\]
with weighted outcomes $Y_i^{TM} = \Di Y_i (1-\propfni)^{1/2} \propfni^{-3/2} + (1-\Di) Y_i \propfni^{1/2}(1-\propfni)^{-3/2}$ has $\wh \gamma = \gamma^* + \op(1)$. 
Then the estimator $\estadj(\wh \gamma)$ is efficient in the sense of achieving the minimal variance $\min_{\gamma} V(\gamma)$. 

\subsection{Non-Interacted Regression Adjustment} \label{section:naive-adjustment}
For completeness, before continuing we describe the asymptotic behavior of the commonly used non-interacted regression estimator under stratified designs.
Let $\estnaive$ be the coefficient on $\Di$ in $Y \sim 1 + D + h$.

\begin{thm} \label{thm:naive}
Suppose Assumptions \ref{assumption:moment-conditions} and \ref{assumption:conditional-variance} hold.
The estimator has representation $\estnaive = \est - \coeffnaive'(\hbarone - \hbarzero) + \Op(n \inv)$.
If $\Dn \sim \localdesigncond(\psi, \propfn)$ then $\rootn(\estnaive - \ate) \convwprocess \normal(0, V)$ with variance 
\begin{align*}
V = \var(\catefn(X)) + E[\var(\balancefn - \coeffnaivepop'h | \psi)] + E\left[\frac{\hk_1(X)}{\propfn} + \frac{\hk_0(X)}{1-\propfn}\right]. 
\end{align*}
The coefficient $\coeffnaivepop = \argmin_{\gamma \in \mr^{\dimh}} \var(f - \gamma'h) $ for target function 
\[
f(x) = \ceffn_1(x) \sqrt{\frac{\propfn}{1-\propfn}} + \ceffn_0(x)\sqrt{\frac{1-\propfn}{\propfn}} 
\]
with $f(x) \not = \balancefn(x)$ in general. 
The fixed effects estimator is efficient if either $\propfn = 1/2$ or $\cov(h, Y(1) - Y(0)) = 0$. 
\end{thm}

Theorem \ref{thm:naive} shows that $\estnaive$ is generally inefficient since it uses the wrong objective function.
In particular, the target function $f(x) \not = \balancefn(x)$ unless $\propfn = 1/2$.
Also, the limiting coefficient $\coeffnaivepop$ minimizes marginal instead of conditional variance.
The results in Section \ref{section:inference} show how to construct asymptotically exact confidence intervals for the $\ate$ using $\estnaive$.

\subsection{Nonlinear Adjustment} \label{appendix:nonlinear-adjustment}

Alternately, we may consider general nonlinear covariate adjustment strategies. 
Let $\wh h(x)$ be a function estimated in some class $\mc H$ and consider the adjusted estimator
\begin{align*}
\estadj(\wh h) = \en\left[\frac{(Y_i - \wh h(X_i))(\Di - \propfni)}{\propfni - \propfni^2} \right].   
\end{align*}
For example, the usual AIPW estimator in Section \ref{section:efficiency-semiparametric-regression} can be shown to take this form.   
Linear adjustment corresponds to the parametric family $\mc H = \{h(x)'\gamma: \gamma \in \mr^{d_h}\}$. 
Similar to \cite{bai2023adjustment}, suppose that for some function $h(X) \in L_2$ the equicontinuity condition holds
\[
\rootn \en\left[\frac{(\wh h - h)(X_i)(\Di - \propfni)}{\propfni - \propfni^2} \right] = \op(1).   
\]
Theorem \ref{thm:adjusted-efficiency} can be extended to show that if $\Dn \sim \localdesigncond(\psi, \propfn(\psi))$ then $\rootn (\estadj(\wh h) - \ate) \convwprocess \normal(0, V(h))$ with asymptotic variance
\[
V(h) = \var(\catefn(X)) + E\left [\var\left (\balancefn - h / \propconstant(\psi) \big | \psi \right ) \right] + E\left[\frac{\hk_1(X)}{\propfn(\psi)} + \frac{\hk_0(X)}{1-\propfn(\psi)}\right] 
\]
for $\propconstant(\psi) = \sqrt{\propfn(\psi) - \propfn(\psi)^2}$.
One natural extension of the current work would be to solve a general version of the optimal adjustment problem over a nonlinear or general nonparametric function class $\mc H$.  
\begin{equation}
\min_{h \in \mc H} E\left [\var\left (\balancefn - h / \propconstant(\psi) \big | \psi \right ) \right]
\end{equation} 

This requires new technical tools, the development of which we leave to future work.

\subsection{Proofs for Section \ref{section:optimal-adjustment}} \label{section:proofs-adjustment}

\begin{proof}[Proof of Theorem \ref{thm:adjusted-efficiency}]
First, note that since $E[|h|_2^2] < \infty$ we may apply Lemma \lemmastochasticbalance \, of \cite{cytrynbaum2023} to show that  
\begin{align*}
\wh \gamma'(\hbarone - \hbarzero) \propconstant &= \wh \gamma'\en\left [\frac{(\Di - \propfn)}{\sqrt{\propfn - \propfn^2}} \hi \right] = \gamma'\en\left [\frac{(\Di - \propfn)}{\sqrt{\propfn - \propfn^2}} \hi \right] + (\wh \gamma - \gamma)'\en\left [\frac{(\Di - \propfn)}{\sqrt{\propfn - \propfn^2}} \hi \right]  \\
&= \gamma'\en\left [\frac{(\Di - \propfn)}{\sqrt{\propfn - \propfn^2}} \hi \right] + \op(\negrootn) = \gamma'(\hbarone - \hbarzero) \propconstant + \op(\negrootn). 
\end{align*}

Define auxiliary potential outcomes $Z(d) = Y(d) - \propconstant \gamma' h(X)$ for $d \in \{0,1\}$ with $Z_i = Z(\Di)$.
Summarizing, we have shown that $\estadj = \bar Z_1 - \bar Z_0 + \op(\negrootn)$.
Observe that $E[Z(d)^2] \lesssim E[Y(d)^2] + \propconstant^2 |\gamma|^2_2 E[|h(X)|_2^2] < \infty$. 
Then we may apply the general version of Theorem \thmclt \, in \cite{cytrynbaum2023} (Equation 3.6).
Setting $q = 1$ and $\psi_1 = \psi_2$ and applying the theorem to the auxiliary potential outcomes $Z(d)$, we have $\rootn(\estadj - \ate) \convwprocess \normal(0, V)$ 
\[
V = \var(\catefn_Z(X)) + E[\var(\balancefn_Z(X; \propfn) | \psi)] + E\left[\frac{\hk_{1, Z}(X)}{\propfn} + \frac{\hk_{0, Z}(X)}{1-\propfn}\right]. 
\]
Calculating, we have $\catefn_Z(X) = E[Z(1) - Z(0) | X] = \catefn(X)$ and 
\[
\balancefn_Z(X) = E[Z(1)|X] \left (\frac{1-\propfn}{\propfn} \right)^{1/2} + \, E[Z(0)|X] \left(\frac{\propfn}{1-\propfn}\right)^{1/2} = \balancefn(X; \propfn) - \gamma'h(X).
\]
Finally, $\hk_{d, Z}(X) = \var(Z(d) | X) = \var(Y(d) | X) = \hk_d(X)$. 
Then the variance $V$ above is
\[
V = \var(\catefn(X)) + E[\var(\balancefn - \gamma'h | \psi)] + E\left[\frac{\hk_1(X)}{\propfn} + \frac{\hk_0(X)}{1-\propfn}\right] 
\]
as claimed.
\end{proof}

\begin{proof}[Proof of Theorem \ref{thm:lin}]
Define $W_i = (1,\hitilde)$.
First consider the regression $Y_i \sim \Di \Wi + (1-\Di) \Wi$, with coefficients $(\wh \gamma_1, \wh \gamma_0)$.
By Frisch-Waugh and orthogonality of regressors, $\wh \gamma_1$ is numerically equivalent to the regression coefficient $Y_i \sim \Di W_i$ and similarly for $\wh \gamma_0$.  
Then consider $Y_i = \Di W_i' \wh \gamma_1 + e_i$ with $\en[e_i (\Di W_i)] = 0$.
Then $\Di Y_i = \Di W_i' \wh \gamma_1 + \Di e_i$ and $\en[\Di e_i (\Di W_i)] = \en[e_i (\Di W_i)] = 0$.
Then $\wh \gamma_1$ can be identified with the regression coefficient of $Y_i \sim W_i$ in the set $\{i: \Di=1\}$.
Let $\wh \gamma_1 = (\wh c_1, \wh \alpha_1)$. 
By the usual OLS formula $\wh c_1 = \en[Y_i|\Di=1] - \wh \alpha_1' \en[\hitilde|\Di=1]$ and $\wh \alpha_1 = \var_n(\hitilde | \Di = 1)\inv \cov_n(\hitilde, Y_i | \Di = 1)$. 
Similar formulas hold for $\Di = 0$ by symmetry.
Next, note that for $m = \dimh + 1$ the original regressors can be written as a linear transformation
\[
\begin{pmatrix} \Di W_i \\ W_i \end{pmatrix} = \begin{pmatrix} I_m & 0 \\ I_m & I_m \end{pmatrix} \begin{pmatrix} \Di W_i \\ (1-\Di) W_i \end{pmatrix}.
\]
Then the OLS coefficients for the original regression $Y_i \sim \Di W_i + W_i$ are given by the change of variables formula
\[
\left(\begin{pmatrix} I_k & 0 \\ I_k & I_k \end{pmatrix}'\right) \inv \begin{pmatrix} \wh \gamma_1 \\ \wh \gamma_0 \end{pmatrix} = \begin{pmatrix} I_k & -I_k \\ 0 & I_k \end{pmatrix} \begin{pmatrix} \wh \gamma_1 \\ \wh \gamma_0 \end{pmatrix} = \begin{pmatrix} \wh \gamma_1 - \wh \gamma_0 \\ \wh \gamma_0 \end{pmatrix}.
\]
In particular, the coefficient on $\Di$ in the original regression is 
\begin{align*}
\estlin = \wh c_1 - \wh c_0 &= \en[Y_i - \wh \alpha_1' \hitilde |\Di=1] - \en[Y_i - \wh \alpha_0' \hitilde |\Di=0] \\
&= \est - \en \left[ \frac{\wh \alpha_1' \hitilde \Di}{\propfn} \right] + \en \left[ \frac{\wh \alpha_0' \hitilde (1-\Di)}{1-\propfn} \right] \\
&= \est - \en \left[ \frac{\wh \alpha_1' \hi (\Di - \propfn)}{\propfn} \right] - \en \left[ \frac{\wh \alpha_0' \hi (\Di - \propfn)}{1-\propfn} \right] \\
&= \est - \left( \wh \alpha_1 (1-p) + \wh \alpha_0 p  \right)' \en \left[ \frac{\hi (\Di - \propfn)}{\propfn(1-\propfn)} \right] \\
&= \est - \left( \wh \alpha_1 \sqrt{\frac{1-p}{\propfn}} + \wh \alpha_0 \sqrt{\frac{p}{1-\propfn}} \right)' (\hbarone - \hbarzero)\propconstant. 
\end{align*}
The second equality since $\en[\Di] = \propfn$ identically.
The third equality by expanding $\Di = \Di - \propfn + \propfn$ and using $\en[\hitilde] = 0$ and $\en[(\Di - \propfn)\en[\hi]] = 0$. 
The fourth equality is algebra and collecting terms.
The fifth equality since $\hbarone - \hbarzero = \en [ \hi (\Di - \propfn) / \propfn(1-\propfn)]$ again using $\en[\Di] = \propfn$ and $\propconstant = \sqrt{\propfn(1-\propfn)}$ by definition.

Next, consider the coefficient $\wh \alpha_1 = \var_n(\hitilde | \Di = 1)\inv \cov_n(\hitilde, Y_i | \Di = 1)$.
We have $\var_n(\hitilde | \Di = 1) = \propfn \inv \en[\Di \hitilde \hitilde'] - \propfn^{-2} \en[\Di \hitilde]\en[\Di \hitilde']$. 
Let $1 \leq t, t' \leq \dimh$. 
Then we may compute $\en[\Di \tilde h_{it} \tilde h_{it'}] = \en[(\Di - \propfn) \tilde h_{it} \tilde h_{it'}] + \propfn \en[\tilde h_{it} \tilde h_{it'}]$. 
Expanding the first term 
\begin{align*}
\en[(\Di - \propfn) \tilde h_{it} \tilde h_{it'}] &= \en[(\Di - \propfn) h_{it} h_{it'}] - \en[h_{it}]\en[(\Di - \propfn) h_{it'}] - \en[h_{it'}]\en[(\Di - \propfn) h_{it}] \\
&+ \en[h_{it'}]\en[h_{it}]\en[\Di-\propfn] = \op(1).    
\end{align*}
The final equality follows since $\en[(\Di - \propfn) h_{it} h_{it'}] = \op(1)$ by applying Lemma \lemmastochasticbalance \, of \cite{cytrynbaum2023}, using that $E[|h_{it} \tilde h_{it'}|] \leq E[|\hi|_2^2] < \infty$, and similarly for the other terms.
By WLLN, we also have $\en[\tilde h_{it} \tilde h_{it'}] \convp \var(h)$.
Then by continuous mapping $\var_n(\hitilde | \Di = 1)\inv = \var(h)\inv + \op(1)$.
Similar reasoning shows $\cov_n(\hitilde, Y_i | \Di = 1) = \cov(\hi, Y_i(1)) + \op(1)$. 

\medskip

Then we have shown $\wh \alpha_1 = \var(h) \inv \cov(h, Y(1)) + \op(1) = \var(h) \inv \cov(h, \ceffn_1) + \op(1)$. 
By symmetry, we also have $\wh \alpha_0 = \var(h) \inv \cov(h, \ceffn_0) + \op(1)$. 
Putting this all together, we have $\wh \alpha_1 \sqrt{\frac{1-p}{\propfn}} + \wh \alpha_0 \sqrt{\frac{p}{1-\propfn}} = \var(h) \inv \cov(h, \balancefn) + \op(1) = \coefflinpop + \op(1)$.
Then by Theorem \ref{thm:adjusted-efficiency}, $\rootn(\estlin - \ate) \convwprocess \normal(0 ,V)$ with
\[
V = \varlimit(\coefflinpop) = \var(\catefn(X)) + E\bigg[\var(\balancefn - \coefflinpop'h | \psi)\bigg] + E\left[\frac{\hk_1(X)}{\propfn} + \frac{\hk_0(X)}{1-\propfn}\right] 
\]
as claimed. 
The claimed representation follows from the change of variables formula above, since $\wh \alpha_1 = \wh a_1 + \wh a_0$ and $\wh \alpha_0 = \wh a_0$.
This finishes the proof.
\end{proof}

\begin{proof}[Proof of Theorem \ref{thm:naive}]
We have $Y_i = \wh c + \estnaive \Di + \coeffnaive' \hi + e_i$ with $\en[e_i(1, \Di, \hi)] = 0$.
By applying Frisch-Waugh twice, we have $\tilde Y_i = \estnaive (\Di - \propfn) + \coeffnaive' \hitilde + e_i$ and $\estnaive = \en[(\check \Di)^2]\inv \en[\check \Di Y_i]$ with partialled treatment $\check \Di = (\Di - \propfn) - (\en[\hitilde \hitilde']\inv \en[\hitilde(\Di - \propfn)])'\hitilde$.
Squaring this expression gives
\begin{align*}
(\check \Di)^2 &= (\Di - \propfn)^2 - 2(\Di - \propfn)(\en[\hitilde \hitilde']\inv \en[\hitilde(\Di - \propfn)])'\hitilde \\
&+ ((\en[\hitilde \hitilde']\inv \en[\hitilde(\Di - \propfn)])'\hitilde)^2 \equiv \eta_{i1} + \eta_{i2} + \eta_{i3}. 
\end{align*}
Using $\en[\hitilde(\Di - \propfn)] = \Op(\negrootn)$ by Lemma \lemmastochasticbalance \, of \cite{cytrynbaum2023} and $\en[\hitilde \hitilde'] \convp \var(h) \succ 0$, we see that $\en[\eta_{i2}] = \Op(n \inv)$ and $\en[\eta_{i3}] = \Op(n \inv)$.
Then we have $\en[(\check \Di)^2] = \en[(\Di - \propfn)^2] + \Op(n \inv) = \propfn-\propfn^2 + \Op(n \inv)$.
Then apparently $\estnaive = (\propfn- \propfn^2)\inv \en[\check \Di Y_i] + \Op(n \inv)$.
Now note that 
\begin{align*}
\en[\check \Di Y_i] &= \en[(\Di - \propfn)Y_i] - \en[(\en[\hitilde \hitilde']\inv \en[\hitilde(\Di - \propfn)])'\hitilde Y_i] \\
&= \en[(\Di - \propfn)Y_i] - \en[(\Di - \propfn) \hitilde]'(\en[\hitilde \hitilde']\inv \en[ \hitilde Y_i]). 
\end{align*}
By using Frisch-Waugh to partial out $\Di - \propfn$ from the original regression, we have $\coeffnaive= \en[\bar h_i \bar h_i'] \inv \en[\bar \hi Y_i]$ with $\bar \hi = \hitilde - (\en[(\Di - \propfn)^2] \inv \en[\hitilde (\Di - \propfn)])(\Di - \propfn)$.
Then using $\en[\hitilde(\Di - \propfn)] = \Op(\negrootn)$ again, we have $\en[\bar \hi \bar \hi'] = \en[\hitilde \hitilde'] + \Op(n \inv)$.
Similarly, $\en[\bar \hi Y_i] = \en[\hitilde Y_i] - \est \en[\hitilde(\Di - \propfn)] = \en[\hitilde Y_i] + \Op(\negrootn)$. 
Then the coefficient $\coeffnaive = \en[\hitilde \hitilde']\inv \en[ \hitilde Y_i] + \Op(\negrootn)$.
Then we have shown that
\begin{align*}
\estnaive &= \est - \en\left [\frac{(\Di - \propfn) \hitilde}{\sqrt{\propfn-\propfn^2}} \right ]'(\en[\hitilde \hitilde']\inv \en[ \hitilde Y_i])(\propfn - \propfn^2)^{-1/2} + \Op(n \inv) \\
&= \est - \en\left [\frac{(\Di - \propfn) \hi}{\sqrt{\propfn-\propfn^2}} \right ]' \coeffnaive (\propfn - \propfn^2)^{-1/2} + \Op(n \inv) \\
&= \est - (\coeffnaive / \propconstant)'(\hbarone - \hbarzero)\propconstant + \Op(n \inv). 
\end{align*}
The second line uses that $\en[(\Di - \propfn)c] = 0$ for any constant. 
This shows the claimed representation.
We have $\en[\hitilde \hitilde'] = \var(h) + \op(1)$.
Note also that $\en[\hitilde Y_i(1) \Di] = \propfn \cov(h, Y(1)) + \op(1)$ and $\en[\hitilde Y_i(0) (1-\Di)] = (1-\propfn) \cov(h, Y(0)) + \op(1)$.
Putting this together, we have shown that 
\begin{align*}
\coeffnaive / \propconstant &= \var(h) \inv \cov \left (h,  \ceffn_1 \sqrt{\frac{\propfn}{1-\propfn}} + \ceffn_0 \sqrt{\frac{1-\propfn}{\propfn}} \right) + \op(1) \\
&= \argmin_{\gamma} \var(f - \gamma'h) + \op(1) = \coeffnaivepop + \op(1).
\end{align*}
Then the first claim follows from Theorem \ref{thm:adjusted-efficiency}.
For the efficiency claims, (a) if $\propfn = 1/2$ and $\psi = 1$, then $f = \balancefn$ and $ \coeffnaivepop = \argmin_{\gamma} \var(f - \gamma'h) = \argmin_{\gamma} E[\var(\balancefn - \gamma'h | \psi)]$. 
For (c), if $\psi = 1$ and $\cov(h, \ceffn_1 - \ceffn_0) = 0$, then we have
\begin{align*}
\cov(h, f) - \cov(h, \balancefn) = \cov\left(h, (\ceffn_1 - \ceffn_0) \frac{2 \propfn - 1}{\sqrt{\propfn(1-\propfn)}} \right) = 0.
\end{align*}
By expanding the variance, we have $\argmin_{\gamma} \var(f - \gamma'h) = \argmin_{\gamma}\var(\balancefn - \gamma'h)$.
If (b) holds, then $\ceffn_1 - \ceffn_0 = 0$ and the same conclusion follows.
This finishes the proof. 
\end{proof}

\begin{proof}[Proof of Theorem \ref{thm:semiparam}]
For any $\gamma \in \mr^{\dimh}$, we have $\argmin_{g \in L_2(\psi)} E[(Y(d) - g(\psi) - \gamma'h)^2] = E[Y(d) - \gamma'h | \psi]$ by standard arguments.
Then the coefficients
\begin{align*}
\gamma_d  &= \argmin_{\gamma \in \mr^{\dimh}} E[(Y(d) - \gamma'h - E[Y(d) - \gamma'h | \psi])^2] = \argmin_{\gamma \in \mr^{\dimh}} E[\var(Y(d) - \gamma'h | \psi)] 
\end{align*}
and $g_d(\psi) = E[Y(d) - \gamma_d'h | \psi]$.
Define $f_d(x) = g_d(\psi) + \gamma_d'h$. 
Then the AIPW estimator
\begin{align*}
\estsemiparam &= \en[f_1(X_i) - f_0(X_i)] + \en\left[\frac{\Di (Y_i - f_1(X_i))}{\propfn}\right] - \en\left[\frac{(1-\Di) (Y_i - f_0(X_i))}{1-\propfn}\right] \\
&= \est - \en\left[f_1(X_i) \frac{(\Di - \propfn)}{\propfn} \right] - \en\left[f_0(X_i) \frac{(\Di - \propfn)}{1-\propfn} \right] \\
&= \est - \en\left[(\Di - \propfn) \left ( \frac{f_1(X_i)}{\propfn} + \frac{f_0(X_i)}{1-\propfn} \right) \right] \\
&= \en\left[\frac{\Di - \propfn}{\propfn-\propfn^2} \left(Y_i - (1-\propfn)f_1(X_i) - \propfn f_0(X_i)  \right) \right]. 
\end{align*}
Let $F(x) = (1+\propfn)f_1(x) + \propfn f_0(x)$.
Then by vanilla CLT we have $\rootn(\estsemiparam - \ate) \convwprocess \normal(0, V)$ with $V = \var \left(\frac{\Di - \propfn}{\propfn-\propfn^2} \left(Y_i - F(X_i) \right) \right) \equiv \var(W_i)$ with $W_i = \frac{\Di - \propfn}{\propfn-\propfn^2} \left(Y_i - F(X_i) \right) - \ate$.
By fundamental expansion of the IPW estimator from \cite{cytrynbaum2023}
\begin{align*}
W_i &= \frac{\Di - \propfn}{\propfn-\propfn^2} \left(Y_i - F(X_i) \right) - \ate = \left[ \frac{\Di \residual^1_i}{\propfn} - \frac{(1-\Di)\residual^0_i}{1-\propfn} \right]  \\
&+ [\catefn(X_i) - \ate]  + \left [\frac{\Di - \propfn}{\sqrt{\propfn - \propfn^2}} \left ((\ceffn_1 - f_1) \sqrt{\frac{1-\propfn}{\propfn}} + (\ceffn_0 - f_0) \sqrt{\frac{\propfn}{1-\propfn}} \right) \right]. 
\end{align*}
By the law of total variance and tower law 
\begin{align*}
\var(W) &= \var(E[W | X]) + E[\var(W | X)] \\
&= \var(E[W | X]) + E[\var(E[W | X,D] | X)] + E[\var(W | X,D)]. 
\end{align*}
From the expansion above, $\var(E[W | X]) = \var(\catefn(X) - \ate) = \var(\catefn(X))$.
Next
\begin{align*}
&E[W | X, D] = [\catefn(X_i) - \ate]  + \left [\frac{\Di - \propfn}{\sqrt{\propfn - \propfn^2}} \left ((\ceffn_1 - f_1) \sqrt{\frac{1-\propfn}{\propfn}} + (\ceffn_0 - f_0) \sqrt{\frac{\propfn}{1-\propfn}} \right) \right] \\
&E[\var(E[W | X, D] | X)] = E\left[\left ((\ceffn_1 - f_1) \sqrt{\frac{1-\propfn}{\propfn}} + (\ceffn_0 - f_0) \sqrt{\frac{\propfn}{1-\propfn}} \right)^2\right] 
\end{align*}
Using the definition of $f_d(x)$ gives
\begin{align*}
&E\left[\left ((\ceffn_1 - \gamma_1'h - E[\ceffn_1 - \gamma_1'h|\psi]) \sqrt{\frac{1-\propfn}{\propfn}} + (\ceffn_0 - \gamma_0'h - E[Y(0) - \gamma_0'h|\psi]) \sqrt{\frac{\propfn}{1-\propfn}} \right)^2\right] \\
&= E\left[\var \left ((\ceffn_1 - \gamma_1'h) \sqrt{\frac{1-\propfn}{\propfn}} + (\ceffn_0 - \gamma_0'h) \sqrt{\frac{\propfn}{1-\propfn}} \bigg | \psi \right)\right] \\
&= E\left[\var \left (\balancefn - \left (\gamma_1\sqrt{\frac{1-\propfn}{\propfn}} + \gamma_0 \sqrt{\frac{\propfn}{1-\propfn}} \right)'h \bigg | \psi \right)\right] = \argmin_{\gamma \in \mr^{\dimh}} E[\var(\balancefn - \gamma'h | \psi)]. 
\end{align*}
The final line by characterization of $\gamma_d$ above and linearity of $Z \to \argmin_{\gamma} E[\var(Z - \gamma'h | \psi)]$.
Finally note that
\begin{align*}
\var(W | X, D) &= E\left[ \left (\frac{\Di \residual^1_i}{\propfn} - \frac{(1-\Di)\residual^0_i}{1-\propfn} \right )^2 \bigg | X, D \right] = E\left[ \frac{\Di (\residual^1_i)^2}{\propfn^2} + \frac{(1-\Di)(\residual^0_i)^2}{(1-\propfn)^2}  \bigg | X_i, \Di \right]  \\
&= \frac{\Di \hk_1(X_i)}{\propfn^2} + \frac{(1-\Di)\hk_0(X_i)}{(1-\propfn)^2}.
\end{align*}
Then $E[\var(W | X, D)] = E\left[\frac{\hk_1(X_i)}{\propfn} + \frac{\hk_0(X_i)}{1-\propfn} \right]$. 
Comparing with Equation \ref{equation:variance-augmented} finishes the proof.
\end{proof}

\subsection{Proofs for Section \ref{section:efficiency-tilted}} \label{section:proofs:efficiency-tilted} 

\begin{proof}[Proof of Theorem \ref{thm:efficiency-tilted}]
By Theorem \ref{thm:lin}, the middle term of the asymptotic variance is $E[\var(\balancefn - \olscoeffbalancepop'h | \psi)]$ with $\olscoeffbalancepop = \var(h) \inv \cov(h, \balancefn)$. 
This is the OLS coefficient from the population regression $\balancefn = a + \beta'h + e = a + \alpha'z + \gamma'w + e$ with $E[e(1,w,z)] = 0$ and $h = (w,z)$.
Denote $\tilde \balancefn = \balancefn - E[\balancefn]$ and similarly for $\tilde w, \tilde z$.
By Frisch-Waugh we have $\tilde \balancefn  = \alpha' \tilde z + \gamma' \tilde w + e$.
Let $\check w = \tilde w - (E[\tilde z \tilde z']\inv E[\tilde z \tilde w'])'\tilde z$. 
Then again by Frisch-Waugh the coefficient of interest is $\gamma = E[\check w \check w']\inv E[\check w \balancefn]$. 
Next, we characterize this coefficient. \\ 

\noindent By assumption, $E[w | \psi] = c + \Lambda z$. 
De-meaning both sides gives $E[\tilde w | \psi] = \Lambda \tilde z$.
Write $\tilde u = \tilde w - E[\tilde w | \psi] = \tilde w - \Lambda \tilde z$ with $E[\tilde u | \psi] = 0$.
Then we have
\begin{align*}
E[\tilde z \tilde w'] = E[\tilde z (\tilde w - E[\tilde w | \psi] + E[\tilde w | \psi])'] =  E[\tilde z \tilde u'] + E[\tilde z \tilde z' \Lambda'] = E[\tilde z \tilde z' ]\Lambda'.
\end{align*}
Then $\check w = \tilde w - (E[\tilde z \tilde z']\inv E[\tilde z \tilde z'] \Lambda')'\tilde z = \tilde w - \Lambda \tilde z= \tilde u$.
We have now shown that
\[
\gamma = E[\tilde u \tilde u']\inv E[\tilde u\balancefn] = E[\var(\tilde w | \psi)] \inv E[\cov(\tilde w, \balancefn | \psi)] = E[\var(w | \psi)] \inv E[\cov(w, \balancefn | \psi)].
\]
In particular, the coefficient $\beta = (\alpha, \gamma)$ is optimal 
\begin{align*}
E[\var(\balancefn - \beta'h | \psi)] &= E[\var(\balancefn - \gamma'w | \psi)] = \min_{\tilde \gamma} E[\var(\balancefn - \tilde \gamma'w | \psi)] \\
&= \min_{\tilde \alpha, \tilde \gamma} E[\var(\balancefn - \tilde \alpha'z - \tilde \gamma'w | \psi)] = \min_{\beta} E[\var(\balancefn - \beta'h | \psi)].
\end{align*}
The second equality since $z = z(\psi)$.
This completes the proof.
\end{proof}

\subsection{Proofs for Section \ref{section:generic-efficiency}} \label{section:proofs-generic-efficiency}

\begin{proof}[Proof of Theorem \ref{thm:naive-fe}]
By Frisch-Waugh $\check Y_i = \estnaivefe \check \Di + \coeffnaivefe' \check \hi + e_i$ with $\check \Di = \Di - k \inv \sum_{j \in \group(i)} \Dj = \Di - \propfn$ and $\hicheck = \hi - k \inv \sum_{j \in \group(i)} \hj$.
Applying Frisch-Waugh again, the estimator is $\estnaivefe = \en[(\bar \Di)^2]\inv \en[\bar \Di Y_i]$ with $\bar \Di = (\Di - \propfn) - (\en[\check \hi \check \hi']\inv \en[\check \hi (\Di - \propfn)])'\check \hi$.  
By Lemma \ref{lemma:partialled-lin} we have $\en[\check \hi \check \hi'] \convp \frac{k-1}{k}E[\var(h| \psi)] \succ 0$, so that $\en[\check \hi \check \hi']\inv = \Op(1)$.
By the definition of stratification, $\en[(\Di - \propfn)\one(\group(i) = g)] = 0$ for all $\group$.
Then defining $\bar h_{\group} \equiv k \inv \sum_{j \in \group} \hj$ we may write 
\begin{align*}
\en[(\Di - \propfn)\check \hi]  &= \en \left[(\Di - \propfn)\left(\hi - \sum_{\group}\one(\group(i) = \group) \bar h_{\group} \right)\right] \\
&= \en[(\Di - \propfn)\hi] = \Op(\negrootn).
\end{align*}
The final equality since $E[|h|_2^2] < \infty$ and by Lemma \lemmastochasticbalance \, of \cite{cytrynbaum2023}.
Then apparently $\en[(\Tilde \Di)^2] = \en[(\Di - \propfn)^2] + \Op(n \inv)$ so that $\en[(\Tilde \Di)^2] \inv = (\propfn - \propfn^2)\inv + \Op(n \inv)$. 
Then we have shown that
\begin{align*}
\estnaivefe &= \frac{\en[(\Di - \propfn)Y_i]}{\propfn - \propfn^2} - \frac{\en[\check \hi (\Di - \propfn)]'\en[\check \hi \check \hi']\inv \en[\check \hi Y_i]}{\propfn - \propfn^2} + \Op(n \inv)  \\
&= \est - (\hbarone - \hbarzero)'\en[\check \hi \check \hi']\inv \en[\check \hi Y_i] + \Op(n \inv). 
\end{align*}
By Lemma \ref{lemma:partialled-lin} we have
\begin{align*}
\en[\check \hi Y_i] &= \en[\check \hi \Di Y_i(1)] + \en[\check \hi (1-\Di) Y_i(0)] \\
&= \frac{\propfn (k-1)}{k} E[\cov(h, Y(1) | \psi)] + \frac{(1-\propfn)(k-1)}{k} E[\cov(h, Y(0) | \psi)] + \op(1) \\
&= \frac{(k-1)}{k} E\left [\cov \left (h, \propfn \cdot \ceffn_1(X) + (1-\propfn) \cdot \ceffn_0(X) | \psi \right) \right] + \op(1). 
\end{align*}
Putting this together, we have $\propconstant \inv \en[\check \hi \check \hi']\inv \en[\check \hi Y_i] \convp E[\var(h | \psi)] \inv E[\cov(h, f | \psi)] = \argmin_{\gamma} E[\var(f - \gamma'h | \psi)] $. 
Similar reasoning shows that $\coeffnaivefe = \en[\check \hi \check \hi']\inv \en[\check \hi Y_i] + \Op(\negrootn)$. 
Then we have representation $\estnaivefe = \est - (\propconstant \inv \coeffnaivefe)'(\hbarone - \hbarzero)\propconstant + \op(\negrootn)$.
The efficiency claims follow identically to the reasoning in Theorem \ref{thm:naive}.
This finishes the proof.
\end{proof}

\begin{proof}[Proof of Theorem \ref{thm:partialled-lin} (Part I)]
Consider the regression $Y_i \sim \Di (1, \hicheck) + (1-\Di) (1, \hicheck)$ with $\hicheck = \hi - k\inv \sum_{j \in \group(i)} \hj$.
Denote the OLS coefficients by $(\wh c_1, \wh \alpha_1)$ and $(\wh c_0, \wh \alpha_0)$ respectively.
By Frisch-Waugh, the coefficient $(\wh c_1, \wh \alpha_1)$ is given by the equation $Y_i = \wh c_1 + \wh \alpha_1' \hicheck + e_i$ with $\en[e_i(1, \hicheck) | \Di=1] = 0$. 
By the usual OLS formula $\wh \alpha_1 = \var_n(\hicheck | \Di=1) \inv \cov_n(\hicheck, Y_i | \Di = 1)$.
Observe that by definition of stratification 
\[
P_n(\group(i)=\group | \Di=1) = \frac{P_n(\Di = 1 | \group(i) = \group) P_n(\group(i)=\group)}{P_n(\Di = 1)} = P_n(\group(i)=\group).
\]
This shows that $\en[\en[\hi | \group(i)] | \Di=1] = \en[\en[\hi | \group(i)]] = \en[\hi]$, so that $\en[\hicheck | \Di=1] = \en[\hi | \Di=1] - \en[\hi] = \en[\propfn \inv (\Di - \propfn) \hi] = \Op(\negrootn)$ as above.
Then we have 
\begin{align*}
\var_n(\hicheck | \Di=1) &= \en[\hicheck \hicheck' | \Di=1] - \en[\hicheck | \Di = 1]\en[\hicheck | \Di = 1]' \\
&= \en[\hicheck \hicheck' | \Di=1] + \Op(n \inv).
\end{align*}
Similarly, $\cov_n(\hicheck, Y_i | \Di = 1) = \en[\hicheck Y_i | \Di = 1] + \Op(\negrootn)$.
Then we have 
\begin{align*}
\wh \alpha_1 &= \en[\hicheck \hicheck' | \Di=1] \inv \en[\hicheck Y_i | \Di = 1] + \Op(\negrootn) \\
&= \frac{k-1}{k} \frac{k}{k-1} E[\var(h | \psi)] \inv E[\cov(h, Y(1) | \psi)] + \op(1)
\end{align*}
by Lemma \ref{lemma:partialled-lin}.
Similarly, $\wh \alpha_0 = E[\var(h | \psi)] \inv E[\cov(h, Y(0) | \psi)] + \op(1)$. 
By the usual OLS formula, the constant term $\wh c_1$ has form $\wh c_1 = \en[Y_i | \Di = 1] - \wh \alpha_1'\en[\hicheck | \Di=1]$ and similarly for $\wh c_0$.
By change of variables used in the proof of Theorem \ref{thm:lin}, our estimator
\begin{align*}
\estaug = \wh c_1 - \wh c_0 &= \en[Y_i | \Di = 1] - \en[Y_i | \Di = 0] - \bigg [\wh \alpha_1'\en[\hicheck | \Di=1] - \wh \alpha_0'\en[\hicheck | \Di=0] \bigg] \\
&= \est - \en \left [\frac{\wh \alpha_1'\hi (\Di - \propfn)}{\propfn} + \frac{\wh \alpha_0'\hi (\Di - \propfn)}{1-\propfn} \right] \\
&= \est - \left [\wh \alpha_1 \sqrt{\frac{1-\propfn}{\propfn}} + \wh \alpha_0 \sqrt{\frac{\propfn}{1-\propfn}} \, \right]' \en \left [\frac{\hi (\Di - \propfn)}{\sqrt{\propfn - \propfn^2}} \right].
\end{align*}
Define $\wh \gamma = \wh \alpha_1 \sqrt{\frac{1-\propfn}{\propfn}} + \wh \alpha_0 \sqrt{\frac{\propfn}{1-\propfn}}$. 
Then by work above 
\begin{align*}
\wh \gamma &= E[\var(h | \psi)] \inv E\left[\cov\left(h, \sqrt{\frac{1-\propfn}{\propfn}}Y(1) +  \sqrt{\frac{\propfn}{1-\propfn}} Y(0)| \psi \right)\right] + \op(1) \\
&= E[\var(h | \psi)] \inv E\left[\cov\left(h, \balancefn | \psi \right)\right] + \op(1) = \argmin_{\gamma} E[\var(\balancefn - \gamma'h | \psi)] + \op(1). 
\end{align*}
Then applying Theorem \ref{thm:adjusted-efficiency} completes the proof.
As before, $\wh \alpha_1 = \wh a_1 + \wh a_0$ and $\wh \alpha_0 = \wh a_0$ by change of variables.
\end{proof}

\begin{proof}[Proof of Theorem \ref{thm:partialled-lin} (Part II)]
Next, we analyze the group OLS estimator.
By Theorem \ref{thm:adjusted-efficiency}, it suffices to show that $\coeffgroupols = \var_g(h_g)\inv \cov_g(h_g, y_g) = \propconstant \cdot  E[\var(h | \psi)] \inv E[\cov(h, \balancefn | \psi)] + \op(1)$. 
For the first term, note that $\eg[\hg] = \Op(\negrootn)$ as above, so that $\var(\hg) = \eg[\hg \hg'] - \eg[h_g]\eg[\hg]' = \eg[h_g \hg'] + \Op(n \inv)$. 
Similarly, $\cov_g(h_g, y_g) = \eg[\hg \yg] + \Op(\negrootn)$. 
Applying Lemma \ref{lemma:groupols} to each component of $\hi \hi'$ shows that
\begin{align*}
\eg[\hg \hg'] = \frac{k}{n} \sum_{\group} \left( k \inv \sum_{i \in \group} \frac{\hi (\Di - p)}{\propfn-\propfn^2} \right) \left( k \inv \sum_{i \in \group} \frac{\hi' (\Di - p)}{\propfn-\propfn^2} \right) = \frac{ k E[\var(h | \psi)]}{a(k-a)} + \op(1).
\end{align*}
Using the fundamental expansion of the IPW estimator, we have
\begin{align*}
\eg[y_g h_g] &= \frac{k}{n} \sum_g \left (k \inv \sum_{i \in \group} \frac{\hi (\Di - p)}{\propfn-\propfn^2} \right) \left (k\inv \sum_{i \in \group} \frac{Y_i (\Di - p)}{\propfn-\propfn^2} \right) \\
&= \frac{k}{n} \sum_g \left (k\inv \sum_{i \in \group} \frac{\hi (\Di - p)}{\propfn-\propfn^2} \right) \left (k\inv \sum_{i \in \group} \catefn(X_i) + \frac{\balancefni(\Di - \propfn)}{\sqrt{\propfn - \propfn^2}} + \frac{\Di \residual^1_i}{\propfn} - \frac{(1-\Di)\residual^0_i}{1-\propfn} \right) \\
&\equiv A_n + B_n + C_n.
\end{align*}
First, note that $A_n = \Op(\negrootn)$ and $C_n = \Op(\negrootn)$ by Lemma \ref{lemma:groupols}.
Moreover, we have
\begin{align*}
B_n &= \frac{k}{n} \sum_g \left (k\inv \sum_{i \in \group} \frac{\hi (\Di - p)}{\propfn-\propfn^2} \right) \left ( k \inv \sum_{i \in \group} \frac{\balancefni(\Di - \propfn)}{\sqrt{\propfn - \propfn^2}} \right) \\
&= \frac{k\sqrt{\propfn-\propfn^2}}{a(k-a)} E[\cov(h, \balancefn | \psi)] + \op(1) = \frac{E[\cov(h, \balancefn | \psi)]}{\sqrt{a(k-a)}} + \op(1).      
\end{align*}
Putting this together, by continuous mapping we have
\begin{align*}
\coeffgroupols = \var_g(h_g)\inv \cov_g(h_g, y_g) &= \frac{a(k-a)}{k} \frac{1}{\sqrt{a(k-a)}} E[\var(h | \psi)]\inv E[\cov(h, \balancefn | \psi)] + \op(1) \\
&= \sqrt{\propfn-\propfn^2} E[\var(h | \psi)]\inv E[\cov(h, \balancefn | \psi)] + \op(1).
\end{align*}
Applying Theorem \ref{thm:adjusted-efficiency} completes the proof. 
\end{proof}

\begin{proof}[Proof of Theorem \ref{thm:partialled-lin} (Part III)]
Finally, we analyze the ToM estimator.
From the work in part I of this proof we have 
\begin{align*}
\coefflinpartial &= \var_n(\hicheck | \Di=1)\inv \cov_n(\hicheck, Y_i | \Di=1) \sqrt{\frac{1-\propfn}{\propfn}} \\
&+ \var_n(\hicheck | \Di=0)\inv\cov_n(\hicheck, Y_i | \Di=0) \sqrt{\frac{\propfn}{1-\propfn}} 
\end{align*}
Comparing with Equation \ref{equation:tom_definition}, it suffices to show that $\var_n(\hicheck | \Di=1)\inv \var_n(\hicheck) = \op(1)$ and $\var_n(\hicheck | \Di=0)\inv \var_n(\hicheck) = \op(1)$.
This follows immediately from Lemma \ref{lemma:partialled-lin}.
Applying Theorem \ref{thm:adjusted-efficiency} completes the proof. 
\end{proof}

\begin{proof}[Proof of Theorem \ref{thm:z-estimators}]
First, consider the fixed effects estimator with 
\[
Y_i = \wh c + \estnaivefez \Di + \coeffnaivefe' \check \hi + \coeffnaivefez'\zi + e_{i,1}.
\]
Note that $\tilde \Di = \Di - \propfn$ and $\hicheck - \en[\hicheck] = \hicheck - (\en[\hi] - \en[\en[\hi | \group_i=\group]) = \hicheck$.
By Frisch-Waugh, we may instead study $Y_i = \estnaivefez (\Di - \propfn) + \coeffnaivefe' \check \hi + \coeffnaivefez'\zitilde + e_{i,2}$.
Let $\wicheck = (\hicheck, \zitilde)$ and $\wi = (\hi, \zi)$.
Then by work in Theorem \ref{thm:naive-fe}, $\estnaivefez = \en[(\bar \Di)^2]\inv \en[\bar \Di Y_i]$ with  
\begin{align*}
\bar \Di = (\Di - \propfn) - (\en[\wicheck \wicheck']\inv \en[\wicheck (\Di - \propfn)])'\wicheck.
\end{align*}
Previous work suffices to show that $\en[\wicheck (\Di - \propfn)] = \Op(\negrootn)$. 
Then as before, $\en[(\bar \Di)^2] \inv = (\propfn - \propfn^2) \inv + \Op(n \inv)$.
Then we have 
\begin{align*}
\estnaivefez &= \est - (\propfn-\propfn^2)\inv(\en[\wicheck \wicheck']\inv \en[\wicheck (\Di - \propfn)])'\en[\wicheck Y_i] \\
&=  \est - (\bar w_1 - \bar w_0)'\en[\wicheck \wicheck']\inv \en[\wicheck Y_i]. 
\end{align*}
The second equality uses $\en[\hicheck(\Di - \propfn)] = \en[\hi(\Di - \propfn)]$ and $\en[\zitilde(\Di - \propfn)] = \en[\zi(\Di - \propfn)]$ as noted before.
This shows the claim about estimator representation. 

Next, consider $\coeffnaivefe$.
Define $g_i = (\Di - \propfn, \zitilde)$. 
Let $\bar \hi = \hicheck - (\en[g_i g_i']\inv \en[g_i \hicheck])'g_i$.
Then by Frisch-Waugh $\coeffnaivefe = \en[\bar \hi \bar \hi']\inv \en[\bar \hi Y_i]$.
Consider $\en[\zitilde \hicheck] = \en[\zi \hicheck]$ since $\en[\hicheck] = 0$.
We have $\en[\zi \hicheck] = \op(1)$ by Lemma \ref{lemma:partialled-lin}.
Then by previous work $\en[g_i \hicheck] = \op(1)$.
Then $\en[\bar \hi \bar \hi'] = \en[\hicheck \hicheck'] + \op(1)$.
Similarly, $\en[\bar \hi Y_i] = \en[\hicheck Y_i] + \op(1)$. 
Then by continuous mapping $\coeffnaivefe = \en[\bar \hi \bar \hi']\inv \en[\bar \hi Y_i] = \en[\hicheck \hicheck']\inv \en[\hicheck Y_i] + \op(1)$, the coefficient from the regression without strata variables $\zi$ included shown in Theorem \ref{thm:naive-fe}. 
Consider the coefficient $\coeffnaivefez$ on $z(\psi)$.
Let $q_i = (\Di - \propfn, \hicheck)$ and $\bar \zi = \zitilde - (\en[q_i q_i']\inv \en[q_i \zitilde])'q_i$.
We just showed that $\en[q_i \zitilde] = \op(1)$.
Then by similar reasoning as above and Frisch-Waugh 
\begin{align*}
\coeffnaivefez &= \en[\bar \zi \bar \zi']\inv \en[\bar \zi Y_i] = \en[\zitilde \zitilde']\inv \en[\zitilde Y_i] + \op(1) \\
&= \var(z)\inv \cov(z, \propfn \ceffn_1 + (1-\propfn) \ceffn_0) + \op(1) = \propconstant \var(z)\inv \cov(z, f) + \op(1). 
\end{align*}
Our work so far also shows that $\en[\wicheck \wicheck'] \convp \diag(\en[\hicheck \hicheck'], \en[\zitilde \zitilde'])$.
Then it's easy to see from our expression for $\estnaivefez$ that we may identify $\coeffnaivefez = \wh \alpha_1 + \op(1)$.
This finishes the proof for $\estnaivefez$. 
The proofs for the modified partialled Lin estimator $\estlinpartialz$ and modified ToM estimators are similar and omitted for brevity.

\end{proof}

\subsection{Proofs for Section \ref{section:inference}} \label{section:proofs-inference}

\begin{proof}[Proof of Theorem \ref{thm:inference}]
Define population augmented potential outcomes $Y^b(d) = Y(d) - \propconstant \gamma' h(X)$ for $d \in \{0,1\}$ with outcomes $Y_i^b = Y_i^b(\Di) = Y_i - \propconstant \gamma' \hi$.
The proof of Theorem \ref{thm:adjusted-efficiency} showed that $\estadj = \bar Y^b_1 - \bar Y^b_0 + \op(\negrootn)$.
Define $\varestone^b$, $\varestzero^b$, and $\varestcross^b$ to be the analogues of $\varestone$, $\varestzero$, and $\varestcross$ substituting $Y_i^b$ for $Y_i^a$. 
By applying Theorem \thminference \, of \cite{cytrynbaum2023} to $\est_b \equiv \bar Y^b_1 - \bar Y^b_0$, we have $\varest_b = V + \op(1)$ for variance estimator   
\[
\varest_b = \var_n \left( \frac{(\Di - \propfn) Y_i^b}{\propfn-\propfn^2} \right) - \varestone^b - \varestzero^b - 2 \varestcross^b.
\]
Then it suffices to show the following claim: $\varest - \varest_b = \op(1)$.
We prove a slight generalization, letting $\hi(d)$ possibly have a potential outcomes structure and setting $\hi = \hi(\Di)$. 
The case with $\hi(1)=\hi(0) = \hi$ is a special case. 

We work term by term. 
Define the weights $\dweighti = (\Di - \propfn) / (\propfn - \propfn^2)$.
Then we have $\var_n(\dweighti Y_i^b) - \var_n(\dweighti Y_i^a) = \en[\dweighti^2 (Y_i^b)^2] - \en[\dweighti Y_i^b]^2 - \en[\dweighti^2 (Y_i^a)^2] + \en[\dweighti Y_i^a]^2$.
We have $\en[\dweighti Y_i^a]^2 - \en[\dweighti Y_i^b]^2 = \ate^2 - \ate^2 + \op(1) = \op(1)$ by previous work. 
Next, we have $|\en[\dweighti^2 (Y_i^b)^2] - \en[\dweighti^2 (Y_i^a)^2]| = |\en[\dweighti^2 (Y_i^b - Y_i^a)(Y_i^b + Y_i^a)]| \lesssim \en[(Y_i^b - Y_i^a)^2]^{1/2}\en[(Y_i^b + Y_i^a)^2]^{1/2}$.
It's easy to see that $\en[(Y_i^b + Y_i^a)^2]^{1/2} = \Op(1)$. 
We have $\en[(Y_i^b - Y_i^a)^2] = \propconstant^2 \en[(\gamma'\hi - \wh \gamma' \hi)^2] = \propconstant^2 (\wh \gamma - \gamma)'\en[\hi \hi'](\wh \gamma - \gamma) = \op(1)$.
This shows that $\var_n(\dweighti Y_i^b) - \var_n(\dweighti Y_i^a) = \op(1)$, completing the proof for the first term.

Next consider $\varestone^b - \varestone$.
We may expand
\begin{align*}
\varestone^b - \varestone = n \inv \sum_{\group \in \groupsetnu_n} \frac{1}{a(\group) - 1} \frac{1-\propfn}{\propfn^2}  \sum_{i \not = j \in \group} \Di \Dj (Y_i^a Y_j^a - Y_i^b Y_j^b). 
\end{align*}
Note that $Y_i^a Y_j^a - Y_i^b Y_j^b = (Y_i^a - Y_i^b)Y_j^a + Y_i^b(Y_j^a - Y_j^b) = \propconstant (\wh \gamma - \gamma)'(\hi Y_j^a + Y_i^b \hj)$.  
Then by triangle inequality and Cauchy-Schwarz 
\begin{align*}
|\varestone^b - \varestone| &= \left |\propconstant (\wh \gamma - \gamma)' n \inv \sum_{\group \in \groupsetnu_n} \frac{1}{a(\group) - 1} \frac{1-\propfn}{\propfn^2}  \sum_{i \not = j \in \group} \Di \Dj (\hi Y_j^a + Y_i^b \hj) \right | \\
&\lesssim |\wh \gamma - \gamma|_2 \left (n \inv \sum_{\group \in \groupsetnu_n} \sum_{i \not = j \in \group} |\hi|_2 |Y_j^a| + |Y_i^b| |\hj|_2 \right )
\end{align*}
Observe that 
\[
\sum_{i \not = j \in \group} |\hi|_2 |Y_j^a| \leq (1/2) \sum_{i \not = j \in \group} |\hi|_2^2 + |Y_j^a|^2 = \frac{k-1}{2} \sum_{i \in \group} |\hi|_2^2 + |Y_i^a|^2
\]
Then since $\groupsetnu_n$ is a partition of $[n]$ we have $|\varestone^b - \varestone| \lesssim |\wh \gamma - \gamma|_2 \en[|\hi|_2^2 + |Y_i^a|^2] = \op(1)\Op(1) = \op(1)$. 
Then by symmetry $\varestzero^b - \varestzero = \op(1)$ as well. 
A similar calculation shows that $\varestcross^b - \varestcross = \op(1)$.  
Then we have shown that $\varest_b - \varest = \op(1)$, which completes the proof. 
\end{proof}

\subsection{Proofs of Noncompliance Theorems} \label{section:proofs-noncompliance}

\begin{proof}[Proof of Theorems \ref{thm:adjusted-efficiency-late}, \ref{thm:generic-efficiency-late}, \ref{thm:inference-late}] 
First we show Theorem \ref{thm:adjusted-efficiency-late}.
Define $\est^W(\alpha) = \bar W_1 - \bar W_0 - \alpha'(\hbarone - \hbarzero)\propconstant$ and similary for $\est^D(\alpha)$.  
We claim that $\wh \tau_{adj} = \est^W(\gamma_W) / \est^D(\gamma_D) + \op(\negrootn)$. 
By algebra, we have 
\[
\wh \tau_{adj} - \frac{\est^W(\gamma_W)}{\est^D(\gamma_D)} = \frac{\est^D(\gamma_D)(\wh \gamma_W - \gamma_W)'(\hbarone - \hbarzero)\propconstant + \est^W(\gamma_W)(\gamma_D - \wh \gamma_D)'(\hbarone - \hbarzero)\propconstant}{\est^D(\gamma_D) \est^D(\wh \gamma_D)}
\]
By Theorem \ref{thm:adjusted-efficiency}, $\est^D(\gamma_D), \est^D(\wh \gamma_D) = \ated + \op(1)$ with $\ated > 0$, so the denominator is $\Op(1)$. 
The numerator is $\op(\negrootn)$ since $\est^D(\gamma_D), \est^W(\gamma_W) = \Op(1)$ and $(\wh \gamma_A - \gamma_A)'(\hbarone - \hbarzero)\propconstant = \op(\negrootn)$ for $A=D, W$ by the first line of the proof of Theorem \ref{thm:adjusted-efficiency}.
Next, recall the potential outcomes $Q(z) = W(z) - \atel D(z)$ and define $\gamma_Q = \gamma_W - \atel \gamma_D$.
Then we have
\begin{align*}
\frac{\est^W(\gamma_W)}{\est^D(\gamma_D)} - \atel = \frac{\est^W(\gamma_W) - \atel \est^D(\gamma_D)}{\est^D(\gamma_D)} = \frac{\est^Q(\gamma_Q)}{\est^D(\gamma_D)}.
\end{align*}
The $\ate$-like quantity $E[Q(1)-Q(0)] = 0$ by definition of $\atel$.
Then by Theorem \ref{thm:adjusted-efficiency}, we have $\rootn \est^Q(\gamma_Q) \convwprocess \normal(0, V_Q)$ with variance
\begin{equation} \label{equation:variance-late}
V_Q = \var(\catefn_Q) + E\bigg[\var(\balancefn_Q - h'\gamma_Q| \psi)\bigg] + E\left[\frac{\hk_{1Q}(X)}{\propfn} + \frac{\hk_{0Q}(X)}{1-\propfn}\right].
\end{equation}
The claim now follows by Slutsky since $\est^D(\gamma_D) = E[D(1)-D(0)] + \op(1)$ so that $\rootn(\wh \tau_{adj} - \atel) = \rootn \est^Q(\gamma_Q) / \est^D(\gamma_D) + \op(1) = \rootn \est^Q(\gamma_Q) / E[D(1) - D(0)] + \op(1)$. 

Next, we prove Theorem \ref{thm:generic-efficiency-late}.
By linearity of the balance function (Equation \ref{equation:balance_function}), we have $\balancefn_Q = \balancefn_W - \atel \balancefn_D$.
The optimal coefficient is $\gamma_Q^* = E[\var(h | \psi)] \inv E[\cov(h, \balancefn_Q | \psi)] = E[\var(h | \psi)] \inv (E[\cov(h, \balancefn_W | \psi)] - \atel E[\cov(h, \balancefn_D | \psi)]) = \gamma_W^* - \atel \gamma_D^*$.  
This shows that $\wh \tau_{adj}$ is efficient if and only if $\gamma_W - \atel \gamma_D = \gamma_W^* - \atel \gamma_D^*$. 
In particular, this holds if $\gamma_W = \gamma_W^*$ and $\gamma_D = \gamma_D^*$.
By the estimator representations in Section \ref{section:generic-efficiency}, the estimator $\est_{k}^W = \bar W_1 - \bar W_0 - \wh \gamma_{W, k}'(\hbarone - \hbarzero)\propconstant$ for $\wh \gamma_{W, k} = \gamma_W^* + \op(1)$ for $k \in \{PL, GO, TM\}$, and similarly for $\est_{k}^D$.  
Then $\estadjlatek$ is efficient for each $k \in \{PL, GO, TM\}$.

Finally, we show Theorem \ref{thm:inference-late}.
With $\gamma_Q = \gamma_W - \atel \gamma_D$, define the ``population'' augmented potential outcomes $Q^b(z) = Q(z) - h'\gamma_Q$ and outcomes $Q_i^b = Q_i - \hi' \gamma_Q$.
Let $\varest_{Q}^a$ denote the bracketed term in Equation \ref{thm:inference}, and let $\varest_Q^b$ denote the bracketed term with $Q_i^a$ replaced by the population version $Q_i^b$. 
Note that we showed above that $\rootn (\bar Q^b_1 - \bar Q^b_0) \convwprocess N(0, V_Q)$. 
Then $\varest_Q^b = V_Q + \op(1)$ by Theorem \ref{thm:inference}.
Then it suffices to show that $\varest_Q^b - \varest_{Q}^a = \op(1)$. 
To see this, note that we may write $Q_i^b = W_i - \beta' S_i$ and $Q_i^a = W_i - \wh \beta' S_i$ with $\wh \beta = \beta + \op(1)$ for $\wh \beta = (\estadjlatek, \wh \gamma_Q)$, $\beta = (\atel, \gamma_Q)$ and $S_i = (\Di, \hi)$. 
Then the fact that $\varest_Q^b - \varest_{Q}^a = \op(1)$ for outcomes of this form and $\wh \beta = \beta + \op(1)$ is exactly what we showed in the main claim in the proof of Theorem \ref{thm:inference}.
This finishes the proof.
\end{proof}

\subsection{Technical Lemmas} \label{section:proofs-lemmas}

\begin{lem}[Conditional Convergence] \label{lemma:conditional_markov}
Let $(\filtrationg)_{n \geq 1}$ and $(A_n)_{n\geq1}$ a sequence of $\sigma$-algebras and RV's.
Then the following results hold 
\begin{enumerate}[label={(\roman*)}, itemindent=.5pt, itemsep=.4pt] 
\item $E[|A_n| | \filtrationg] = \op(1) / \Op(1) \implies A_n = \op(1)/\Op(1)$. 
\item $\var(A_n| \filtrationg) = \op(c_n^2) / \Op(c_n^2) \implies A_n - E[A_n | \filtrationg] = \op(c_n)/\Op(c_n)$ for any positive sequence $(c_n)_n$.
\item If $(A_n)_{n\geq1}$ has $A_n \leq \Bar{A} < \infty$ $\filtrationg$-a.s. $\forall n$ and $A_n = \op(1)$ $\implies$ $E[|A_n| | \filtrationg] = \op(1)$. 
\end{enumerate}
\end{lem}

\noindent See Appendix C of \cite{cytrynbaum2023} for the proof.

\begin{lem} \label{lemma:product-inequalities}
Let $(a_i), (b_i), (c_i)$ be positive scalar arrays for $i \in I$ for some index set $I$.
Then we have $\sum_{\substack{i,j,s \in I \\ i \not =j, j \not = s}} a_i b_j c_s \leq 3 \sum_{i \in I} (a_i^3 + b_i^3 + c_i^3)$. 
\end{lem}

\begin{proof}
Note that by AM-GM inequality and Jensen, for non-negative $x,y,z$ we have $xyz \leq ((1/3)(x + y + z))^3 \leq (1/3)(x^3 + y^3 + z^3)$.
Applying this gives 
\begin{align*}
\sum_{\substack{i,j,s \\ i \not =j, j \not = s}} a_i b_j c_s &\leq \left (\sum_i a_i \right) \left (\sum_j b_j \right) \left (\sum_s c_s \right) \\
&\leq (1/3) \left [ \left (\sum_i a_i \right)^3 +  \left (\sum_j b_j \right)^3 + \left (\sum_s c_s \right)^3 \right] \leq 3 \sum_i (a_i^3 + b_i^3 + c_i^3).
\end{align*}
\end{proof}

\begin{lem}[Group OLS] \label{lemma:groupols}
Let $h, w: \xspace \to \mr$. 
Denote $\hi = h(X_i)$ and $\wi = w(X_i)$ and suppose $E[\hi| \psii=\psi]$ and $E[\wi | \psii = \psi]$ are Lipschitz continuous.
Suppose $E[\hi^4] < \infty$ and $E[\wi^4] < \infty$.
Let $\residual_i^d = Y_i(d) - \ceffn_d(X_i)$ for $d \in \{0,1\}$.
Then we have
\begin{align*}
A_n &= n \inv \sum_g \left ( k \inv \sum_{i \in \group} \frac{\hi (\Di - p)}{\propfn-\propfn^2} \right) \left ( k \inv \sum_{i \in \group} \frac{w_i (\Di - p)}{\propfn-\propfn^2} \right) = \frac{E[\cov(h, w | \psi)]}{a(k-a)} + \op(1). \\   
B_n &= n \inv \sum_g \left ( k \inv \sum_{i \in \group} \frac{\hi (\Di - p)}{\propfn-\propfn^2} \right) \left ( k \inv \sum_{i \in \group} w_i \right) = \Op(\negrootn). \\ 
C_n &= \sum_g \left (k\inv \sum_{i \in \group} \frac{\hi (\Di - p)}{\propfn-\propfn^2} \right) \left (k\inv \sum_{i \in \group} \frac{\Di \residual^1_i}{\propfn} - \frac{(1-\Di)\residual^0_i}{1-\propfn} \right) = \Op(\negrootn).
\end{align*}
\end{lem}
\begin{proof}
Define $\hbaronegroup = a \inv \sum_{i \in \group} \hi \one(\Di = 1)$, $\hbarzerogroup = (k-a) \inv \sum_{i \in \group} \hi \one(\Di = 0)$, and $\wbar = k \inv \sum_{i \in \group} \wi$.
Recall that $\group \in \sigma(\psin, \permn)$ for each $\group$ and $\Dn \in \sigma(\psin, \permn, \grouprand)$ for an exogenous variable $\grouprand \indep (\Xn, Y(0)_{1:n}, Y(1)_{1:n})$ used to randomize treatments. 
Notice that $k \inv \sum_{i \in \group} \frac{\hi (\Di - p)}{\propfn-\propfn^2} = \hbaronegroup - \hbarzerogroup$.
First consider $B_n$.
By Lemma \lemmadesignproperties \, of \cite{cytrynbaum2023}, we have $E[B_n | \Xn, \permn] = 0$.
Next, we have
\begin{align*}
E[B_n^2 | \Xn, \permn] &= E\left[n^{-2} \sum_{\group, \group'} \left ( k \inv \sum_{i \in \group} \frac{\hi (\Di - p)}{\propfn-\propfn^2} \right) \left ( k \inv \sum_{i \in \group'} \frac{\hi (\Di - p)}{\propfn-\propfn^2} \right) \wbar \bar{w}_{\group'} \bigg | \Xn, \permn \right ] \\
&= E\left[n^{-2} \sum_{\group} \left ( k \inv \sum_{i \in \group} \frac{\hi (\Di - p)}{\propfn-\propfn^2} \right)^2  \wbar^2  \bigg | \Xn, \permn \right ]. 
\end{align*}
The second equality follows by Lemma \lemmadesignproperties \, of \cite{cytrynbaum2023}, since $\cov(\Di, \Dj | \Xn, \permn) = 0$ if $i,j$ are in different groups. 
We may calculate
\begin{align*}
&E\left[\left ( k \inv \sum_{i \in \group} \frac{\hi (\Di - p)}{\propfn-\propfn^2} \right)^2  \bigg | \Xn, \permn \right ] = \frac{1}{k^2(\propfn-\propfn^2)^2} \sum_{i \in \group} \hi^2 \var(\Di | \Xn, \permn) \\
&+ \frac{1}{k^2(\propfn-\propfn^2)^2}\sum_{i \not = j \in \group} \hi \hj \cov(\Di, \Dj| \Xn, \permn) = \frac{1}{k^2(\propfn-\propfn^2)} \left [\sum_{i \in \group} \hi^2 - (k-1)\inv \sum_{i \not = j \in \group} \hi \hj  \right ].  
\end{align*}
Note that $\sum_{i \not = j \in \group} |\hi \hj| \leq \left (\sum_{i \in \group} |\hi| \right)^2 = k^2 \left (k \inv \sum_{i \in \group} |\hi| \right)^2 \leq k\sum_{i \in \group} |\hi|^2$. 
The final inequality by Jensen.
Then by triangle inequality, a simple calculation gives 
\[
\frac{1}{k^2} \left |\sum_{i \in \group} \hi^2 - (k-1)\inv \sum_{i \not = j \in \group} \hi \hj  \right| \leq \frac{1}{k^2} \frac{2k-1}{k-1} \sum_{i \in \group} \hi^2 \leq 3k^{-2} \sum_{i \in \group} \hi^2.
\]
Then continuing from above
\begin{align*}
E[B_n^2 | \Xn, \permn] &\lesssim k^{-2}n^{-2} \sum_{\group} \left (\sum_{i \in \group} \hi^2 \right) \left (\sum_{i \in \group} \wi \right)^2 \leq \frac{1}{kn^2} \sum_{\group} \left (\sum_{i \in \group} \hi^2 \right) \left (\sum_{i \in \group} \wi^2 \right) \\
&\leq \frac{1}{2kn^2} \sum_{\group} \left [\left (\sum_{i \in \group} \hi^2 \right)^2 + \left(\sum_{i \in \group} \wi^2 \right)^2 \right] = (2n) \inv \en[\hi^4 + \wi^4] = \Op(n\inv). 
\end{align*}
The second inequality follows from Jensen, and the third by Young's inequality.
The first equality by Jensen and final equality by our moment assumption.
Then by Lemma \ref{lemma:conditional_markov}, $B_n = \Op(\negrootn)$. 

Next, consider $A_n$.
Using the within-group covariances above, we compute
\begin{align*}
E[A_n | \Xn, \permn] &= \frac{1}{nk^2(\propfn-\propfn^2)^2} \sum_g \sum_{i, j \in \group}  \cov(\Di, \Dj | \Xn, \permn) \hi \wj \\
&= \frac{1}{nk^2(\propfn-\propfn^2)^2} \sum_{\group} \left (\sum_{i \in \group}  (\propfn-\propfn^2) \hi \wi - \sum_{i \not = j \in \group} \frac{a(k-a)}{k^2(k-1)} \hi \wj \right) \\
&= \frac{1}{k^2(\propfn-\propfn^2)} \left( \en[\hi \wi] - \frac{1}{n(k-1)} \sum_{\group} \sum_{i \not = j \in \group} \hi \wj \right). 
\end{align*}
Define $\reswi = \wi - E[\wi | \psii]$ and $\reshi = \hi - E[\hi | \psii]$.
Consider the second term.
We have
\begin{align*}
n \inv \sum_{\group} \sum_{i \not = j \in \group} \hi \wj = n \inv \sum_{\group} \sum_{i \not = j \in \group} (E[\hi | \psii] + \reshi)(E[\wj | \psij] + \reswj) \equiv \sum_{l=1}^4 A_{n, l}.
\end{align*}
First, note that for any scalars $a_i b_j + a_j b_i = a_i b_i + a_j b_j + (a_i - a_j)(b_j - b_i)$. 
Then we have
\begin{align*}
A_{n, 1} &\equiv n \inv \sum_{\group} \sum_{i \not = j \in \group} E[\hi | \psii] E[\wj | \psij] = n \inv \sum_{\group} \sum_{i < j \in \group} E[\hi | \psii] E[\wj | \psij] + E[\hj | \psij] E[\wi | \psii] \\ 
&= n \inv \sum_{\group} \sum_{i < j \in \group} E[\hi | \psii] E[\wi | \psii] + E[\hj | \psij] E[\wj | \psij] \\
&+ n \inv \sum_{\group} \sum_{i < j \in \group} (E[\hi | \psii] - E[\hj | \psij])(E[\wj | \psij] - E[\wi | \psii]) \equiv B_{n,1} + C_{n,1}. 
\end{align*}
By counting ordered tuples $(i,j)$, it's easy to see that 
\begin{align*}
B_{n, 1} &= n \inv \sum_{\group} \sum_{i \in \group} (k-1) E[\hi | \psii] E[\wi | \psii] = (k-1) \en[E[\hi | \psii] E[\wi | \psii]] \\
&= (k-1) E[E[\hi | \psii] E[\wi | \psii]] + \op(1) = (k-1) (E[\hi \wi] - E[\reshi \reswi]) + \op(1). 
\end{align*}
For the second term, by our Lipschitz assumptions we have $|C_{n,1}| \lesssim n \inv \sum_{\group} \sum_{i < j \in \group} |\psii - \psij|_2^2 = \op(1)$.
Next, claim that $A_{n,l} = \op(1)$ for $l = 2,3,4$.
For instance, we have 
\begin{align*}
E[A_{n,2} | \psin, \permn] = n \inv \sum_{\group} \sum_{i \not = j \in \group} E\left[E[\hi | \psii] \reswj | \psin, \permn \right] = 0. 
\end{align*}
Since $E[\reswj | \psin, \permn] = E[\reswj | \psij] = 0$ by Lemma \lemmaconditionalmoments \, of \cite{cytrynbaum2022local}.
Moreover, we have 
\begin{align*}
E[A^2_{n,2} | \psin, \permn] = n^{-2}  \sum_{\group, \group'} \sum_{i \not = j \in \group} \sum_{s \not = t \in \group'} E[\hi | \psii] E[h_s | \psi_s] E[\reswj u_t | \psin, \permn ]. 
\end{align*}
For $j \not = t$, we have $E[\reswj u_t | \psin, \permn ] = E[\reswj | \psij] E[u_t | \psi_t] = 0$ by Lemma \lemmaconditionalmoments \, of the paper above.
Since the groups $\group$ are disjoint, and using $E[\reswj^2 | \psin, \permn ] = E[\reswj^2 | \psij]$
\begin{align*}
E[A^2_{n,2} | \psin, \permn] &= n^{-2}  \sum_{\group} \sum_{\substack{i,j,s \in \group \\ i \not =j, j \not = s}} E[\hi | \psii] E[h_s | \psi_s] E[\reswj^2 | \psij] \\
&\leq 3 n^{-2} \sum_{\group} \sum_{i \in \group} 2E[\hi | \psii]^3 + E[\reswi^2 | \psii]^3 \\
&= 3n\inv \en[2E[\hi | \psii]^3 + E[\reswi^2 | \psii]^3] = \Op(n\inv).  
\end{align*}
Then we have shown $A_{n,2} = \Op(\negrootn)$ by Lemma \ref{lemma:conditional_markov}. 
The proof for $l = 3,4$ is almost identical. 
Summarizing, the work above has shown that
\begin{align*}
E[A_n | \Xn, \permn] &= \frac{1}{k^2(\propfn-\propfn^2)} \left( \en[\hi \wi] - \frac{1}{k-1} (k-1) (E[\hi \wi] - E[\reshi \reswi]) \right) + \op(1) \\
&= \frac{1}{k^2(\propfn-\propfn^2)} E[\reshi \reswi] + \op(1) = \frac{E[\cov(h, w | \psi)]}{a(k-a)} + \op(1). 
\end{align*}
Next, we claim that $\var(A_n | \Xn, \permn) = \op(1)$. 
Define $\Delta_{h, \group} = k \inv \sum_{i \in \group} \frac{\hi (\Di - p)}{\propfn-\propfn^2}$, then
\begin{align*}
\var(A_n | \Xn, \permn) = n^{-2} \sum_{\group, \group'} \cov \left (\Delta_{h, \group} \Delta_{w, \group},  \Delta_{h, \group'} \Delta_{w, \group'} | \Xn, \permn \right).
\end{align*}
Note that $\Delta_{h, g} \Delta_{w, g} \indep \Delta_{h, \group'} \Delta_{w, \group'} | \Xn, \permn$ for $\group \not = \group'$, since treatment assignments are (conditionally) independent between groups. 
Then the on-diagonal terms are
\begin{align*}
\var(A_n | \Xn, \permn) &= n^{-2} \sum_{\group} \var \left ( \left (k \inv \sum_{i \in \group} \frac{\hi (\Di - p)}{\propfn-\propfn^2} \right) \left (k \inv \sum_{i \in \group} \frac{\wi (\Di - p)}{\propfn-\propfn^2} \right) \bigg | \Xn, \permn \right ) \\
&= n^{-2}k^{-4}(\propfn-\propfn)^{-4} \sum_{\group} \var \left (\sum_{i,j \in \group} \hi \wj (\Di - \propfn)(\Dj - \propfn) \bigg | \Xn, \permn \right ). 
\end{align*}
The inner variance term can be expanded as 
\begin{align*}
\sum_{i,j \in \group} \sum_{s,t \in \group} \hi \wj h_s w_t \cov \left ((\Di - \propfn)(\Dj - \propfn), (D_s - \propfn)(D_t - \propfn) \bigg | \Xn, \permn \right ).
\end{align*}
We have $|\cov ((\Di - \propfn)(\Dj - \propfn), (D_s - \propfn)(D_t - \propfn) | \Xn, \permn )| \leq 2$ since $|(\Di - \propfn)| \leq 1$ for all $i \in [n]$.
Using Lemma \lemmaproductdiagonalization \, in \cite{cytrynbaum2022local}, the previous display is bounded above by $\sum_{i,j \in \group} \sum_{s,t \in \group} |\hi \wj h_s w_t| \cdot 2 \leq 2k^3 \sum_{i \in \group} (\hi^4 + \wi^4)$.
Putting this all together, 
\begin{align*}
\var(A_n | \Xn, \permn) &\leq 2 n^{-2}k^{-4}(\propfn-\propfn)^{-4} k^3 \sum_{\group}\sum_{i \in \group} (\hi^4 + \wi^4) \\
&= 2n^{-1}k^{-1}(\propfn-\propfn)^{-4} \en[\hi^4 + \wi^4] = \Op(n \inv)
\end{align*}
By conditional Markov, this shows that $A_n - E[A_n | \Xn, \permn] = \Op(\negrootn)$.
Then we have shown that $A_n = \frac{E[\cov(h, w | \psi)]}{a(k-a)} + \op(1)$. 

Finally, we consider $C_n$. 
Note that $\group, \Dn \in \sigma(\Xn, \permn, \grouprand)$ and $E[\residual_i^d | \Xn, \permn, \grouprand] = E[\residual_i^d | X_i] = 0$ for $d = 0,1$ by Lemma \lemmaconditionalmoments \, of \cite{cytrynbaum2022local}, so we have $E[C_n | \Xn, \permn, \grouprand] = 0$. 
Next, we claim that $E[C^2_n | \Xn, \permn, \grouprand] = \Op(n \inv)$. 
Note that $C^2_n$ can be written
\begin{align*}
\frac{1}{n^2 k^4} \sum_{\group, \group'} \left ( \sum_{i, j \in \group} \sum_{s, t \in \group'} \frac{\hi (\Di - p)}{\propfn-\propfn^2}  \left (\frac{\Dj \residual^1_j}{\propfn} - \frac{(1-\Dj)\residual^0_j}{1-\propfn} \right) \frac{h_s (D_s - p)}{\propfn-\propfn^2}  \left (\frac{D_t \residual^1_t}{\propfn} - \frac{(1-D_t)\residual^0_t}{1-\propfn} \right) \right). 
\end{align*}
We have $E[\residual^d_j \residual^{d'}_t | \Xn, \permn, \grouprand] = E[\residual^d_j | X_j] E[\residual^{d'}_t | X_t] = 0$ for any $j \not = t$ by Lemma \lemmaconditionalmoments \, of \cite{cytrynbaum2022local}.
By group disjointness, the term $E[C_n^2 | \Xn, \permn, \grouprand]$ simplifies to
\begin{align*}
\frac{1}{n^2 k^4} \sum_{\group} \left ( \sum_{i, j, s \in \group} \frac{\hi (\Di - p)}{\propfn-\propfn^2}  \frac{h_s (D_s - p)}{\propfn-\propfn^2}  E\left[\left (\frac{\Dj \residual^1_j}{\propfn} - \frac{(1-\Dj)\residual^0_j}{1-\propfn} \right)^2 \bigg | \Xn, \permn, \grouprand \right] \right). 
\end{align*}
We have $E[(\residual_i^d)^2 | \Xn, \permn, \grouprand] = E[(\residual_i^d)^2 | X_i] = \hk_d(X_i)$.
Then by Young's inequality and Lemma \lemmaconditionalmoments \, of the paper above 
\begin{align*}
E\left[\left (\frac{\Dj \residual^1_j}{\propfn} - \frac{(1-\Dj)\residual^0_j}{1-\propfn} \right)^2 \bigg | \Xn, \permn, \grouprand \right] \leq 2(\propfn \wedge (1-\propfn))\inv (\hk_1(X_j) + \hk_0(X_j)). 
\end{align*}
Taking the absolute value of the second to last display and using triangle inequality gives the upper bound
\begin{align*}
&2[n^2 k^4 (\propfn-\propfn^2)^2 (\propfn \wedge (1-\propfn))]\inv \sum_{\group} \left ( \sum_{i, j, s \in \group} |\hi h_s| (\hk_1(X_j) + \hk_0(X_j)) \right) \\
\lesssim \, &n^{-2} \sum_{\group} \left ( \sum_{i, j, s \in \group} |\hi h_s|^2 + (\hk_1(X_j) + \hk_0(X_j))^2 \right) \\
\leq \, &n^{-1}k^2 \en[(\hk_1(X_i) + \hk_0(X_i))^2] + n^{-2}k \sum_{\group} \sum_{i, s \in \group} |\hi h_s|^2. 
\end{align*}
By Young's inequality and assumption $E[\en[(\hk_1(X_i) + \hk_0(X_i))^2]] \leq 2 E[\hk_1(X_i)^2 + \hk_0(X_i)^2] < \infty$.
For the second term, using Jensen we have 
\begin{align*}
 n\inv \sum_{\group} \sum_{i, s \in \group} |\hi h_s|^2 = n \inv \sum_{\group} \left (\sum_{i \in \group} |\hi|^2 \right)^2 \leq k n\inv \en[\hi^4] = \Op(1).   
\end{align*}
Then we have shown that $E[C_n^2 | \Xn, \permn, \grouprand] = \Op(n \inv)$, so by conditional Markov inequality in Lemma \ref{lemma:conditional_markov}, $C_n = \Op(\negrootn)$.
This finishes the proof.
\end{proof}

\begin{lem}[Partialled Lin] \label{lemma:partialled-lin}
Under assumptions, $\en[\hicheck \zi] = \op(1)$.
Also, we have 
\begin{align*}
\en[\Di \hicheck \hicheck'] &= \frac{\propfn (k-1)}{k} E[\var(h | \psi)] + \op(1) \quad \en[\hicheck \hicheck'] = \frac{k-1}{k} E[\var(h | \psi)] + \op(1) \\
\en[\Di \hicheck Y_i] &= \frac{\propfn (k-1)}{k} E[\cov(h, \ceffn_1 | \psi)] + \op(1) \\
\en[(1-\Di) \hicheck Y_i] &= \frac{(1-\propfn) (k-1)}{k} E[\cov(h, \ceffn_0 | \psi)] + \op(1).
\end{align*}
\end{lem}

\begin{proof}
First, observe that 
\[
\hicheck = \hi - k\inv \sum_{j \in \group(i)} \hj = \frac{k-1}{k} \cdot \hi - k \inv \sum_{j \in \group(i) \setminus \{i\}} \hj = k \inv \sum_{j \in \group(i) \setminus \{i\}} (\hi - \hj). 
\]
Note that $\en[\Di \hicheck \hicheck] = \en[(\Di - \propfn) \hicheck \hicheck] + \propfn \en[\hicheck \hicheck]$.
We claim that $\en[(\Di - \propfn) \hicheck \hicheck] = \Op(\negrootn)$.
For $1 \leq t,t' \leq \dimh$, by Lemma \lemmastochasticbalance \, of \cite{cytrynbaum2022local} \, and Cauchy-Schwarz we have $\var(\rootn \en[(\Di - \propfn) \hicheckt \hichecktprime] | \Xn, \permn) \leq 2 \en[\hicheckt^2 \hichecktprime^2] \leq 2 \en[\hicheckt^4]^{1/2} \en[\hichecktprime^4]^{1/2}$.
Next, note that by Jensen's followed by Young's inequality 
\begin{align*}
\hicheckt^4 &= \frac{(k-1)^4}{k^4} \left (\frac{1}{k-1} \sum_{j \in \group(i) \setminus \{i\}} (\hit - \hjt) \right)^4 \leq \frac{(k-1)^3}{k^4} \sum_{j \in \group(i) \setminus \{i\}} (\hit - \hjt)^4 \\
&\leq 8\frac{(k-1)^3}{k^4} \sum_{j \in \group(i) \setminus \{i\}} (\hit^4 + \hjt^4) \leq 8\frac{(k-1)^3}{k^4} \left ((k-1)\hit^4 +  \sum_{j \in \group(i) \setminus \{i\}} \hjt^4 \right).
\end{align*}
By counting, we have $\en \left[\sum_{j \in \group(i) \setminus \{i\}} \hjt^4 \right] = (k-1)\en[\hit^4]$.
Putting this all together, $\en[\hicheckt^4] \lesssim \en[\hit^4] = \Op(1)$.
Then $\var(\rootn \en[(\Di - \propfn) \hicheckt \hichecktprime] | \Xn, \permn) = \Op(1)$ so that $\en[(\Di - \propfn) \hicheckt \hichecktprime] = \Op(\negrootn)$ by Lemma \ref{lemma:conditional_markov}. 
Then it suffices to show the claim for $\en[\hicheck \hicheck]$.
Let $f_{it} = E[h_t(X_i) | \psii]$ and write $\hit = f_{it} + u_{it}$. 
Then we have
\begin{align*}
\en[\hicheckt \hichecktprime] = &\frac{1}{nk^2} \sum_i \left (\sum_{j \in \group(i) \setminus \{i\}} \hit - \hjt \right) \left (\sum_{l \in \group(i) \setminus \{i\}} \hitprime - \hltprime \right) \\
= \, &\frac{1}{nk^2} \sum_i \Di \sum_{j,l \in \group(i) \setminus \{i\}} (\hit - \hjt) (\hitprime - \hltprime). 
\end{align*}
We can expand the expression above as
\begin{align*}
&\frac{1}{nk^2} \sum_i \sum_{j,l \in \group(i) \setminus \{i\}} \bigg [(f_{it} - f_{jt})(f_{it'} - f_{lt'}) + (f_{it} - f_{jt})(u_{it'} - u_{lt'}) \\
&+ (u_{it} - u_{jt})(f_{it'} - f_{lt'}) + (u_{it} - u_{jt})(u_{it'} - u_{lt'}) \bigg ] \equiv A_n + B_n + C_n + D_n. 
\end{align*}
First consider $A_n$. 
By the Lipschitz assumption in \ref{assumption:moment-conditions} and Young's inequality 
\begin{align*}
|A_n| &\leq \frac{1}{nk^2} \sum_i \sum_{j,l \in \group \setminus \{i\}} |f_{it} - f_{jt}||f_{it'} - f_{lt'}| \lesssim \frac{1}{nk^2} \sum_i \sum_{j,l \in \group \setminus \{i\}} |\psii - \psij|_2|\psi_i - \psi_l|_2 \\
&\leq \frac{2}{nk^2} \sum_i \sum_{j,l \in \group \setminus \{i\}} (|\psii - \psij|_2^2 + |\psi_i - \psi_l|_2^2) = \frac{4(k-1)}{nk^2} \sum_{\group} \sum_{i,j \in \group} |\psii - \psij|_2^2 = \op(1). 
\end{align*}
The second to last equality by counting and the final equality by Assumption \ref{equation:homogeneity}.
Next consider $B_n$.
Note that each $\group \in \sigma(\psin, \permn)$ and $E[u_{it} | \psin, \permn] = E[u_{it} | \psii] = 0$, so $E[B_n | \psin, \permn] = 0$.
We can rewrite the sum 
\[
\sum_i \sum_{j,l \in \group \setminus \{i\}} (f_{it} - f_{jt})(u_{it'} - u_{lt'}) = \sum_{\group} \sum_{\substack{i, j,l \in \group \\ j,l \not = i}} (f_{it} - f_{jt})(u_{it'} - u_{lt'}).
\]
Then we may compute $\var(\rootn B_n | \psin, \permn) = E[n B_n^2 | \psin, \permn]$ as follows.
By Lemma \lemmaconditionalmoments \, of \cite{cytrynbaum2022local}, $E[u_{it'}u_{jt'} | \psin, \permn] = 0$ for any $\group(i) \not = \group(j)$, so we only consider the diagonal 
\begin{align*}
0 &\leq \frac{1}{nk^4} \sum_{\group} \sum_{\substack{i, j,l \in \group \\ j,l \not = i}} \sum_{\substack{a,b,c \in \group \\ b,c \not = a}}E[(f_{it} - f_{jt})(f_{at} - f_{bt})(u_{it'} - u_{lt'})(u_{at'} - u_{ct'}) | \psin, \permn] \\
&\leq n \inv \sum_{\group} \sum_{\substack{i, j,l \in \group \\ j,l \not = i}} \sum_{\substack{a,b,c \in \group \\ b,c \not = a}}|f_{it} - f_{jt}||f_{at} - f_{bt}| |E[(u_{it'} - u_{lt'})(u_{at'} - u_{ct'}) | \psin, \permn]| \\
&\lesssim n \inv \sum_{\group} \max_{i,j \in \group} |\psii - \psij|_2^2 \sum_{\substack{i, j,l \in \group \\ j,l \not = i}} \sum_{\substack{a,b,c \in \group \\ b,c \not = a}}|E[(u_{it'} - u_{lt'})(u_{at'} - u_{ct'}) | \psin, \permn]|.
\end{align*}
Next, by Lemma \lemmaconditionalmoments \, of \cite{cytrynbaum2022local}, $E[(u_{it'} - u_{lt'})(u_{at'} - u_{ct'}) | \psin, \permn]$ is 
\begin{align*}
&\delta_{ai} E[u_{it'}^2|\psii] - \delta_{la} E[u_{at'}^2|\psi_{a}] - \delta_{ci} E[u_{it'}^2 | \psii] + \delta_{lc} E[u_{lt'}^2 | \psi_l].
\end{align*}
Applying the triangle inequality and summing out using this relation, the above is
\begin{align*}
&\leq \frac{4k(k-1)^3}{n} \sum_{\group} \max_{i,j \in \group} |\psii - \psij|_2^2 \sum_{i \in \group} E[u_{it'}^2| \psii] \\
&\lesssim n \inv \sum_{\group} \left (\max_{i,j \in \group} |\psii - \psij|_2^4 + \sum_{i \in \group} E[u_{it'}^2| \psii]^2 \right) \\
&\leq n \inv \sum_{\group} \diam(\supp(\psi))^2 \sum_{i,j \in \group} |\psii - \psij|_2^2 + \en[E[u_{it'}^2| \psii]^2]. 
\end{align*}
We claim that $E[u_{it'}^4] < \infty$. 
Note that $E[u_{it'}^4] = E[(h_{it'} - f_{it'})^4] \leq 8E[h_{it'}^4] + 8E[f_{it'}^4]$ by Young's inequality.
We have $E[h_{it'}^4] < \infty$ by assumption.
Note that $E[f_{it'}^4] \leq C_f |\psii|^4 \leq C_f \diam(\supp(\psi))^4 < \infty$ by Assumption \ref{assumption:moment-conditions}, with Lipschitz constant $C_f$.
Then $E[u_{it'}^4] < \infty$, so $E[\en[E[u_{it'}^2|\psii]^2]] = E[E[u_{it'}^2|\psii]^2] \leq E[u_{it'}^4] < \infty$.
The inequality follows by conditional Jensen and tower law.
Then $\en[E[u_{it'}^2|\psii]^2 = \Op(1)$ by Markov inequality. 
Then using Assumption \ref{equation:homogeneity} in the display above, we have shown $E[nB_n^2|\psin, \permn] = \Op(1)$ and by Lemma \ref{lemma:conditional_markov} we have shown $B_n = \Op(\negrootn)$.
We have $C_n = \Op(\negrootn)$ by symmetry.
Finally, consider $D_n$. 
By Lemma \lemmaconditionalmoments \, of \cite{cytrynbaum2022local} compute $E[(u_{it} - u_{jt})(u_{it'} - u_{lt'}) | \psin, \permn] = E[u_{it}u_{it'}| \psii] + E[u_{jt} u_{jt'} | \psij] \delta_{jl}$ for $j,l \not = i$.
Then we calculate  
\begin{align*}
&E[D_n | \psin, \permn] = \frac{1}{nk^2} \sum_i \sum_{j,l \in \group(i) \setminus \{i\}}  E[u_{it}u_{it'}| \psii] +  E[u_{jt} u_{jt'} | \psij] \one(j=l) \\
= \, &\frac{1}{nk^2} \sum_i (k-1)^2  E[u_{it}u_{it'}| \psii] + \frac{1}{nk^2} \sum_i \sum_{j \in \group(i) \setminus \{i\}}   E[u_{jt} u_{jt'} | \psij]  \\
= \, &\frac{(k-1)^2}{nk^2} \sum_i E[u_{it}u_{it'}| \psii] + \frac{k-1}{nk^2} \sum_i E[u_{it}u_{it'}| \psii] = \frac{k(k-1)}{nk^2} \sum_i E[u_{it}u_{it'}| \psii]. 
\end{align*}
Now $E[E[u_{it}u_{it'}| \psii]^2] \leq E[u_{it}^2 u_{it'}^2] \leq 2E[u_{it}^4] + 2E[u_{it'}^4] < \infty$ by Jensen, tower law, Young's, and work above.
Then by Chebyshev $\frac{(k-1)}{nk} \sum_i E[u_{it}u_{it'}| \psii] = \frac{k-1}{k} E[u_{it}u_{it'}] + \Op(\negrootn) = \frac{k-1}{k} E[\cov(h_{it}, h_{it'} | \psii)] + \Op(\negrootn)$.
Then we have shown $E[D_n | \psin, \permn] = \frac{k-1}{k} E[\cov(h_{it}, h_{it'} | \psii)] + \Op(\negrootn)$.
Next, we claim that $\var(\rootn D_n | \psin, \permn) = \Op(1)$.
Following the steps above for $B_n$ replacing terms shows that $\var(\rootn D_n | \psin, \permn)$ is 
\begin{align*}
0 &\leq \frac{1}{nk^4} \sum_{\group} \sum_{\substack{i, j,l \in \group \\ j,l \not = i}} \sum_{\substack{a,b,c \in \group \\ b,c \not = a}} \cov((u_{it} - u_{jt})(u_{it'} - u_{lt'}), (u_{at} - u_{bt})(u_{at'} - u_{ct'}) | \psin, \permn).
\end{align*}
For any variables $A,B$, $|\cov(A, B)| \leq |E[AB]| + |E[A]E[B]| \leq 2 |A|_2 |B|_2$ by Cauchy-Schwarz and increasing $L_p(\mathbb P)$ norms.
By Young's inequality, $(a-b)^4 \leq 8(a^4 + b^4)$ for any $a,b \in \mr$.
Then using these facts  
\begin{align*}
&|\cov((u_{it} - u_{jt})(u_{it'} - u_{lt'}), (u_{at} - u_{bt})(u_{at'} - u_{ct'}) | \psin, \permn)| \\
\leq \, &2 E[(u_{it} - u_{jt})^2(u_{it'} - u_{lt'})^2| \psin, \permn]^{1/2} E[(u_{at} - u_{bt})^2(u_{at'} - u_{ct'})^2 | \psin, \permn]^{1/2} \\
\leq \, &4 E[(u_{it} - u_{jt})^2(u_{it'} - u_{lt'})^2| \psin, \permn] + 4E[(u_{at} - u_{bt})^2(u_{at'} - u_{ct'})^2 | \psin, \permn] \\
\leq \, & 2 E[(u_{it} - u_{jt})^4 + (u_{it'} - u_{lt'})^4| \psin, \permn] + 2E[(u_{at} - u_{bt})^4 + (u_{at'} - u_{ct'})^4 | \psin, \permn] \\    
\leq \, & 16(E[u_{it}^4 + u_{jt}^4 + u_{it'}^4 + u_{lt'}^4| \psin, \permn] + E[u_{at}^4 + u_{bt}^4 + u_{at'} + u_{ct'}^4 | \psin, \permn]) \\    
= \, & 16(2 E[u_{it}^4 | \psii] + E[u_{jt}^4 | \psij] +  E[u_{lt'}^4| \psi_l] + 2E[u_{at}^4| \psi_a] + E[u_{bt}^4 | \psi_b] + E[u_{ct'}^4 | \psi_c]). 
\end{align*}
Plugging this bound in above and summing out gives
\begin{align*}
\var(\rootn D_n | \psin, \permn) \leq \frac{32k^5}{nk^4} \sum_{\group} \sum_{i \in \group} E[u_{it}^4 | \psii] \asymp \en[E[u_{it}^4 | \psii]] = \Op(1). 
\end{align*}
The final equality by Markov since $E[u_{it}^4] < \infty$.
Then by conditional Markov \ref{lemma:conditional_markov} we have $D_n = \frac{k-1}{k} E[\cov(h_{it}, h_{it'} | \psii)] + \Op(\negrootn)$.
Since $t,t'$ were arbirary, this shows $\en[\hicheck \hicheck'] = E[\var(h | \psi)] + \op(1)$. \\ 

\noindent Next, consider $\en[\Di \hicheck Y_i] = \en[(\Di - \propfn) \hicheck Y_i(1)] + \propfn \en[\hicheck Y_i(1)]$.
We claim that $\en[(\Di - \propfn) \hicheck Y_i(1)] = \Op(\negrootn)$.
For $1 \leq t \leq \dimh$, by Lemma \lemmastochasticbalance \, of \cite{cytrynbaum2023}, and Cauchy-Schwarz
\begin{align*}
\var(\rootn \en[(\Di - \propfn) \hicheckt Y_i(1)] | \Xn, Y(1)_{1:n}, \permn) \leq 2 \en[\hicheckt^2 Y_i(1)^2] \leq 2 \en[\hicheckt^4]^{1/2} \en[Y_i(1)^4]^{1/2}.
\end{align*}
Note that $\en[Y_i(1)^4] = \Op(1)$ by Markov inequality and Assumption \ref{assumption:moment-conditions} and $\en[\hicheckt^4] = \Op(1)$ was shown above.
Then by Lemma \ref{lemma:conditional_markov} (conditional Markov), this shows the claim.
Then it suffices to analyze $\en[\hicheck Y_i(1)]$. 
Let $g_i = E[Y_i(1) | \psii]$ and $v_i = Y_i(1) - g_i$ with $E[v_i | \psii] = 0$.
Then as above we may expand
\begin{align*}
\en[\hicheck Y_i(1)] &= \frac{1}{nk} \sum_i \left (\sum_{j \in \group(i) \setminus \{i\}} f_{it} - f_{jt} + u_{it} - u_{jt} \right) (g_i + v_i) \\
&= \frac{1}{nk} \sum_i \sum_{j \in \group(i) \setminus \{i\}} (f_{it} - f_{jt})g_i + (f_{it} - f_{jt})v_i + (u_{it} - u_{jt})g_i + (u_{it} - u_{jt})v_i \\
&\equiv H_n + J_n + K_n + L_n.
\end{align*}
First consider $H_n$.
By Assumption \ref{assumption:moment-conditions}, $\psi \to g(\psi)$ is continuous and $\supp(\psi) \sub \bar B(0, K)$ compact, so $\sup_{\psi \in \bar B(0, K)} |g(\psi)| \equiv K' < \infty$ and $|g_i| \leq K'$ a.s.
Then we have
\begin{align*}
|H_n| \lesssim n \inv \sum_i \sum_{j \in \group(i) \setminus \{i\}} |\psii - \psij|_2 |g_i| \lesssim n \inv \sum_{\group} \sum_{i,j \in \group} |\psii - \psij|_2 = \op(1). 
\end{align*}
For the final equality, note that here we have the unsquared norm, different from Assumption \ref{equation:homogeneity}.
Proposition \propbalancinghigher \, of \cite{cytrynbaum2022local} showed that this quantity is also $\op(1)$.
By substituting $z_i$ for $g_i$, which satisfies the same conditions, this also shows that $\en[\zi \hicheck'] = \op(1)$.
The proof that $J_n, K_n = \Op(\negrootn)$ are similar to the terms $B_n, C_n$ above.
Next, consider $L_n$.
We have 
\begin{align*}
E[L_n | \psin, \permn] &= \frac{1}{nk} \sum_i \sum_{j \in \group(i) \setminus \{i\}} E[(u_{it} - u_{jt})v_i | \psin, \permn] \\
&= \frac{1}{nk} \sum_i \sum_{j \in \group(i) \setminus \{i\}} E[u_{it}v_i |\psii] = \frac{k-1}{k} \en[E[u_{it}v_i |\psii]] \\
&= \frac{k-1}{k} E[\cov(h_{it}, Y_i(1)| \psii)] + \Op(\negrootn). 
\end{align*}
The second equality follows since $j \not = i$ and by Lemma \lemmaconditionalmoments \, of \cite{cytrynbaum2022local}.
The third equality by counting.
For the last equality, note that by Jensen, tower law, Young's inequality $E[E[u_{it}v_i |\psii]^2] \leq E[u_{it}^2 v_i^2] \leq (1/2) (E[u_{it}^4] + E[v_i^4])$. 
We showed $E[u_{it}^4] < \infty$ above and a similar proof applies to $v_i$. 
Then the final equality above follows by Chebyshev.
The proof that $\var(L_n | \psin, \permn) = \Op(\negrootn)$ is similar to our analysis of $D_n$ above.
Then we have shown $\en[\Di \hicheck Y_i] = \propfn \frac{k-1}{k} E[\cov(h, Y(1) | \psi)] + \op(1)$.
The conclusion for $\en[(1-\Di)\hicheck Y_i]$ follows by symmetry. 
This finishes the proof.
\end{proof}

\clearpage

\clearpage

\end{document}